\documentclass[aps,prb,reprint,10pt,superscriptaddress,floatfix,nofootinbib]{revtex4-2}
\usepackage[T1]{fontenc}
\usepackage[utf8]{inputenc}
\usepackage{lmodern}
\usepackage{amsmath,amssymb,bm}
\usepackage{graphicx}
\usepackage{booktabs}
\usepackage{placeins}
\usepackage{hyperref}
\usepackage{orcidlink}
\usepackage{microtype}
\graphicspath{{figures/}}
\hypersetup{colorlinks=true,linkcolor=blue,citecolor=blue,urlcolor=blue}
\newcommand{\dd}{\mathrm{d}}
\newcommand{\ee}{\mathrm{e}}
\newcommand{\ii}{\mathrm{i}}

\newcommand{\zhat}{\hat{\bm z}}
\newcommand{\phihat}{\hat{\bm\varphi}}

\newcommand{\eps}{\varepsilon}

\newcommand{\Ham}{H_{\mathrm{AM}}}
\newcommand{\Hti}{H_{\mathrm{TI}}}
\newcommand{\PhiAB}{\Phi_{\mathrm{AB}}}
\newcommand{\phiAB}{\phi_{\mathrm{AB}}}
\newcommand{\mstar}{m^{\ast}}

\begin{document}

\title{Finite-width altermagnetic quantum rings: angular-channel hybridization and Aharonov--Bohm physics}

\author{Edilberto O. Silva\orcidlink{0000-0002-0297-5747}}
\email[Corresponding author: ]{edilberto.silva@ufma.br}
\affiliation{Programa de P\'os-Gradua\c c\~ao em F\'{\i}sica \& Coordena\c c\~ao do Curso de F\'{\i}sica -- Bacharelado, Universidade Federal do Maranh\~ao, 65085-580 S\~ao Lu\'{\i}s, Maranh\~ao, Brazil}

\date{October 2026}

\begin{abstract}
We study a finite-width two-dimensional quantum ring with a quadratic
$d$-wave altermagnetic term, Tan--Inkson confinement, a perpendicular
magnetic field, and independently tunable Aharonov--Bohm (AB) flux.
The altermagnetic interaction mixes $m$ and $m\pm2$ angular channels
while preserving their parity. Circular symmetry renders the crystal
orientation spectrally irrelevant and makes the two spin blocks exactly isospectral. The resulting same-parity anticrossings are resolved at low filling, and multichannel mixing is found near the Fermi level of highly occupied rings. Independent converged calculations for $N_e=1400$ and $1800$ establish the reconstructed Fermi edge and magnetic response in a finite field window and the persistent current over a full AB flux period. The analytic conventional reference separates radial-subband depopulation (dHvA-type cycles) from finer finite-ring shell oscillations; we do not assign a global Fourier spectrum to unverified altermagnetic field scans. The few-electron thermal response is obtained from an exact fixed-particle-number canonical partition function, avoiding finite-system errors of a fixed-mean-number approximation. The circular model supports an equilibrium charge current but no persistent spin current.
\end{abstract}


\maketitle

\section{Introduction}
\label{sec:intro}

Magnetic compensation is attractive for spintronic devices because it suppresses stray fields without necessarily suppressing electronic spin responses. Antiferromagnetic systems have shown electrical switching, N\'eel-order torques, anomalous Hall effects, and magnetoresistive readout \cite{Jungwirth2016NatNano,Baltz2018RMP,Marti2014NatMater,Zelezny2014PRL,Wadley2016Science,Manchon2019RMP,Nakatsuji2015Nature,Nayak2016SciAdv,Kiyohara2016PRA,Nagaosa2010RMP,Sinova2015RMP,Chen2014PRL,Smejkal2017PRL,Smejkal2022NRMHall}. These observations make clear that electronic transport depends on the symmetries of the magnetic structure, not simply on its net moment.

Altermagnets provide a distinct example. Their collinear, compensated spin sublattices are connected by crystal rotations, allowing spin-split nonrelativistic bands despite zero net magnetization. The splitting changes sign across symmetry-related momentum directions and can have $d$-, $g$-, or $i$-wave character. Its symmetry description draws on work on compensated spin-split magnets, spin currents, multipolar order, and spin groups \cite{Naka2019NatCommun,Smejkal2020SciAdv,Ahn2019PRB,Yuan2020PRB,Hayami2020PRB,Ma2021NatCommun,Mazin2021PNAS,Liu2022PRXSpinGroup,Smejkal2022PRXPhase,Smejkal2022PRXLandscape,Mazin2022PRX,GonzalezHernandez2021PRL,Smejkal2022PRXTMR,Shao2021NatCommun,Bhowal2024PRX,McClarty2024PRL,Jaeschke2024PRB,Das2024PRL,Liu2024PRLTwisted,Bai2024AFM}. The leading splitting is nonrelativistic: unlike Rashba or Dresselhaus coupling, it does not require spin--orbit interaction, although such interaction can add further effects.

Experiments have begun to identify the relevant band structures. RuO$_2$ was an early focus of proposed altermagnetic transport, although its bulk magnetic order remains debated \cite{Berlijn2017PRL,Zhu2019PRL,Feng2022NatElectron,Bai2022PRL,Karube2022PRL,Hiraishi2024PRL,Liu2024PRLRuO2,Fedchenko2024SciAdv,Smolyanyuk2024PRB}. Momentum-resolved measurements have provided evidence in MnTe and CrSb, while metallic room-temperature $d$-wave behavior has been reported in $\mathrm{KV_2Se_2O}$ \cite{Krempasky2024Nature,Reimers2024NatCommun,Lee2024PRL,Osumi2024PRB,Zeng2024AdvSci,Ding2024PRL,Jiang2025NPhys}. Other studies address Hall responses and the search for suitable compounds \cite{Reichlova2024NatCommun,Fender2025JACS,Song2025NRM,Jungwirth2025Newton,Jungwirth2026Nature,Jungwirth2026NPhys}. These bulk results raise a question that depends explicitly on device geometry: what becomes of the momentum-dependent splitting when motion is confined?

Confinement replaces the continuous set of momentum directions by a discrete collection of orbital channels. Recent studies have explored altermagnetic models in twisted structures, two-dimensional layers, spin pumps, topological systems, junctions, quantum dots, proximity-coupled devices, and van der Waals structures \cite{He2023PRL,Sheoran2024PRM,Hodt2024PRB,Ma2024PRB,Lv2025SciRep,Soori2026PRM,AbadilloUriel2026arXiv,Zhu2026PRL,Chen2026NPhysTwisted}. In a ring, the momentum direction changes around the orbit, and the magnetic flux through the central opening can be varied separately from a uniform field acting on the electrons. This setting allows us to identify which angular channels are coupled by the altermagnetic term.

The flux response of mesoscopic rings is well established. The Aharonov--Bohm (AB) effect connects an enclosed magnetic flux to a quantum phase, and differentiating the equilibrium many-body energy with respect to that flux gives the persistent current \cite{AharonovBohm1959,Buttiker1983}. Experiments in metallic and semiconductor rings resolved AB oscillations and persistent currents \cite{Webb1985PRL,Levy1990PRL,Chandrasekhar1991PRL,Timp1987PRL,Timp1989PRB,Mailly1993PRL,Liu1993PRB,Liu1994PRB}. The effects of disorder, interactions, spin, geometric phases, and multichannel motion have since been investigated in detail \cite{Loss1990PRL,Stern1992PRL,Aronov1993PRL,Qian1994PRL,Yi1997PRB,Morpurgo1998PRL,Yau2002PRL,Rabaud2001PRL,Bluhm2009PRL,Bleszynski2009Science}. These measurements and models provide an orbital reference against which the effects of altermagnetic coupling can be compared.

In semiconductor devices, confinement and electron number can be controlled more readily. Spectroscopy and transport experiments have resolved discrete ring states, angular-momentum transitions, and few-electron charging \cite{Lorke2000,Fuhrer2001Nature,Warburton2000Nature,Kleemans2007PRL,Tarucha1996PRL,Keyser2003PRL,Pedersen2000PRB}. Calculations using continuum models, density-functional theory, exact diagonalization, and effective one-dimensional descriptions have examined shell rearrangements, interaction effects, edge states, and the crossover between dots and annuli \cite{Viefers2004,Reimann2002RMP,Maksym1996PRB,Koskinen2001PRB,Loss1991PRB,Reimann1999PRB,Reimann2000PRB,Yannouleas2000PRL,Dineykhan1997PRB,Manninen2001PRB,Chakraborty1994PRB,Kusmartsev1994PRB,Viefers2000PRB}. For an annulus in a homogeneous magnetic field, Yerin \emph{et al.} investigated the emergence and disappearance of persistent currents as the system evolves between dot-like and annular regimes, including the multichannel response of a finite-width Corbino geometry \cite{YerinGusyninSharapovVarlamov2021PRB}. Their analysis underlines the role of radial structure even before a spin-dependent anisotropy is introduced. Pereira and co-workers examined how controlled curvature modifies electronic states, magnetization, and persistent currents in quantum dots and finite-width rings \cite{PereiraAndradeSilva2019AnnPhys,PereiraAndradeSilva2021PhysE,PereiraAndradeSilva2022FewBody}. In related ring models, rotation changes the orbital spectrum and its magnetic response \cite{PereiraCunhaSilva2022FewBody,PereiraSilva2023AnnPhys}. A recent calculation by Pereira, Azevedo, and Silva treats the rotating Tan--Inkson ring and follows the Fermi energy, magnetization, and persistent current \cite{PereiraAzevedoSilva2026PhysE}. An overview of the electronic properties of quantum rings is given by Pereira, de Lira, and Silva \cite{PereiraLiraSilva2024Encyclopedia}. These studies show that the orbital response can change substantially through geometry or inertia, without the spin-dependent anisotropy considered here.

Spin interferometry offers a complementary comparison. Studies of Rashba-coupled rings, Aharonov--Casher phases, and spin-dependent persistent currents showed how a momentum-dependent spin interaction can affect interference around a closed orbit \cite{Meijer2002PRB,Frustaglia2004PRB,Molnar2004PRB,Splettstoesser2003PRB,SoumaNikolic2005PRL,Konig2006PRL,Grbic2007PRL}. The coupling studied here differs in two respects: it is even in momentum and conserves $\sigma_z$ in the minimal collinear Hamiltonian. We therefore look for a restructuring of angular-momentum channels, rather than spin precession of the Rashba type.

We use the finite-width Tan--Inkson potential because the orbital problem is exactly solvable when the altermagnetic coupling vanishes \cite{TanInkson1996,TanInkson1999}. Its analytic states give a convenient basis for isolating the new channel couplings. Related work on annular confinement has documented changes in the Fermi edge and the AB- and dHvA-type magnetic responses under controlled curvature and effective-mass anisotropy \cite{PereiraAndradeSilva2021PhysE,LiraPereiraSilva2024PhysE}. Rotation and edge-state studies provide further points of comparison \cite{PereiraSilva2023AnnPhys,LimaAzevedoPereira2023AnnPhys,PereiraAzevedoSilva2026PhysE}. Those models modify orbital motion through geometry, effective mass, or inertia; the $d$-wave term used below instead correlates anisotropy with spin. The distinction matters even when the two spin sectors turn out to have identical spectra in a circular ring.

Our question is how a quadratic $d$-wave altermagnetic term changes the spectrum, magnetization, and equilibrium persistent current of a finite-width ring when the homogeneous field and the AB flux are independently controlled. To our knowledge, this specific combination has not been treated previously. In the circular geometry the perturbation couples $m$ to $m\pm2$, conserving angular parity, while the two spin sectors remain isospectral. We follow the resulting hybridization from isolated low-energy anticrossings to the many-electron Fermi edge. The latter also allows us to distinguish the broad response associated with radial-subband depopulation from the faster flux-scale oscillations. Open-ring conductance is not calculated here; that extension requires contacts and a transport formalism, as in prior studies of geometry-controlled magnetoresistance in two-dimensional rings \cite{LiraSilvaSantangelo2026PhysE}.

\section{Physical model and continuum Hamiltonian}
\label{sec:model}

\subsection{Gate-defined annular confinement in a two-dimensional altermagnet}

We consider an effectively two-dimensional itinerant altermagnetic layer in the $xy$ plane. The mobile carriers belong to the electronic bands of the altermagnet itself; they are not introduced as an independent electron gas placed inside a separate magnetic medium. Lateral electrostatic or structural confinement restricts the low-energy carriers to an annular region centered at the origin. Within an effective-mass description, this confinement is modeled by the Tan--Inkson potential \cite{TanInkson1996}
\begin{equation}
V_{\mathrm{TI}}(r)=\frac{a_1}{r^2}+a_2 r^2-V_0,
\qquad a_1>0,\quad a_2>0.
\label{eq:VTI}
\end{equation}
The minimum follows from
\begin{equation}
\frac{\dd V_{\mathrm{TI}}}{\dd r}
=-\frac{2a_1}{r^3}+2a_2r=0,
\end{equation}
which gives
\begin{equation}
r_0=\left(\frac{a_1}{a_2}\right)^{1/4}.
\label{eq:r0}
\end{equation}
At this radius,
\begin{equation}
V_{\mathrm{TI}}(r_0)=2\sqrt{a_1a_2}-V_0.
\end{equation}
We therefore choose
\begin{equation}
V_0=2\sqrt{a_1a_2},
\label{eq:V0}
\end{equation}
so that the bottom of the annular confinement is at zero energy. The curvature at the minimum is
\begin{equation}
V_{\mathrm{TI}}''(r_0)=8a_2,
\end{equation}
and it is convenient to introduce the radial confinement frequency
\begin{equation}
\omega_0=\sqrt{\frac{8a_2}{\mstar}}.
\label{eq:omega0}
\end{equation}
Near $r=r_0$, the potential therefore has the harmonic form
\begin{equation}
V_{\mathrm{TI}}(r)
=\frac{1}{2}\mstar\omega_0^2(r-r_0)^2
+\mathcal O\!\left[(r-r_0)^3\right].
\end{equation}
The associated oscillator length $\ell_0=\sqrt{\hbar/(\mstar\omega_0)}$ sets the characteristic radial width of the lowest confined states.

A uniform magnetic field is applied perpendicular to the layer,
\begin{equation}
\bm B=B\zhat,
\end{equation}
and an additional thin flux tube carrying flux $\PhiAB$ threads the center of the annulus. For the electronic region outside the flux tube, we use the azimuthal gauge
\begin{equation}
\bm A(r)=A_\varphi(r)\phihat,
\qquad
A_\varphi(r)=\frac{Br}{2}+\frac{\PhiAB}{2\pi r}.
\label{eq:Aphi}
\end{equation}
Throughout the paper, $e>0$ denotes the elementary charge, so the electron charge is $-e$ and the kinetic momentum is
\begin{equation}
\bm\pi=-\ii\hbar\bm\nabla+e\bm A.
\label{eq:pi}
\end{equation}
The orbital Tan--Inkson Hamiltonian is then
\begin{equation}
\Hti=\frac{\bm\pi^2}{2\mstar}+V_{\mathrm{TI}}(r).
\label{eq:HTI}
\end{equation}
The $a_1/r^2$ term suppresses the wave function near the origin, which is consistent with treating the AB flux tube as lying inside the inaccessible central region.

\subsection{Minimal $d$-wave altermagnetic band Hamiltonian}
\label{subsec:AMcart}

Let $(x',y')$ denote the principal crystal axes of the $d$-wave altermagnetic dispersion. We take these axes to be rotated by a fixed angle $\theta_A$ relative to the device axes $(x,y)$. The corresponding kinetic-momentum components are
\begin{align}
\pi_{x'}&=\pi_x\cos\theta_A+\pi_y\sin\theta_A,
\label{eq:pixp}\\
\pi_{y'}&=-\pi_x\sin\theta_A+\pi_y\cos\theta_A.
\label{eq:piyp}
\end{align}
The minimal nonrelativistic $d$-wave altermagnetic term is taken as
\begin{equation}
\Ham=\frac{\beta}{2\mstar}
\left(\pi_{x'}^2-\pi_{y'}^2\right)\sigma_z,
\label{eq:HAMprincipal}
\end{equation}
where the dimensionless parameter $\beta$ measures the strength of the altermagnetic anisotropy. This is the gauge-covariant ring version of the continuum $d$-wave Hamiltonian used in Ref.~\cite{Lv2025SciRep}. In the field-free and unconstrained limit, $\bm A=0$ and $V_{\mathrm{TI}}=0$, the sum of the isotropic kinetic term and Eq.~\eqref{eq:HAMprincipal} reproduces Eq.~(1) of that work after the identifications $B_{\mathrm{Lv}}=\hbar^2/(2\mstar)$, $M=\beta B_{\mathrm{Lv}}$, $\beta_M=\theta_A$, and $\sigma=\pm1$ with the eigenvalues of $\sigma_z$. Thus $\beta=M/B_{\mathrm{Lv}}$ is dimensionless. The present ring problem then follows from the minimal-coupling replacement $\hbar\bm k\rightarrow\bm\pi$ together with the Tan--Inkson confinement. The full single-particle Hamiltonian is
\begin{equation}
H=\Hti+\Ham.
\label{eq:Htotal}
\end{equation}
At this stage we deliberately omit Rashba and Dresselhaus spin--orbit couplings, Zeeman coupling, disorder, impurities, and electron--electron interactions to isolate the consequences of the nonrelativistic altermagnetic term.

The physical meaning of $\beta$ is transparent when $\bm A=0$. Since $[H,\sigma_z]=0$, a spin eigenstate $\sigma_z\chi_s=s\chi_s$, with $s=\pm1$, has the dispersion
\begin{equation}
\eps_s(\bm k)=\frac{\hbar^2}{2\mstar}
\left[(1+s\beta)k_{x'}^2+(1-s\beta)k_{y'}^2\right].
\label{eq:dispersion}
\end{equation}
Thus the two spin sectors have orthogonal anisotropies,
\begin{equation}
m_{x',s}=\frac{\mstar}{1+s\beta},
\qquad
m_{y',s}=\frac{\mstar}{1-s\beta}.
\label{eq:principal_masses}
\end{equation}
Within the quadratic continuum model, $|\beta|<1$ ensures a positive-definite kinetic energy in both principal directions. Equation~\eqref{eq:principal_masses} also makes clear why a circular trajectory is nontrivial: as the local momentum turns around the ring, its direction continually changes relative to the fixed altermagnetic principal axes.

Because $\pi_x$ and $\pi_y$ do not commute in a perpendicular magnetic field, the rotation of Eq.~\eqref{eq:HAMprincipal} must be carried out without replacing operator products by ordinary numbers. Squaring Eqs.~\eqref{eq:pixp} and \eqref{eq:piyp} gives
\begin{align}
\pi_{x'}^2
&=\cos^2\theta_A\,\pi_x^2
+\sin^2\theta_A\,\pi_y^2
+\sin\theta_A\cos\theta_A\{\pi_x,\pi_y\},\\
\pi_{y'}^2
&=\sin^2\theta_A\,\pi_x^2
+\cos^2\theta_A\,\pi_y^2
-\sin\theta_A\cos\theta_A\{\pi_x,\pi_y\},
\end{align}
where $\{A,B\}=AB+BA$. Therefore
\begin{equation}
\pi_{x'}^2-\pi_{y'}^2
=\cos(2\theta_A)(\pi_x^2-\pi_y^2)
+\sin(2\theta_A)\{\pi_x,\pi_y\},
\label{eq:rotated_difference}
\end{equation}
and
\begin{equation}
\Ham=\frac{\beta}{2\mstar}
\left[
\cos(2\theta_A)(\pi_x^2-\pi_y^2)
+\sin(2\theta_A)\{\pi_x,\pi_y\}
\right]\sigma_z.
\label{eq:HAMcartesian}
\end{equation}
This expression is exact and retains the operator ordering required by the magnetic field.

\section{Exact Tan--Inkson basis with magnetic field and AB flux}
\label{sec:taninkson}

Before including altermagnetism, it is useful to solve the $\beta=0$ problem in the same sign conventions that will be used below. Define the flux quantum and dimensionless AB flux by
\begin{equation}
\Phi_0=\frac{h}{e},
\qquad
\phiAB=\frac{\PhiAB}{\Phi_0},
\label{eq:fluxquantum}
\end{equation}
and the signed cyclotron frequency by
\begin{equation}
\omega_c=\frac{eB}{\mstar}.
\label{eq:omegac}
\end{equation}
Using Eq.~\eqref{eq:Aphi}, the orbital Hamiltonian can be written as
\begin{align}
\Hti={}&-\frac{\hbar^2}{2\mstar}
\left(\frac{\partial^2}{\partial r^2}
+\frac{1}{r}\frac{\partial}{\partial r}\right)
+\frac{1}{2\mstar r^2}
\Bigg(-\ii\hbar\frac{\partial}{\partial\varphi}
\notag\\& +\hbar\phiAB+\frac{\mstar\omega_c r^2}{2}\Bigg)^2+\frac{a_1}{r^2}+a_2r^2-V_0.
\label{eq:HTIpolar}
\end{align}
For a conventional angular state
\begin{equation}
\Psi_{nm}^{(0)}(r,\varphi)
=\frac{1}{\sqrt{2\pi}}R_{nm}^{(0)}(r)\ee^{\ii m\varphi},
\qquad m\in\mathbb Z,
\label{eq:TIansatz}
\end{equation}
the radial operator becomes
\begin{align}
H_m^{(0)}=&-\frac{\hbar^2}{2\mstar}
\left(\frac{\dd^2}{\dd r^2}+\frac{1}{r}\frac{\dd}{\dd r}\right)
+\frac{\hbar^2}{2\mstar r^2}(m+\phiAB)^2
\nonumber\\
&+\frac{a_1}{r^2}
+\left(a_2+\frac{\mstar\omega_c^2}{8}\right)r^2
+\frac{\hbar\omega_c}{2}(m+\phiAB)-V_0.
\label{eq:H0m}
\end{align}
Introduce
\begin{equation}
\Omega=\sqrt{\omega_0^2+\omega_c^2},
\label{eq:Omega}
\end{equation}
and
\begin{equation}
\lambda_m=
\sqrt{(m+\phiAB)^2+\frac{2\mstar a_1}{\hbar^2}}.
\label{eq:lambdam}
\end{equation}
The exact energy levels are then
\begin{equation}
E_{nm}^{(0)}
=\hbar\Omega
\left[n+\frac{\lambda_m+1}{2}\right]
+\frac{\hbar\omega_c}{2}(m+\phiAB)-V_0,
\label{eq:TIenergy}
\end{equation}
with $n=0,1,2,\ldots$, which reduces to the Tan--Inkson spectrum in the present notation \cite{TanInkson1996}. Defining
\begin{equation}
\rho=\frac{\mstar\Omega}{2\hbar}r^2,
\label{eq:rho}
\end{equation}
the normalized radial functions may be written as
\begin{equation}
R_{nm}^{(0)}(r)
=\mathcal N_{nm}\,
\rho^{\lambda_m/2}\ee^{-\rho/2}
L_n^{(\lambda_m)}(\rho),
\label{eq:TIradial}
\end{equation}
with
\begin{equation}
\mathcal N_{nm}
=\left[
\frac{\mstar\Omega}{\hbar}
\frac{n!}{\Gamma(n+\lambda_m+1)}
\right]^{1/2},
\label{eq:TInorm}
\end{equation}
so that $\int_0^\infty r\dd r\,|R_{nm}^{(0)}(r)|^2=1$.

Equations~\eqref{eq:TIenergy}--\eqref{eq:TInorm} will serve two purposes. First, they provide an exact reference against which all $\beta\neq0$ results must be tested. Second, the states $|n,m,s\rangle$ form a physically adapted basis in which the altermagnetic Hamiltonian is sparse.

\section{Polar form of the altermagnetic operator}
\label{sec:polar}

\subsection{Covariant angular ladder operators}

Define the kinetic-momentum combinations
\begin{equation}
\pi_{\pm}=\pi_x\pm\ii\pi_y.
\label{eq:pipm}
\end{equation}
They satisfy
\begin{align}
\pi_+^2&=\pi_x^2-\pi_y^2+\ii\{\pi_x,\pi_y\},\\
\pi_-^2&=\pi_x^2-\pi_y^2-\ii\{\pi_x,\pi_y\}.
\end{align}
Substitution into Eq.~\eqref{eq:rotated_difference} gives
\begin{equation}
\pi_{x'}^2-\pi_{y'}^2
=\frac{1}{2}
\left(
\ee^{-2\ii\theta_A}\pi_+^2
+\ee^{2\ii\theta_A}\pi_-^2
\right),
\label{eq:rotated_pipm}
\end{equation}
and hence
\begin{equation}
\Ham=\frac{\beta}{4\mstar}
\left(
\ee^{-2\ii\theta_A}\pi_+^2
+\ee^{2\ii\theta_A}\pi_-^2
\right)\sigma_z.
\label{eq:HAMpipm}
\end{equation}

For the azimuthal gauge of Eq.~\eqref{eq:Aphi}, direct transformation of the Cartesian derivatives gives
\begin{align}
\pi_+
&=-\ii\hbar\ee^{\ii\varphi}
\left(
\frac{\partial}{\partial r}
+\frac{\ii}{r}\frac{\partial}{\partial\varphi}
-\frac{eA_\varphi(r)}{\hbar}
\right),
\label{eq:pipluspolar}\\
\pi_-
&=-\ii\hbar\ee^{-\ii\varphi}
\left(
\frac{\partial}{\partial r}
-\frac{\ii}{r}\frac{\partial}{\partial\varphi}
+\frac{eA_\varphi(r)}{\hbar}
\right).
\label{eq:piminuspolar}
\end{align}
It is useful to define
\begin{equation}
\eta(r)=\frac{erA_\varphi(r)}{\hbar}
=\phiAB+\frac{\mstar\omega_c}{2\hbar}r^2
\label{eq:eta}
\end{equation}
and the first-order radial operators
\begin{align}
\mathcal D_m^{(+)}
&=\frac{\dd}{\dd r}-\frac{m+\eta(r)}{r},
\label{eq:Dplus}\\
\mathcal D_m^{(-)}
&=\frac{\dd}{\dd r}+\frac{m+\eta(r)}{r}.
\label{eq:Dminus}
\end{align}
Their action on an angular channel is
\begin{align}
\pi_+\left[R_m(r)\ee^{\ii m\varphi}\right]
&=-\ii\hbar\ee^{\ii(m+1)\varphi}
\mathcal D_m^{(+)}R_m(r),
\label{eq:piplusaction}\\
\pi_-\left[R_m(r)\ee^{\ii m\varphi}\right]
&=-\ii\hbar\ee^{\ii(m-1)\varphi}
\mathcal D_m^{(-)}R_m(r).
\label{eq:piminusaction}
\end{align}
Applying the operators once more yields
\begin{align}
\pi_+^2\left[R_m\ee^{\ii m\varphi}\right]
&=-\hbar^2\ee^{\ii(m+2)\varphi}
\mathcal D_{m+1}^{(+)}\mathcal D_m^{(+)}R_m,
\label{eq:piplus2action}\\
\pi_-^2\left[R_m\ee^{\ii m\varphi}\right]
&=-\hbar^2\ee^{\ii(m-2)\varphi}
\mathcal D_{m-1}^{(-)}\mathcal D_m^{(-)}R_m.
\label{eq:piminus2action}
\end{align}
The angular change is therefore not an assumption but an immediate consequence of the covariant differential operators:
\begin{equation}
\Ham:\qquad m\longrightarrow m\pm2.
\label{eq:selection}
\end{equation}
The result remains valid for arbitrary $B$, $\PhiAB$, and $\theta_A$.

\subsection{Coupled radial equations}

Because $[H,\sigma_z]=0$ in the minimal model, the spin label $s=\pm1$ remains a good quantum number. The appropriate wave function is not a single angular harmonic but the expansion
\begin{equation}
\Psi_s(r,\varphi)
=\frac{1}{\sqrt{2\pi}}
\sum_{m=-\infty}^{\infty}
R_{m,s}(r)\ee^{\ii m\varphi}\chi_s,
\qquad
\sigma_z\chi_s=s\chi_s.
\label{eq:generalexpansion}
\end{equation}
Substitution of Eqs.~\eqref{eq:piplus2action} and \eqref{eq:piminus2action} into the Schr\"odinger equation gives, for each angular channel $m$,
\begin{align}
0={}&\left(H_m^{(0)}-E\right)R_{m,s}
\nonumber\\
&-\frac{s\beta\hbar^2}{4\mstar}
\ee^{-2\ii\theta_A}
\mathcal D_{m-1}^{(+)}\mathcal D_{m-2}^{(+)}R_{m-2,s}
\nonumber\\
&-\frac{s\beta\hbar^2}{4\mstar}
\ee^{2\ii\theta_A}
\mathcal D_{m+1}^{(-)}\mathcal D_{m+2}^{(-)}R_{m+2,s}.
\label{eq:coupledradial}
\end{align}
Equation~\eqref{eq:coupledradial} is the central differential system of the finite-width altermagnetic ring. In the limit $\beta\to0$, all off-diagonal angular couplings vanish and each equation reduces independently to the exactly solvable Tan--Inkson radial problem of Sec.~\ref{sec:taninkson}.

The Hamiltonian is Hermitian on the physical domain of normalizable annular states. This is transparent from the Cartesian form, where $\pi_x$ and $\pi_y$ are Hermitian covariant momenta and $\pi_{x'}^2-\pi_{y'}^2$ is Hermitian. Equivalently, with the planar inner product $\int_0^{2\pi}\dd\varphi\int_0^\infty r\dd r$, the operators obey $\pi_+^\dagger=\pi_-$ when boundary terms vanish. Therefore, the two terms in Eq.~\eqref{eq:HAMpipm} are mutual Hermitian conjugates, including their phases $\ee^{\mp2\ii\theta_A}$.

\section{Symmetry structure of the circular altermagnetic ring}
\label{sec:symmetries}

\subsection{Angular parity}

The $\Delta m=\pm2$ rule implies that the angular Hilbert space splits into two disconnected chains,
\begin{align}
\mathcal H_{\mathrm e}&=\{\ldots,-4,-2,0,2,4,\ldots\},\\
\mathcal H_{\mathrm o}&=\{\ldots,-3,-1,1,3,\ldots\}.
\end{align}
This separation is the representation of an exact twofold rotational symmetry. Defining
\begin{equation}
\mathcal P_\pi=\exp\left(-\frac{\ii\pi L_z}{\hbar}\right),
\label{eq:Ppi}
\end{equation}
we have
\begin{equation}
\mathcal P_\pi\ee^{\ii m\varphi}=(-1)^m\ee^{\ii m\varphi},
\end{equation}
and
\begin{equation}
[H,\mathcal P_\pi]=0.
\label{eq:paritycommutator}
\end{equation}
Thus $m$ itself ceases to be a good quantum number for $\beta\neq0$, but $m\bmod2$ remains conserved. Crossings between states belonging to opposite angular-parity sectors are symmetry protected, while states in the same sector can hybridize.

A useful perturbative consequence follows immediately. For an isolated unperturbed Tan--Inkson state $|n,m,s\rangle$,
\begin{equation}
\langle n,m,s|\Ham|n,m,s\rangle=0,
\label{eq:firstorderzero}
\end{equation}
because the altermagnetic operator changes $m$ by two units. Generic nondegenerate energy shifts therefore begin at second order in $\beta$. In contrast, whenever two unperturbed levels with $\Delta m=\pm2$ become degenerate or nearly degenerate as $B$ or $\PhiAB$ is varied, the same matrix element acts already at first order within the degenerate subspace and opens an avoided crossing linear in $|\beta|$. This identifies the microscopic mechanism by which the $d$-wave order reconstructs the AB spectrum.

\subsection{Why the crystal orientation is not spectral in a perfect circular ring}
\label{subsec:theta}

Although $\theta_A$ appears explicitly in Eqs.~\eqref{eq:HAMcartesian} and \eqref{eq:HAMpipm}, it cannot affect the energy spectrum of a perfectly circular closed ring. Let
\begin{equation}
U(\theta_A)=\exp\left(-\frac{\ii\theta_A L_z}{\hbar}\right).
\label{eq:Utheta}
\end{equation}
The Tan--Inkson Hamiltonian, including the rotationally symmetric gauge of Eq.~\eqref{eq:Aphi}, satisfies
\begin{equation}
[\Hti,U(\theta_A)]=0.
\end{equation}
Moreover,
\begin{equation}
U(\theta_A)\pi_\pm U^\dagger(\theta_A)
=\ee^{\mp\ii\theta_A}\pi_\pm.
\end{equation}
Consequently,
\begin{equation}
H(\theta_A)
=U(\theta_A)H(0)U^\dagger(\theta_A).
\label{eq:thetaunitary}
\end{equation}
All values of $\theta_A$ are therefore unitarily equivalent and have identical spectra. The angle becomes observable only when another element of the device fixes a spatial direction, as occurs for an elliptic confinement, crystallographically anisotropic boundary, impurity configuration, or attached leads. For the circular closed ring considered here, we may set $\theta_A=0$ in numerical calculations without loss of generality after retaining the general derivation above.

\subsection{Exact spin degeneracy of the minimal circular model}
\label{subsec:spindeg}

A second unitary relation has an important physical consequence. In a fixed spin sector,
\begin{equation}
H_s=\Hti+s\frac{\beta}{2\mstar}
(\pi_{x'}^2-\pi_{y'}^2).
\label{eq:Hs}
\end{equation}
A quarter-turn orbital rotation changes the sign of the $d$-wave form factor,
\begin{equation}
U\left(\frac{\pi}{2}\right)
(\pi_{x'}^2-\pi_{y'}^2)
U^\dagger\left(\frac{\pi}{2}\right)
=-(\pi_{x'}^2-\pi_{y'}^2),
\end{equation}
while leaving $\Hti$ invariant. Therefore
\begin{equation}
H_{-s}=U\left(\frac{\pi}{2}\right)
H_s
U^\dagger\left(\frac{\pi}{2}\right).
\label{eq:spinunitary}
\end{equation}
The two spin blocks are exactly isospectral,
\begin{equation}
E_{\nu,+}=E_{\nu,-},
\label{eq:spindeg}
\end{equation}
within the present circular model, even when the orbital magnetic field and AB flux are nonzero. The paired states need not have identical spatial probability distributions; rather, their orbital textures are related by a rotation of $\pi/2$. If $U(\alpha)|n,m\rangle=e^{-im\alpha}|n,m\rangle$ is our rotation convention, paired eigenvectors can be chosen such that
\begin{equation}
C_{nm}^{(\nu,-s)}=e^{-im\pi/2}C_{nm}^{(\nu,s)},
\qquad W_m^{(\nu,-s)}=W_m^{(\nu,s)}.
\label{eq:spin_weights_relation}
\end{equation}
The phase convention may be reversed with the convention for $U$, but equality of the channel weights is independent of that choice. Within accidentally degenerate subspaces, the relation applies to suitably paired bases, rather than to arbitrary numerical eigenvectors.

Equation~\eqref{eq:spindeg} sharply distinguishes altermagnetic channel hybridization from a Zeeman-like splitting. In thermal equilibrium with a common chemical potential for the two spin sectors, it also implies equal spin-resolved free energies and hence vanishing persistent spin current in the ideal circular model. A finite equilibrium spin imbalance or spin persistent current therefore requires an additional term that breaks the unitary equivalence in Eq.~\eqref{eq:spinunitary}, such as geometric anisotropy, Zeeman coupling, spin--orbit coupling, or spin-dependent contacts. This conclusion will provide a useful symmetry benchmark for the numerical calculations.

\subsection{What remains specifically altermagnetic in the circular ring}
\label{subsec:what_remains}

The circular geometry makes $\theta_A$ absent from the spectrum [Eq.~\eqref{eq:thetaunitary}] and forces the two spin blocks to be isospectral [Eq.~\eqref{eq:spindeg}]. We must therefore distinguish this model from a ring with a spin-independent $d$-wave orbital perturbation.

The $d$-wave form factor fixes the selection rule $\Delta m=\pm2$ and separates the even and odd angular-parity sectors; neither property follows from an isotropic term. The $\sigma_z$ factor adds a spin-dependent structure to this orbital mixing. Equation~\eqref{eq:principal_masses} assigns orthogonal mass anisotropies to the two spin sectors, whose spatial orbital textures are related by a quarter-turn rotation. This equivalence fixes their energies and angular-channel probabilities, but not the relative phases of the angular coefficients. The textures are related, not identical in a fixed device frame.

The perfectly circular ring is a symmetry reference, not a model of spin-polarized transport. Equilibrium observables determined solely by its spectrum cannot distinguish it from a spin-degenerate orbital model with the same energies. What identifies the altermagnetic term is the relation between spin and the orientation of its momentum-space anisotropy. Elliptic boundaries or contacts may lift the quarter-turn equivalence, but symmetry breaking alone does not establish a measurable spin splitting or polarization. Such responses require the explicit perturbation and its matrix elements. The couplings calculated here set the corresponding orbital scale.

\section{Tan--Inkson basis representation and numerical strategy}
\label{sec:basis}

We expand an eigenstate in the exact $\beta=0$ basis,
\begin{equation}
|\Psi_s\rangle
=\sum_{n=0}^{\infty}\sum_{m=-\infty}^{\infty}
C_{nm}^{(s)}|n,m,s\rangle.
\label{eq:basisexpansion}
\end{equation}
The diagonal matrix elements are
\begin{equation}
\langle n',m',s|\Hti|n,m,s\rangle
=E_{nm}^{(0)}\delta_{nn'}\delta_{mm'}.
\label{eq:HTImatrix}
\end{equation}
The altermagnetic matrix elements vanish unless $m'=m\pm2$,
\begin{equation}
\langle n',m',s|\Ham|n,m,s\rangle=0,
\qquad m'\neq m\pm2.
\label{eq:sparse_selection}
\end{equation}
For $m'=m+2$,
\begin{align}
&\langle n',m+2,s|\Ham|n,m,s\rangle
=-\frac{s\beta\hbar^2}{4\mstar}
\ee^{-2\ii\theta_A}
\nonumber\\
&\times\int_0^\infty r\dd r\,
R_{n',m+2}^{(0)}(r)
\mathcal D_{m+1}^{(+)}\mathcal D_m^{(+)}
R_{nm}^{(0)}(r),
\label{eq:Mplus}
\end{align}
where the exact Tan--Inkson radial functions may be chosen real. Similarly,
\begin{align}
&\langle n',m-2,s|\Ham|n,m,s\rangle
=-\frac{s\beta\hbar^2}{4\mstar}
\ee^{2\ii\theta_A}
\nonumber\\
&\times\int_0^\infty r\dd r\,
R_{n',m-2}^{(0)}(r)
\mathcal D_{m-1}^{(-)}\mathcal D_m^{(-)}
R_{nm}^{(0)}(r).
\label{eq:Mminus}
\end{align}
Hermiticity requires the corresponding matrix elements to be mutual conjugates, providing a direct numerical diagnostic.

The selection rule makes the Hamiltonian sparse and allows the even and odd angular sectors to be diagonalized independently. We truncate the basis at $0\le n\le n_{\max}$ and $|m|\le m_{\max}$ and test the convergence of the resulting spectra and observables with respect to both cutoffs. As an independent benchmark, we also solve selected cases using a two-dimensional real-space discretization. Agreement between the two methods helps distinguish physical anticrossings from truncation artifacts.

To resolve the orbital reconstruction we calculate the eigenenergies $E_\nu(B,\PhiAB,\beta)$ and the angular-channel weights
\begin{equation}
W_m^{(\nu,s)}=\sum_n\left|C_{nm}^{(\nu,s)}\right|^2,
\label{eq:weights}
\end{equation}
and real-space probability densities $|\Psi_{\nu,s}(r,\varphi)|^2$. Once the converged spectrum is established, the equilibrium grand potential at fixed chemical potential may be written as
\begin{equation}
\Omega_{\mathrm G}
=-k_{\mathrm B}T\sum_\nu
\ln\left[1+\ee^{-(E_\nu-\mu)/(k_{\mathrm B}T)}\right],
\label{eq:grandpotential}
\end{equation}
where the index $\nu$ in Eq.~\eqref{eq:grandpotential} includes both spin states. The corresponding derivatives are
\begin{equation}
M=-\frac{\partial\Omega_{\mathrm G}}{\partial B},
\qquad
I_{\mathrm c}=-\frac{\partial\Omega_{\mathrm G}}{\partial\PhiAB}.
\label{eq:MI}
\end{equation}
The calculations reported below use a fixed particle number instead. Section~\ref{sec:thermodynamics} therefore works with the Helmholtz free energy, for which the same thermodynamic derivatives are taken at fixed $N_e$.
For the ideal circular Hamiltonian of Eq.~\eqref{eq:Htotal}, Eq.~\eqref{eq:spindeg} predicts equal spin-resolved contributions and therefore zero equilibrium persistent spin current. This exact result will serve as a stringent internal check before any symmetry-breaking extensions are introduced.

\section{Numerical spectrum and independent validation}
\label{sec:numerics}

\subsection{Dimensionless formulation and convergence}
\label{subsec:dimensionless}

For the numerical calculation, it is convenient to measure length in units of the radial oscillator length
\begin{equation}
\ell_0=\sqrt{\frac{\hbar}{\mstar\omega_0}},
\end{equation}
and energy in units of $\hbar\omega_0$. We introduce
\begin{equation}
x=\frac{r}{\ell_0},
\qquad
b=\frac{\omega_c}{\omega_0},
\qquad
\kappa=\frac{r_0^2}{2\ell_0^2}
=\sqrt{\frac{2\mstar a_1}{\hbar^2}}.
\label{eq:dimensionless_parameters}
\end{equation}
The exact Tan--Inkson energies of Eq.~\eqref{eq:TIenergy} then become
\begin{align}
\eps_{nm}^{(0)}
\equiv\frac{E_{nm}^{(0)}}{\hbar\omega_0}
={}&\sqrt{1+b^2}\left(n+\frac{1}{2}\right)
\notag\\
&+\frac{\sqrt{1+b^2}}{2}
\sqrt{(m+\phiAB)^2+\kappa^2}
\notag\\
&+\frac{b}{2}(m+\phiAB)-\frac{\kappa}{2}.
\label{eq:dimensionless_TI_energy}
\end{align}
Thus $\kappa$ controls the ring radius relative to its radial width. Unless stated otherwise, the calculations below use $\kappa=4.5$, corresponding to $r_0=3\ell_0$. This choice represents a finite-width annulus while keeping the radial and angular level spacings simultaneously visible.

Although the calculation is carried out entirely in these dimensionless variables, it is useful to fix the physical scales once. The radial quantum is $\hbar\omega_0=\hbar^2/(\mstar\ell_0^2)$. For an effective mass $\mstar=0.5\,m_e$, representative of a $d$-band conductor, and $\ell_0=10$~nm---which corresponds to $r_0=100$~nm at $\kappa=50$ and $r_0=30$~nm at $\kappa=4.5$---one obtains $\hbar\omega_0\simeq1.5$~meV, or about $18$~K. On this scale the low-energy $m=-1\leftrightarrow+1$ gap of Sec.~\ref{subsec:ABspectrum} is $\simeq90\,\mu$eV ($\simeq1.1$~K), the near-Fermi couplings of Sec.~\ref{subsec:nearEFmultilevel} are $\simeq0.4$--$0.5$~meV ($\simeq5$--$6$~K), and the reduced temperatures $\tau=0.01$--$0.03$ used below correspond to $0.2$--$0.5$~K, a subkelvin regime relevant to measurements of mesoscopic persistent currents.

The coupling $\beta$ also admits a useful, but model-dependent, interpretation. From Eq.~\eqref{eq:dispersion}, the spin splitting along a principal axis is $\Delta E=2\beta E_{\rm kin}$, where $E_{\rm kin}$ denotes the isotropic kinetic contribution at the same momentum, measured from the effective band minimum. Thus $\beta=\Delta E/(2E_{\rm kin})$ within the quadratic $d$-wave model. Momentum-resolved measurements on CrSb report spin splittings reaching $0.6$~eV \cite{Reimers2024NatCommun} and $0.93$~eV \cite{Ding2024PRL}. CrSb, however, has a different ($g$-wave) symmetry, and neither its binding energy nor its measured splitting can be identified without further band analysis with the $E_{\rm kin}$ entering our effective Hamiltonian. For illustration only, inserting a hypothetical kinetic scale of $1.2$~eV would give $\beta\simeq0.25$--$0.39$; these numbers are not a material-specific determination of $\beta$ for a $d$-wave ring. The $d$-wave metal $\mathrm{KV_2Se_2O}$ provides a more directly relevant symmetry example \cite{Jiang2025NPhys}, but a quantitative mapping to the present effective mass and coupling remains to be established. Accordingly, $\beta=0.20$ is used as a representative strong-coupling benchmark, not as an experimentally extracted or universally conservative parameter. The smaller values $\beta=0.002$--$0.05$ examined in Sec.~\ref{subsec:nearEFmultilevel} probe the weakly coupled regime.

The matrix elements in Eqs.~\eqref{eq:Mplus} and \eqref{eq:Mminus} are evaluated by Gauss--Legendre quadrature in the dimensionless radial coordinate. For the representative low-energy tests in this subsection, spectra were calculated with $0\le n\le10$, $|m|\le16$, 360 radial nodes, and a radial integration interval $0<x<13$. These cutoffs describe the low-energy benchmark only, not the wide-annulus, many-electron calculations in Secs.~\ref{subsec:nearEFmultilevel} and~\ref{subsec:highfill_flux}. Those calculations involve much larger negative angular quantum numbers at finite field. Increasing the low-energy quadrature to 520 nodes and $x_{\max}=15$ changes the first six eigenvalues by less than $10^{-14}\hbar\omega_0$ for the representative parameter set considered below. We independently reconstructed the high-filling matrix problem with analytic Laguerre derivatives, Gauss--Legendre quadrature, and a real symmetric banded eigensolver. Its separately documented convergence tests and limitations are reported in Sec.~\ref{subsec:independent_highfill_validation}. The original figure-data archive does not, by itself, provide the production solver and therefore cannot be regarded as an independent convergence certificate for every point in the many-electron sweeps.

Table~\ref{tab:convergence} quantifies the basis convergence at the generic point $\beta=0.20$, $\phiAB=0.23$, $b=0.25$, $\theta_A=0.31$, and $s=+1$. The reference calculation uses $n_{\max}=14$ and $m_{\max}=18$. The quantity $\delta_{10}$ is the maximum absolute change among the ten lowest eigenvalues, while $\delta_0$ denotes the ground-state change. Angular convergence is already essentially saturated at $m_{\max}=12$, whereas radial convergence is slower; this difference is taken into account in all quantitative results reported below.

\begin{table}[t]
\caption{Convergence of the Tan--Inkson basis at $\beta=0.20$, $\phiAB=0.23$, $b=0.25$, $\theta_A=0.31$, and $s=+1$. Errors are in units of $\hbar\omega_0$.}
\label{tab:convergence}
\centering
\begin{tabular}{ccc}
\toprule
cutoff & $\delta_{10}$ & $\delta_0$\\
\midrule
$n_{\max}=2$  & $6.75\times10^{-4}$ & $1.76\times10^{-4}$\\
$n_{\max}=4$  & $7.97\times10^{-5}$ & $2.75\times10^{-5}$\\
$n_{\max}=6$  & $1.64\times10^{-5}$ & $7.46\times10^{-6}$\\
$n_{\max}=8$  & $4.95\times10^{-6}$ & $2.59\times10^{-6}$\\
$n_{\max}=10$ & $1.70\times10^{-6}$ & $9.65\times10^{-7}$\\
$n_{\max}=12$ & $5.09\times10^{-7}$ & $3.05\times10^{-7}$\\
\midrule
$m_{\max}=6$  & $8.28\times10^{-3}$ & $2.25\times10^{-7}$\\
$m_{\max}=8$  & $8.61\times10^{-5}$ & $2.09\times10^{-9}$\\
$m_{\max}=10$ & $3.95\times10^{-7}$ & $1.06\times10^{-11}$\\
$m_{\max}=12$ & $1.06\times10^{-9}$ & $3.16\times10^{-14}$\\
$m_{\max}=14$ & $2.02\times10^{-12}$ & $3.44\times10^{-15}$\\
\bottomrule
\end{tabular}
\end{table}

\subsection{Exact numerical checks of the symmetry analysis}
\label{subsec:numerical_checks}

Before interpreting the reconstructed spectrum, we required four independent checks. First, setting $\beta=0$ must reproduce Eq.~\eqref{eq:dimensionless_TI_energy}. Second, changing $\theta_A$ must leave the spectrum invariant for the circular closed ring. Third, the two spin blocks must be isospectral according to Eq.~\eqref{eq:spindeg}. Finally, the complete spectrum must be periodic under insertion of one flux quantum,
\begin{equation}
E_\nu(\phiAB+1)=E_\nu(\phiAB),
\label{eq:flux_periodicity}
\end{equation}
where the angular labels are correspondingly relabeled by a large gauge transformation.

For the same generic parameter set used in Table~\ref{tab:convergence}, the maximum deviations among the ten lowest levels are listed in Table~\ref{tab:symmetrychecks}. All four tests are satisfied at the level of floating-point precision. In particular, flux periodicity requires the union of the even- and odd-$m$ blocks because insertion of one flux quantum shifts $m$ by one and therefore interchanges the two angular-parity sectors.

\begin{table}[t]
\caption{Maximum numerical deviations among the ten lowest eigenvalues for exact analytical checks. Energies are in units of $\hbar\omega_0$.}
\label{tab:symmetrychecks}
\centering
\begin{tabular}{lc}
\toprule
check & maximum deviation\\
\midrule
$\beta=0$ vs. exact Tan--Inkson spectrum & $0$\\
$\theta_A=0$ vs. $\theta_A=0.713$ & $9.7\times10^{-15}$\\
$s=+1$ vs. $s=-1$ & $0$\\
$\phiAB$ vs. $\phiAB+1$ & $4.7\times10^{-15}$\\
\bottomrule
\end{tabular}
\end{table}

\subsection{Reconstruction of the Aharonov--Bohm spectrum}
\label{subsec:ABspectrum}

\begin{figure*}[t]
\includegraphics[width=\textwidth]{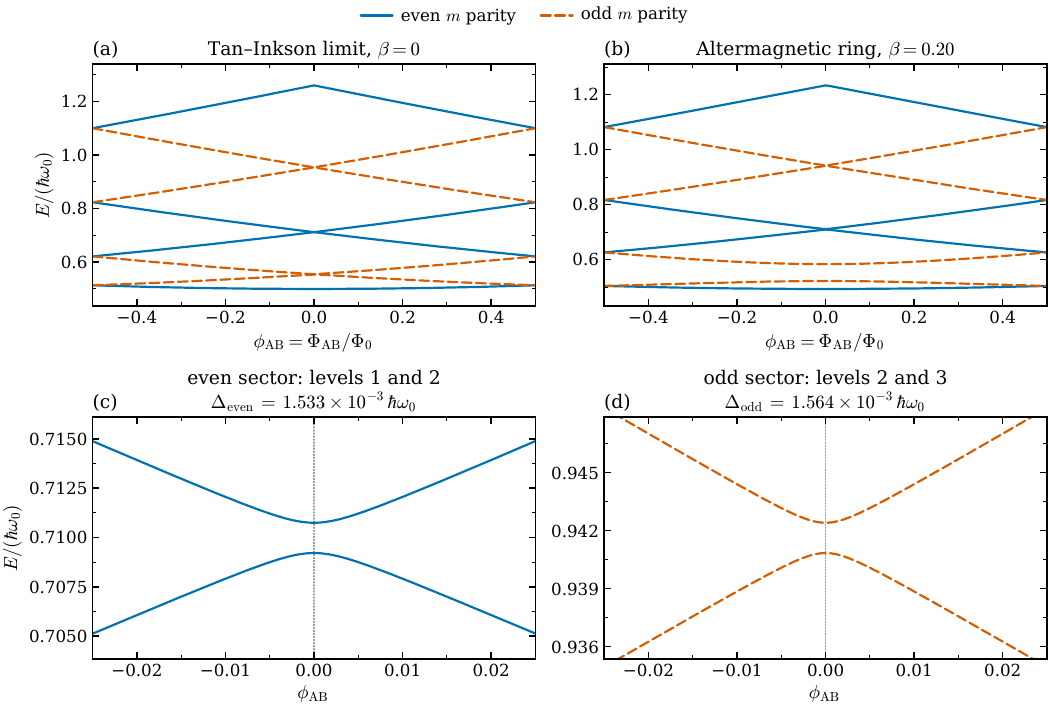}
\caption{Low-energy Aharonov--Bohm spectrum for $\kappa=4.5$ and $B=0$. (a) Conventional Tan--Inkson ring, $\beta=0$. (b) Altermagnetic ring with $\beta=0.20$. Blue solid and orange dashed curves denote the even- and odd-$m$ parity sectors, respectively. Crossings between channels connected by $\Delta m=\pm2$ are converted into anticrossings, whereas crossings between opposite angular-parity sectors remain symmetry protected. Panels (c) and (d), obtained from a 401-point local scan over $-0.06\le\phiAB\le0.06$, resolve the two smaller same-parity gaps at $\phiAB=0$: $1.533\times10^{-3}\hbar\omega_0$ in the even sector and $1.564\times10^{-3}\hbar\omega_0$ in the odd sector. Energies are shown in units of $\hbar\omega_0$.}
\label{fig:ABspectrum}
\end{figure*}

Figure~\ref{fig:ABspectrum} compares the low-energy AB spectral flow at $B=0$ for the conventional Tan--Inkson limit and for a representative altermagnetic coupling $\beta=0.20$. Because of the exact spin degeneracy, only the $s=+1$ block is shown. Solid and dashed curves denote the even- and odd-$m$ sectors, respectively.

At $\beta=0$, the two lowest odd-sector branches originate from the degenerate $n=0$, $m=-1$ and $n=0$, $m=+1$ Tan--Inkson states at $\phiAB=0$. Their unperturbed energy is
\begin{equation}
\eps_{0,\pm1}^{(0)}
=\frac{1}{2}\left[1+\sqrt{1+\kappa^2}\right]-\frac{\kappa}{2},
\end{equation}
which gives $\eps_{0,\pm1}^{(0)}=0.554886$ for $\kappa=4.5$. Since these channels differ by exactly two units of angular momentum, the altermagnetic term couples them directly. At $\beta=0.20$, the degeneracy is replaced by the pair
\begin{equation}
\eps_1=0.522039,
\qquad
\eps_2=0.583165,
\end{equation}
corresponding to an avoided-crossing gap
\begin{equation}
\Delta_{-1,+1}=0.06113\,\hbar\omega_0.
\label{eq:numerical_gap_beta02}
\end{equation}
The same mechanism repeats throughout the AB spectrum whenever unperturbed branches in the same angular-parity sector approach one another. The dense local scan displayed in Fig.~\ref{fig:ABspectrum}(c,d) resolves two additional small anticrossings at $\phiAB=0$: the even-sector levels 1 and 2 are separated by $1.53334\times10^{-3}\hbar\omega_0$, while the odd-sector levels 2 and 3 are separated by $1.56416\times10^{-3}\hbar\omega_0$. These gaps are roughly forty times smaller than the prominent low-energy $m=-1\leftrightarrow+1$ gap and were therefore barely visible on the scale of the original full-spectrum panel. In contrast, an even-sector state and an odd-sector state cannot couple within the minimal Hamiltonian. The smallest residual separations found at opposite-parity crossings among the first four branches are $1.1$--$2.4\times10^{-15}\hbar\omega_0$, i.e. at machine precision, providing a direct numerical verification of the exact $\mathcal P_\pi$ symmetry.

The gap at $\phiAB=0$ can be understood quantitatively from degenerate perturbation theory. Let
\begin{equation}
\mathcal I_{-1,+1}
=\int_0^\infty x\dd x\,
 u_{0,+1}(x)
 \widetilde{\mathcal D}_{0}^{(+)}
 \widetilde{\mathcal D}_{-1}^{(+)}u_{0,-1}(x),
\label{eq:dimensionless_overlap}
\end{equation}
where the tilde denotes the dimensionless form of the radial operators in Eq.~\eqref{eq:Dplus}. For $\kappa=4.5$ and $B=\phiAB=0$, direct quadrature gives
\begin{equation}
\mathcal I_{-1,+1}=-0.608465.
\end{equation}
The two-state first-order gap is therefore
\begin{equation}
\frac{\Delta_{-1,+1}^{(1)}}{\hbar\omega_0}
=\frac{|\mathcal I_{-1,+1}|}{2}|\beta|
=0.304233|\beta|.
\label{eq:linear_gap}
\end{equation}
Direct comparison with the full basis diagonalization gives excellent agreement over the perturbative interval. At $\beta=0.20$, the first-order value is $0.06085\hbar\omega_0$, only about $0.5\%$ below the fully converged gap in Eq.~\eqref{eq:numerical_gap_beta02}. This near-linear behavior demonstrates that the principal low-energy spectral signature is controlled directly by the $\Delta m=2$ matrix element rather than by a uniform spin splitting.

\subsection{Angular-channel content of the hybridized states}
\label{subsec:weights_results}

The avoided crossing is accompanied by a direct reconstruction of the orbital wave function. Evaluating the angular weights of Eq.~\eqref{eq:weights} for the two split odd-sector states at $\phiAB=0$ and $\beta=0.20$, the lower state has
\begin{equation}
W_{-1}=W_{+1}=0.49937,
\end{equation}
while the upper state has
\begin{equation}
W_{-1}=W_{+1}=0.49942.
\end{equation}
The residual weight is mainly carried by $m=\pm3$ and is below $1.3\times10^{-3}$ in total for either state. Thus the low-energy anticrossing is, to very high accuracy, a symmetric/antisymmetric hybridization of the two angular channels selected by the $d$-wave rule.

\subsection{Near-Fermi anticrossings and multilevel reconstruction at high filling}
\label{subsec:nearEFmultilevel}

The low-energy anticrossing discussed above is intentionally close to an isolated two-level problem. The situation near the Fermi edge of a many-electron ring is more demanding because several radial families and many angular channels can lie within the same small energy window. To resolve this regime without confusing altermagnetic hybridization with an arbitrary level crossing, we use the wide-annulus geometry $\kappa=50$ and fix $\phiAB=0.25$. Two representative conventional crossings close to the occupied edge are
\begin{equation}
(n,m)=(5,-86)\leftrightarrow(5,-84),
\qquad N_e=1400,
\label{eq:cross1400}
\end{equation}
and
\begin{equation}
(n,m)=(6,-85)\leftrightarrow(6,-83),
\qquad N_e=1800.
\label{eq:cross1800}
\end{equation}
They occur at $b_c\simeq1.6949$ and $b_c\simeq1.6749$, respectively, and satisfy the exact altermagnetic selection rule $\Delta m=2$.

\begin{figure*}[t]
\includegraphics[width=\textwidth]{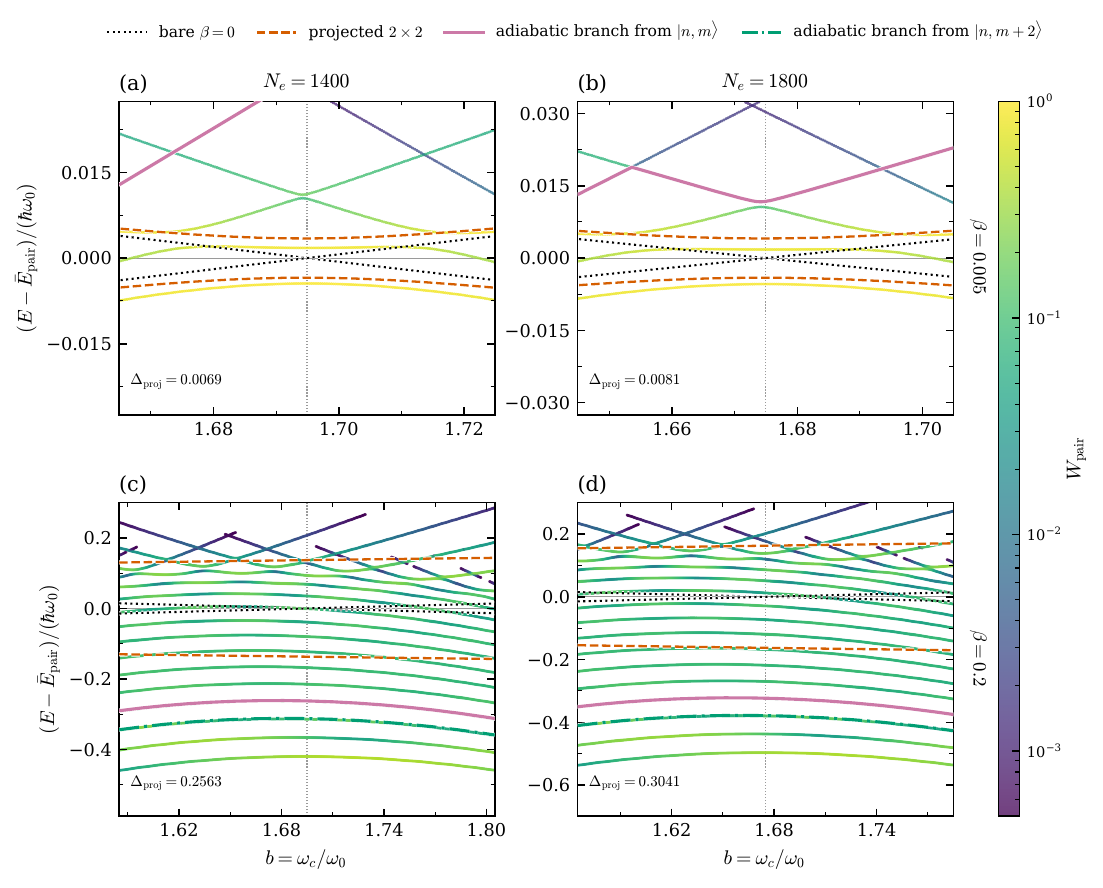}
\caption{Near-Fermi spectral reconstruction at $\kappa=50$ and $\phiAB=0.25$. Panels (a) and (b) show representative $\Delta m=2$ crossings for $N_e=1400$ and $1800$ at weak altermagnetic coupling $\beta=0.005$. The conventional bare branches, the projected $2\times2$ model, the full-spectrum spectral weight, and adiabatically tracked branches are shown together. Panels (c) and (d) display the corresponding full spectra at $\beta=0.20$, where the selected pair is embedded in a dense manifold, and the response becomes strongly multilevel. The plotted energy is measured relative to the local mean of the selected bare pair, $\bar E_{\rm pair}(b)=[E^{(0)}_{n,m}(b)+E^{(0)}_{n,m+2}(b)]/2$; at $b=b_c$ this reference coincides with the bare crossing energy $E_c$. The tracked weak-coupling branches are used only as a guide to the redistribution of character because local near-degeneracies can make their identity ambiguous at isolated field points.}
\label{fig:final_reconstruction}
\end{figure*}

For a selected bare pair $|a\rangle=|n,m\rangle$ and $|b\rangle=|n,m+2\rangle$, the projected Hamiltonian is
\begin{equation}
H_{\mathrm{proj}}=
\begin{pmatrix}
E_a^{(0)} & V_{ab}\\
V_{ab}^{\ast} & E_b^{(0)}
\end{pmatrix},
\qquad
V_{ab}=\langle a|H_{\mathrm{AM}}|b\rangle.
\label{eq:Hproj_highfill}
\end{equation}
At the bare crossing, the intrinsic two-level splitting is therefore
\begin{equation}
\Delta_{\mathrm{proj}}=2|V_{ab}|.
\label{eq:Delta_proj}
\end{equation}
Figure~\ref{fig:final_reconstruction} compares this projected scale with the full high-filling spectrum. At weak coupling, $\beta=0.005$, the projected branches identify the local avoided-crossing scale, but the exact spectral weight already samples additional nearby eigenstates. At $\beta=0.20$, the same energy window becomes a broad bundle of reconstructed levels. Thus the high-filling problem evolves continuously from weak resonant hybridization to a genuinely multilevel regime rather than remaining a collection of independent two-level anticrossings.

The projected splitting is proportional to $|\beta|$ because the off-diagonal matrix element in Eq.~\eqref{eq:Hproj_highfill} is linear in $\beta$. Its coefficients are
\begin{equation}
\frac{\Delta_{\mathrm{proj}}}{\hbar\omega_0}
=1.3709|\beta|,
\qquad N_e=1400,
\label{eq:gap_slope_1400}
\end{equation}
and
\begin{equation}
\frac{\Delta_{\mathrm{proj}}}{\hbar\omega_0}
=1.6286|\beta|,
\qquad N_e=1800,
\label{eq:gap_slope_1800}
\end{equation}
The nonperturbative effect at larger coupling is the participation of additional nearby levels in the exact eigenvectors, not a nonlinear dependence of the projected matrix element on $\beta$.

To quantify how the bare pair is redistributed over the exact eigenstates, we define
\begin{equation}
W_{\mathrm{pair},j}
=|\langle a|\Psi_j\rangle|^2
+|\langle b|\Psi_j\rangle|^2.
\label{eq:Wpair}
\end{equation}
Completeness requires
\begin{equation}
\sum_j W_{\mathrm{pair},j}=2,
\label{eq:Wpair_sum}
\end{equation}
provided the spectral window contains all eigenstates carrying appreciable weight from the two-dimensional bare subspace. A complementary measure is the participation number of an exact eigenstate expanded in the Tan--Inkson basis,
\begin{equation}
N_{\mathrm{part}}
=\frac{1}{\displaystyle\sum_i|c_i|^4}.
\label{eq:Npart}
\end{equation}
A basis state has $N_{\mathrm{part}}=1$, an ideal equal-weight two-state hybrid has $N_{\mathrm{part}}=2$, and progressively larger values indicate fragmentation over more basis channels. In the strong-coupling regime $\beta=0.20$, the participation numbers are smooth and well resolved throughout the selected near-Fermi window. For $N_e=1400$, the two tracked branches span $N_{\mathrm{part}}=11.84$--$12.45$ and $10.64$--$11.19$, respectively; for $N_e=1800$, the corresponding intervals are $13.42$--$14.19$ and $12.06$--$12.76$. These values quantify the broad multichannel fragmentation visible in Fig.~\ref{fig:final_reconstruction}. At weak coupling, some very narrow participation peaks are limited by the field grid and coincide with points where branch tracking is locally ambiguous. We therefore do not assign quantitative significance to those peak heights. The discussion instead rests on the full spectral-weight map, the adaptive completeness sum in Eq.~\eqref{eq:Wpair_sum}, and the well-resolved strong-coupling reconstruction.

The identities of the branches are followed adiabatically through the overlap criterion
\begin{equation}
k_{j+1}=\underset{k}{\operatorname{arg\,max}}
\left|\langle\Psi_{k_j}(b_j)|\Psi_k(b_{j+1})\rangle\right|^2.
\label{eq:adiabatic_tracking}
\end{equation}
At strong coupling, the overlap tracking is stable: for $\beta=0.20$ the minimum step overlap is about $0.9999$ for both fillings. At weak coupling, by contrast, isolated near-degeneracies can reduce the maximum-overlap continuation to roughly $0.50$--$0.59$, so the label of an individual tracked branch is locally ambiguous. The tracked curves in Fig.~\ref{fig:final_reconstruction} are therefore used qualitatively rather than to define the gap or the degree of fragmentation. The bare-state character can be depleted before the nominal two-level crossing when the selected branch encounters spectator levels in the dense high-filling spectrum. This confirms that the selected pair is not spectrally isolated. In the adaptive spectral-weight calculation, the retained window grows to several hundred exact eigenstates at stronger coupling while the accumulated pair weight remains at or above about $1.99$, close to the completeness value in Eq.~\eqref{eq:Wpair_sum}.

\subsection{Independent two-dimensional real-space benchmark}
\label{subsec:realspace}

To verify that the anticrossings are not artifacts of the Tan--Inkson basis, we performed an independent Cartesian finite-difference calculation for $B=\PhiAB=\theta_A=0$. In the dimensionless coordinates $x=X/\ell_0$ and $y=Y/\ell_0$, the fixed-spin Hamiltonian becomes
\begin{align}
\frac{H_s}{\hbar\omega_0}
={}&-\frac{1}{2}(1+s\beta)\frac{\partial^2}{\partial x^2}
-\frac{1}{2}(1-s\beta)\frac{\partial^2}{\partial y^2}
\notag\\
&+\frac{\kappa^2}{2(x^2+y^2)}
+\frac{x^2+y^2}{8}-\frac{\kappa}{2}.
\label{eq:FDHamiltonian}
\end{align}
We discretized Eq.~\eqref{eq:FDHamiltonian} on the square $[-8,8]\times[-8,8]$ with Dirichlet boundaries and a second-order central-difference Laplacian. Calculations with $N=141$, 181, 221, and 261 grid points along each direction were extrapolated linearly in the squared mesh spacing $h^2$, as required by the discretization error. No Tan--Inkson radial function, angular decomposition, or $\Delta m=2$ selection rule is built into this calculation.

Table~\ref{tab:FDbenchmark} compares the extrapolated real-space values with the high-cutoff basis calculation for $\kappa=4.5$ and $\beta=0.20$. The agreement is at the $10^{-7}\hbar\omega_0$ level for all six states, providing an independent validation of both the sparse matrix elements and the resulting low-energy spectrum.

\begin{table}[t]
\caption{Independent real-space benchmark for $\kappa=4.5$, $\beta=0.20$, and $B=\PhiAB=0$. The finite-difference (FD) values are extrapolated to $h^2\to0$. Energies are in units of $\hbar\omega_0$.}
\label{tab:FDbenchmark}
\centering
\begin{tabular}{cccc}
\toprule
$\nu$ & TI basis & FD extrap. & FD$-$TI\\
\midrule
1 & 0.49293843 & 0.49293825 & $-1.8\times10^{-7}$\\
2 & 0.52203861 & 0.52203853 & $-8.3\times10^{-8}$\\
3 & 0.58316489 & 0.58316467 & $-2.2\times10^{-7}$\\
4 & 0.70921377 & 0.70921384 & $ 7.0\times10^{-8}$\\
5 & 0.71074531 & 0.71074517 & $-1.4\times10^{-7}$\\
6 & 0.94084772 & 0.94084792 & $ 2.0\times10^{-7}$\\
\bottomrule
\end{tabular}
\end{table}

These calculations establish the numerical part of the spectral problem independently of any thermodynamic postprocessing. In particular, the AB-periodic reconstruction, the protected even--odd crossings, the $\Delta m=2$ anticrossings, the exact spin isospectrality, and the angular-channel hybridization can all be tested before derivatives of the spectrum are used to calculate magnetization or persistent current.

\subsection{Independent high-filling spot checks}
\label{subsec:independent_highfill_validation}

To validate the large-occupation calculation without relying on the original
plotting data, we reconstructed the matrix elements of
Eqs.~\eqref{eq:Mplus} and~\eqref{eq:Mminus} with analytic derivatives of
associated Laguerre polynomials and independently diagonalized the two
angular-parity blocks using a real symmetric banded solver. The tests
below use $\kappa=50$, $\beta=0.20$, and 400 Gauss--Legendre nodes.
The radial interval is $0<x<30$ for the first two field values and
$0<x<35$ for the strongest field. The radial basis consists of $n=0,\ldots,n_{\max}$; for a
finite magnetic field the angular interval is asymmetric and explicitly
includes the strongly shifted negative-$m$ channels.

At $(b,\phiAB)=(0.8,0)$, an interval
$-150\le m\le120$ gives energies stable when the radial cutoff is
increased from $n_{\max}=21$ to $25$: the many-body energies change by
$2.7\times10^{-8}$ and $4.7\times10^{-7}\,\hbar\omega_0$ for
$N_e=1400$ and $1800$, respectively. Enlarging the angular interval to
$-175\le m\le135$ leaves the corresponding values unchanged at the
reported precision. At $(b,\phiAB)=(1.6749,0.25)$, an angular interval
$-285\le m\le120$ is needed; increasing $n_{\max}$ from 21 to 25
changes both energies by less than $10^{-9}\,\hbar\omega_0$ in the
numerical tests. The resulting one-spin Fermi edges and spin-degenerate
ground-state energies are listed in Table~\ref{tab:independent_highfill}.
At $(b,\phiAB)=(6.5,0.25)$ the same procedure requires
$-1050\le m\le100$ and gives stable energies upon increasing
$n_{\max}$ from 11 to 13, with changes below
$2\times10^{-7}\,\hbar\omega_0$; an independent check of the
angular boundary at a smaller radial cutoff gives identical occupied
energies for $m_{\min}=-1000$ and $-1250$ at the displayed precision.
These tests check representative parameter points, not every entry of
the archived field and flux sweeps.

\begin{table*}[t]
\caption{Independent high-filling spot checks for $\kappa=50$ and
$\beta=0.20$. The occupied levels include both exactly degenerate
spin sectors. Energies are in units of $\hbar\omega_0$; the cutoff
values listed in the text were independently varied.}
\label{tab:independent_highfill}
\centering
\begin{tabular}{ccccc}
\toprule
$b$ & $\phi_{\rm AB}$ & $N_e$ & $E_{\rm GS}/(\hbar\omega_0)$ & $E_F/(\hbar\omega_0)$ \\
\midrule
0.8000 & 0.00 & 1400 & 8973.103607 & 10.55011507 \\
0.8000 & 0.00 & 1800 & 13569.858403 & 12.41162647 \\
1.6749 & 0.25 & 1400 & 8994.294556 & 10.61174016 \\
1.6749 & 0.25 & 1800 & 13591.947532 & 12.44882614 \\
6.5000 & 0.25 & 1400 & 9433.875240 & 10.44046847 \\
6.5000 & 0.25 & 1800 & 13925.063085 & 12.11333068 \\
\bottomrule
\end{tabular}
\end{table*}

The convergence tests demonstrate that cutting off the negative
angular-momentum channels prematurely shifts the many-body energy,
even when the lowest few eigenvalues appear stable. Because
magnetization and persistent current are derivatives of this energy,
the response analysis below is restricted to independently checked
field and flux windows. The exact $\beta=0$ reference and the
projected matrix elements in Eqs.~\eqref{eq:gap_slope_1400}
and~\eqref{eq:gap_slope_1800} were also evaluated separately.

\FloatBarrier

\section{Persistent current and orbital magnetization}
\label{sec:thermodynamics}

\subsection{Fixed-particle-number formulation}
\label{subsec:fixedN}

We use the single-particle spectrum to calculate equilibrium
currents and magnetization for an \emph{isolated ring with exactly $N_e$
electrons}. The distinction between a fixed integer particle number and a
fixed \emph{mean} occupation is important for the four- and six-electron
systems considered below. Denoting the one-spin orbital energies by
$\eps_\nu=E_\nu/(\hbar\omega_0)$, we introduce the reduced temperature
\begin{equation}
\tau=\frac{k_{\rm B}T}{\hbar\omega_0}.
\label{eq:tau_canonical}
\end{equation}
Each orbital has two spin states with the same energy as a consequence
of Eq.~\eqref{eq:spindeg}. For noninteracting fermions the exact canonical
partition function is the coefficient of $z^{N_e}$ in
\begin{equation}
\mathcal Z_{N_e}(b,\phiAB;\tau)
=[z^{N_e}]\prod_\nu
\left(1+z\,e^{-\eps_\nu/\tau}\right)^2,
\label{eq:canonical_Z}
\end{equation}
where $[z^{N_e}]$ means coefficient extraction, not evaluation at a
particular fugacity. The dimensionless Helmholtz free energy is
\begin{equation}
\mathcal F_{N_e}=-\tau\ln\mathcal Z_{N_e}.
\label{eq:F_canonical}
\end{equation}
This expression sums only configurations containing exactly $N_e$
electrons and incorporates the spin degeneracy without replacing it
by an average occupation. Numerically, we shift each spin-orbital energy
by $\eps_{\min}$, put $w_j=\exp[-(\eps_j-\eps_{\min})/\tau]$,
and evaluate the elementary symmetric polynomial through
\begin{equation}
Q_k^{(j)}=Q_k^{(j-1)}+w_j Q_{k-1}^{(j-1)},
\quad Q_0^{(0)}=1,
\label{eq:canonical_recursion}
\end{equation}
with $Q_k^{(0)}=0$ for $k>0$. The index $j$ runs over the
\emph{spin orbitals}, so each spatial eigenenergy is inserted twice;
$\mathcal F_{N_e}=N_e\eps_{\min}-\tau\ln Q_{N_e}$.
This recursion is numerically stable for the low fillings examined here.

At zero temperature and even electron number, the exact ground-state
energy is obtained by filling the lowest $N_e/2$ orbitals twice:
\begin{equation}
\mathcal F_{N_e}(T=0)=2\sum_{\nu=1}^{N_e/2}\eps_\nu.
\label{eq:F_zeroT}
\end{equation}
This remains the ground-state energy even at an accidental orbital
degeneracy; one-sided response derivatives are required at a cusp.
For the many-electron calculations, the occupied edge $E_F$ is fixed
by the level-counting condition
\begin{equation}
N_e=2\sum_\nu\Theta(E_F-E_\nu),
\label{eq:fermi_step}
\end{equation}
with the usual treatment of a partially filled degenerate multiplet.
In a spectral gap the zero-temperature chemical potential need not be
unique; our reported Fermi edge is simply the highest occupied
single-particle energy. Earlier studies used this occupation picture to
trace Fermi-level rearrangements in quantum rings
\cite{PereiraAndradeSilva2021PhysE,PereiraSilva2023AnnPhys,LiraPereiraSilva2024PhysE,PereiraAzevedoSilva2026PhysE}.
Related state-space analyses of a quantum dot can be found in
Ref.~\cite{LiraSilva2023PhysE}.

The often-used alternative of solving
$N_e=2\sum_\nu f_{\rm FD}(\eps_\nu-\bar\mu)$ and evaluating
$\Omega_{\rm G}+\bar\mu N_e$ fixes only the \emph{mean} number in a
grand-canonical ensemble. It is not identical to
Eq.~\eqref{eq:F_canonical} for a finite system and is used below solely
to quantify the size of that approximation; all quoted finite-$T$
responses use the strict canonical result.

It is convenient to introduce the natural current and orbital-moment scales
\begin{equation}
I_0=\frac{\hbar\omega_0}{\Phi_0}
=\frac{e\omega_0}{2\pi},
\qquad
\mu_{\mathrm B}^{\ast}=\frac{e\hbar}{2\mstar}.
\label{eq:response_scales}
\end{equation}
Since $\phiAB=\PhiAB/\Phi_0$ and $b=\omega_c/\omega_0=eB/(\mstar\omega_0)$, the fixed-$N_e$ equilibrium responses are
\begin{equation}
\frac{I_{\mathrm c}}{I_0}
=-\frac{\partial\mathcal F}{\partial\phiAB},
\label{eq:I_dimensionless}
\end{equation}
and
\begin{equation}
\frac{M}{\mu_{\mathrm B}^{\ast}}
=-2\frac{\partial\mathcal F}{\partial b}.
\label{eq:M_dimensionless}
\end{equation}
At finite temperature these derivatives act on the canonical free energy
of Eq.~\eqref{eq:F_canonical}, so only fixed-$N_e$ configurations are
included. At zero temperature the response is piecewise smooth and can
jump where the occupied ground state changes.

The exact spin isospectrality has an immediate thermodynamic consequence. For every parameter set in the circular minimal model,
\begin{equation}
I_{\uparrow}=I_{\downarrow}=\frac{I_{\mathrm c}}{2},
\qquad
M_{\uparrow}=M_{\downarrow}=\frac{M}{2}.
\label{eq:spin_responses_equal}
\end{equation}
Therefore, any equilibrium persistent spin current defined from the difference of the two spin-resolved charge currents vanishes identically. A nonzero spin response requires an additional ingredient that breaks the unitary equivalence between the two spin blocks.

\subsection{Thermodynamic validation}
\label{subsec:thermo_validation}

We checked the thermodynamic derivatives separately from the spectral calculations of Sec.~\ref{sec:numerics}. For $\beta=0$, differentiating Eq.~\eqref{eq:dimensionless_TI_energy} at fixed occupation gives a closed expression for both $I_{\mathrm c}$ and $M$. At $N_e=6$, $\phiAB=0.23$, and $b=0$, the current obtained from the numerical free-energy derivative agrees with the analytical Tan--Inkson value to $6.0\times10^{-13} I_0$. At $\phiAB=0.17$ and $b=0.19$, the corresponding magnetization residual is $2.6\times10^{-12}\mu_{\mathrm B}^{\ast}$.

For the finite-temperature calculations, we evaluated the exact
canonical partition function in Eq.~\eqref{eq:canonical_Z} rather than
assuming a fixed mean number. As an independent implementation test,
direct enumeration of all four-electron configurations in a truncated
14-spin-orbital spectrum at $\tau=0.02$ agrees with the elementary
symmetric-polynomial recursion in free energy to
$4.4\times10^{-16}\hbar\omega_0$. Doubling the density of the
finite-difference stencil changes the canonical $I_c$ and $M$ by
less than $2\times10^{-10}$ in the tests at $\tau=0.01$--$0.08$;
expanding the one-particle basis changes the corresponding responses
by less than $4\times10^{-6}$ in their natural units. The symmetry
relations in Table~\ref{tab:thermo_checks} are also satisfied in the
canonical ensemble.

\begin{table}[t]
\caption{Independent checks for the strictly canonical response.
The three finite-temperature symmetry residuals use $\beta=0.20$,
$N_e=6$, and $\tau=0.03$; exact $\beta=0$ reference tests are at
$T=0$. Residuals are given in the indicated dimensionless response
units.}
\label{tab:thermo_checks}
\centering
\begin{tabular}{lc}
\toprule
check & absolute residual\\
\midrule
Tan--Inkson current at $T=0$ & $6.0\times10^{-13}$\\
Tan--Inkson magnetization at $T=0$ & $2.6\times10^{-12}$\\
$I_c(\phiAB)+I_c(-\phiAB)$ at $b=0$ & $1.1\times10^{-10}$\\
$I_c(\phiAB+1)-I_c(\phiAB)$ & $1.1\times10^{-10}$\\
$M(b)+M(-b)$ at $\phiAB=0$ & $2.0\times10^{-10}$\\
\bottomrule
\end{tabular}
\end{table}

A separate high-filling spot check implements the physical operator
derivatives $\partial_b H$ and $\partial_{\phi_{\rm AB}}H$ directly in
the exact Tan--Inkson basis; these contain non-diagonal radial terms
even when the unperturbed Hamiltonian is diagonal. At $\kappa=50$,
$\beta=0.20$, $b=0.8$, and $\phiAB=0$, the resulting Hellmann--Feynman
magnetizations $M/\mu_B^*$ are $0.257225$ and $-16.335359$ for
$N_e=1400$ and $1800$, respectively; the corresponding currents
$I_c/I_0$ are $0.208231$ and $-0.053934$. Direct central differences
of the independently reconstructed many-body energies, with steps
$\delta b=10^{-5}$ and $\delta\phiAB=10^{-4}$, agree with these
operator-expectation values within $3.4\times10^{-5}\,\mu_B^*$
and $3.4\times10^{-7}\,I_0$, respectively. At Fermi-level
rearrangements the energy is not differentiable and the appropriate
one-sided limits must be examined. In particular, a magnetization
estimated using the much coarser archived field grid can differ
substantially from the local derivative; smooth-looking interpolations
of these sparse points should not be interpreted as pointwise
convergence certificates.

\subsection{Persistent current: removal of the open-shell discontinuity}
\label{subsec:persistent_current}

The effect of an $m\leftrightarrow m\pm2$ anticrossing on the persistent current is clearest at an open shell. We take $N_e=4$ and $B=0$. In the conventional Tan--Inkson limit, two electrons occupy the $m=0$ orbital and the remaining two electrons enter the degenerate $m=\pm1$ manifold at zero AB flux. Infinitesimally away from $\phiAB=0$, one member of this doublet becomes the lower orbital, so the many-body ground-state energy has a cusp. The corresponding zero-temperature persistent current has the one-sided limits
\begin{equation}
\lim_{\phiAB\to0^{\pm}}\frac{I_{\mathrm c}}{I_0}
=\pm\frac{1}{\sqrt{1+\kappa^2}},
\label{eq:TI_current_jump}
\end{equation}
where the positive (negative) sign corresponds to approaching zero flux from above (below). For $\kappa=4.5$, the magnitude of each one-sided limit is $0.21693$; the discontinuity between the two limits is $0.43386$.

\begin{figure}[t]
\includegraphics[width=\columnwidth]{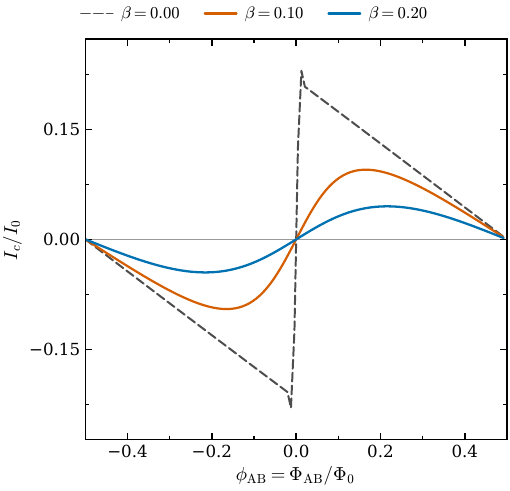}
\caption{Zero-temperature persistent charge current for $N_e=4$, $\kappa=4.5$, and $B=0$. The conventional Tan--Inkson open shell has a zero-flux current discontinuity. The altermagnetic $m=-1\leftrightarrow+1$ anticrossing removes the cusp in the many-body energy and converts the jump into a continuous flux response.}
\label{fig:persistent_beta}
\end{figure}

The altermagnetic term couples precisely these two orbitals because they differ by two angular units. The same gap discussed in Sec.~\ref{subsec:ABspectrum} therefore removes the open-shell cusp and makes the fixed-$N_e$ free energy differentiable at zero flux. Figure~\ref{fig:persistent_beta} shows the resulting current. Already at $\beta=0.10$ the discontinuity is replaced by a smooth sign change through $I_{\mathrm c}=0$, while $\beta=0.20$ produces a much broader rounding region. At $\phiAB=0.05$, the current decreases from $0.19545 I_0$ in the Tan--Inkson limit to $0.01721 I_0$ at $\beta=0.20$. At $\phiAB=0.25$, the reduction is from $0.10893 I_0$ to $0.04401 I_0$.

The current traces the one-particle anticrossing directly. Its suppression and smooth passage through zero flux arise from mixing the two same-parity angular channels, without invoking interactions or spin--orbit coupling. Changing only the radius of a circular, spin-independent Tan--Inkson ring would not remove the $m=\pm1$ degeneracy at zero flux. The effect is largest when the Fermi level intersects a pair connected by $\Delta m=\pm2$.

\subsection{Finite-temperature suppression in the canonical ensemble}
\label{subsec:temperature_current}

The low-energy angular level spacings of a finite-width ring can be
much smaller than the radial confinement quantum. As a result, a
modest reduced temperature is sufficient to populate configurations
with different persistent currents. In a four-electron ring at
$\kappa=4.5$, $b=0$, $\phiAB=0.25$, and $\beta=0.20$ the exactly
canonical calculation gives
\begin{equation}
\frac{I_c}{I_0}=0.04401,\quad 0.04382,\quad 0.03371
\end{equation}
for $\tau=0$, $0.01$, and $0.02$, respectively. The last value is
substantially larger than the fixed-mean-number result, $0.01308$,
because a four-electron system cannot exchange electrons with a
reservoir. The difference is not a numerical error: the two ensembles
sum different many-electron configurations.

\begin{table}[t]
\caption{Persistent current $I_c/I_0$ at $b=0$, $\phiAB=0.25$,
$\kappa=4.5$, and $\beta=0.20$ in the strict canonical ensemble and
in the fixed-mean-number approximation. The canonical values are used
throughout the discussion of finite-temperature properties.}
\label{tab:canonical_compare}
\centering
\begin{tabular}{cccc}
\toprule
$N_e$ & $\tau$ & canonical & fixed mean $N_e$\\
\midrule
4 & 0.01 & 0.04382 & 0.03923\\
4 & 0.02 & 0.03371 & 0.01308\\
4 & 0.03 & 0.01359 & $-0.00242$\\
6 & 0.01 & $-0.14286$ & $-0.12743$\\
6 & 0.02 & $-0.11308$ & $-0.06433$\\
6 & 0.03 & $-0.06418$ & $-0.02236$\\
\bottomrule
\end{tabular}
\end{table}

Table~\ref{tab:canonical_compare} shows that the discrepancy is
also sizeable for six electrons. Nevertheless, the exact canonical
response is progressively reduced at higher temperatures, consistent
with thermal mixing of states carrying different angular currents.
We use the same canonical procedure for the field-dependent orbital
magnetization. At $b=0.15$ and $\phiAB=0.25$, for example, the
four-electron altermagnetic magnetization changes from
$-0.72444\mu_{\rm B}^{\ast}$ at $T=0$ to
$-0.65681\mu_{\rm B}^{\ast}$ at $\tau=0.02$, instead of the
fixed-mean-number estimate $-0.58744\mu_{\rm B}^{\ast}$.

\subsection{Orbital magnetization and field-driven rearrangements}
\label{subsec:magnetization_results}

A uniform perpendicular field probes the same reconstructed spectrum through a different derivative. For $N_e=4$ at $\phiAB=0.25$, the conventional zero-temperature magnetization exhibits linear segments and abrupt changes associated with field-driven rearrangements of occupied angular orbitals. The $d$-wave coupling shifts these transitions and rounds symmetry-allowed crossings into anticrossings. Consequently, the magnetization can change by order unity in units of $\mu_{\mathrm B}^{\ast}$ even for $\beta=0.20$; at $b=0.15$, for example, it changes from $-2.284\mu_{\mathrm B}^{\ast}$ at $\beta=0$ to $-0.724\mu_{\mathrm B}^{\ast}$ at $\beta=0.20$. At finite temperature, the canonical ensemble averages over more than one
many-body occupation pattern, reducing the sharpness of the magnetic
response away from level crossings. As with the persistent current,
this response probes the field dependence of the same reconstructed
spectrum, but it is the derivative with respect to the homogeneous
field rather than the flux tube.

The response is also filling dependent because the altermagnetic correction is largest when the Fermi level lies close to a symmetry-selected anticrossing. For even fillings between two and sixteen electrons at $\kappa=4.5$, $\phiAB=0.25$, and $B=0$, the current retains the usual shell-dependent sign alternation, while the largest relative altermagnetic correction in this range occurs at $N_e=4$. At more nearly closed shells, the changes of several occupied angular channels partially compensate.

\FloatBarrier

\section{Many-electron magnetic oscillations and flux response}
\label{sec:manyelectron_response}

\subsection{Fermi edge and the de Haas--van Alphen benchmark}
\label{subsec:manyelectron}

We used $\kappa=4.5$ for the few-electron spectra, where the radial and angular spacings are readily resolved. At fillings of order $10^3$, however, that geometry ceases to describe a well-developed annulus.  At $N_e=1800$ and $B=0$, for example, the conventional Tan--Inkson Fermi edge is $E_F=19.4074\,\hbar\omega_0$ at $\kappa=4.5$.  The turning points of the \emph{bare} confinement potential at this energy are approximately $x_-=0.685$ and $x_+=13.145$, where $x=r/\ell_0$.  Thus, the energetically accessible region almost reaches the origin and no longer represents the annular regime for which the small-$\kappa$ calculation was introduced.

We first establish a separate $\beta=0$ reference for the many-electron calculation.  A scan in $\kappa$ shows that $\kappa=50$, corresponding to $r_0=10\ell_0$, provides a useful compromise between a well-resolved central hole and a sufficient number of occupied radial subbands.  For $N_e=1800$ at zero field one finds $E_F=12.5694\,\hbar\omega_0$; the same bare-potential diagnostic gives $x_-=6.173$ and $x_+=16.200$.  The occupied energy window remains well separated from the origin while supporting many angular states and several radial families.  The analogous $N_e=1400$ calculation gives $E_F=10.6780\,\hbar\omega_0$.

Increasing $\kappa$ moves the occupied region outward and reduces the excessive mixing of central and annular scales that occurs for the few-electron value $\kappa=4.5$. Very large $\kappa$, however, lowers the number of occupied radial families and shortens the field interval over which successive radial-subband depopulations can be resolved. At $\kappa=50$, the central region remains depleted while several radial subbands are occupied, allowing the slow subband-depopulation cycles to be separated from the faster angular shell oscillations.

At $\beta=0$, the large-filling problem is especially clean because Eq.~\eqref{eq:dimensionless_TI_energy} is exact and no matrix diagonalization is required.  Introducing
\begin{equation}
q=\sqrt{1+b^2},
\qquad
\mathcal A_{nm}=n+\frac{1}{2}+\frac{\lambda_m}{2},
\end{equation}
the field derivative of one orbital is
\begin{equation}
\frac{\partial\eps_{nm}^{(0)}}{\partial b}
=\frac{b}{q}\,\mathcal A_{nm}
+\frac{m+\phiAB}{2}.
\label{eq:TI_field_derivative}
\end{equation}
For the spin-degenerate minimal model, the zero-temperature magnetization can therefore be evaluated without a finite-difference derivative,
\begin{equation}
\frac{M}{\mu_B^\ast}
=-4\sum_{(n,m)\in\mathrm{occ}}
\left[
\frac{b}{q}\,\mathcal A_{nm}
+\frac{m+\phiAB}{2}
\right],
\label{eq:TI_exact_many_M}
\end{equation}
where the occupied orbitals are selected by the Heaviside constraint in Eq.~\eqref{eq:fermi_step}.  This expression is important numerically because it removes differentiation noise from the conventional reference curve.

\begin{figure*}[t]
\includegraphics[width=\textwidth]{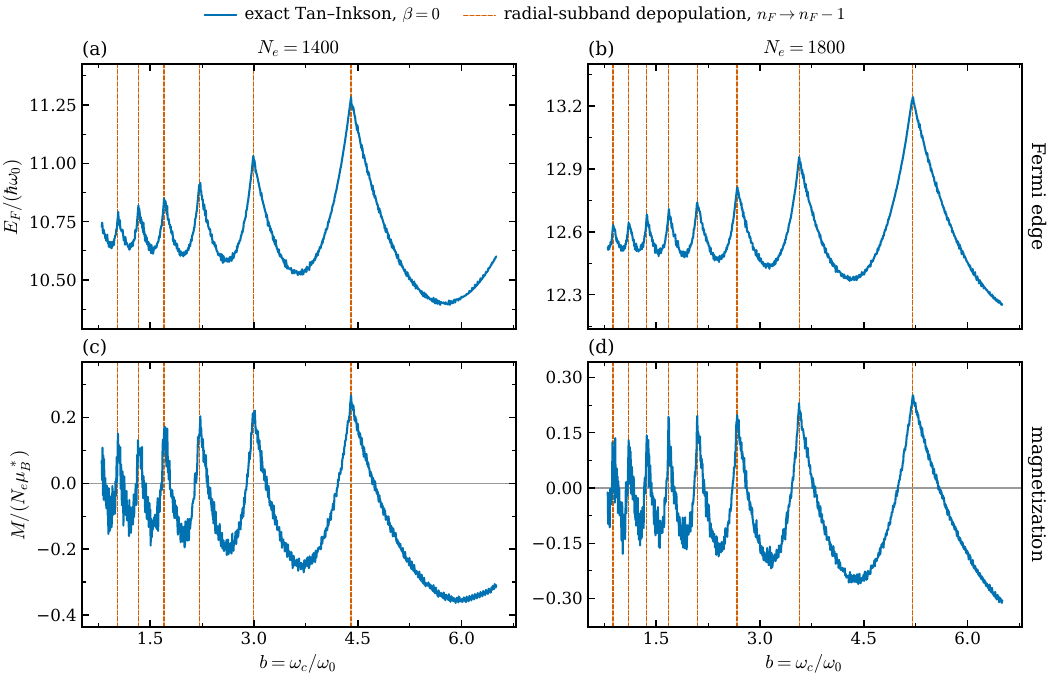}
\caption{Many-electron conventional Tan--Inkson benchmark at $\kappa=50$, $\phiAB=0$, and $T=0$. Panels (a) and (b) show the Fermi edge for $N_e=1400$ and $1800$, respectively, determined from the Heaviside particle-number constraint. Panels (c) and (d) show the exact orbital magnetization from Eq.~\eqref{eq:TI_exact_many_M}, normalized per electron. Vertical dashed lines mark the successive depopulation of the highest occupied radial subband. The large oscillatory cycles track these depopulation events, whereas the finer structure at lower field is produced by discrete angular-level rearrangements within an unchanged set of radial subbands.}
\label{fig:dhva_fillings}
\end{figure*}

Figure~\ref{fig:dhva_fillings} displays the Fermi edge and magnetization for $N_e=1400$ and $1800$ at $\kappa=50$ and $\phiAB=0$.  We denote by $n_F(B)$ the largest radial quantum number represented among the occupied orbitals.  The vertical dashed lines mark fields at which $n_F$ decreases by one.  Rather than reading these positions directly from the finite field grid, we bracket each transition on that grid and then refine it by bisection of the exact $\beta=0$ Tan--Inkson spectrum.  For $N_e=1800$, the depopulation fields within the plotted window are $n_F=9\to8$ at $b=0.876397$, $8\to7$ at $b=1.101657$, $7\to6$ at $b=1.363855$, $6\to5$ at $b=1.680882$, $5\to4$ at $b=2.091900$, $4\to3$ at $b=2.668256$, $3\to2$ at $b=3.564332$, and $2\to1$ at $b=5.197626$.  The corresponding $N_e=1400$ fields in the same window are $b=1.032799$, $1.331376$, $1.700457$, $2.209241$, $2.989827$, and $4.401964$.  We note that $n_F(b)$ is not strictly monotonic at the resolution of the exact spectrum.  For $N_e=1800$, a scan finer than the field grid used here reveals a narrow re-entrant window near $b\simeq1.361$ in which the highest occupied radial subband is briefly repopulated before emptying definitively; the value quoted above is the final crossing.  Such re-entrance is expected whenever several radial families are nearly degenerate at the Fermi edge of a wide annulus, and it does not affect the broad cycles discussed below.  The one-to-one correlation between the large magnetization cycles and radial-subband depopulation motivates their identification as a de Haas--van Alphen-type finite-ring response. Pereira and co-workers found related behavior in curved quantum rings \cite{PereiraAndradeSilva2021PhysE}, while thermomagnetic studies of confined electron gases by Pereira and Silva connected magnetic oscillations to subband depopulation and to rotational changes of the spectrum \cite{PereiraSilva2024QuantumRep,PereiraSilva2025AnnPhysThermo}. Those works concern different confinement or inertial mechanisms; the present calculation isolates the additional mixing induced by the altermagnetic term.

The two field scales come from different parts of the occupied spectrum. At low field, especially for $b\lesssim1.3$, crossings between discrete angular states cause rapid fluctuations of the Fermi edge and magnetization. These are finite-size shell oscillations, not dHvA oscillations in themselves. The broad cycles are different: every major peak is tied to the loss of the highest occupied radial subband, $n_F\to n_F-1$. This is the finite-ring analogue of the level-depopulation mechanism underlying the de Haas--van Alphen effect.  The coexistence of a dHvA-type envelope with a faster Aharonov--Bohm-type component in a confined electron system was established by Sivan and Imry for singly connected quantum dots \cite{SivanImry1988PRL}, and related AB- and dHvA-type responses were found in finite-width rings with curvature, rotation, and anisotropic effective mass \cite{TanInkson1999,PereiraAndradeSilva2021PhysE,PereiraSilva2023AnnPhys,LiraPereiraSilva2024PhysE,PereiraAzevedoSilva2026PhysE}. The analogy should not be pushed to exact Onsager periodicity: the present annulus is finite, smoothly confined, and treated at fixed particle number. We therefore use \emph{dHvA-type envelope} for the broad radial-subband cycles and reserve \emph{shell structure} for the faster angular rearrangements.

\subsection{Converged local high-filling response in a homogeneous field}
\label{subsec:highfill_AM}

We next examine a restricted field window where the altermagnetic
high-occupation spectrum can be recomputed independently rather than
inferred from archived, incompletely documented response traces. For
$\kappa=50$, $\beta=0.20$, $\phiAB=0$, and $N_e=1400$ or $1800$,
we evaluated the ground-state energy and Fermi edge at 61 points over
$0.77\le b\le0.83$. The reconstructed basis uses
$n=0,\ldots,21$ and $-175\le m\le135$, with 400 Gauss--Legendre
nodes and radial interval $0<x<30$. At $b=0.77$ and $0.83$, raising
the radial cutoff to $n_{\max}=25$ changes the $N_e=1800$
many-body energy by less than $6\times10^{-7}\hbar\omega_0$.
Extending the angular interval to $-195\le m\le155$ leaves the
reported energies unchanged at this precision. This test provides a
controlled basis for the finite-window results; it does not establish
uniform convergence of earlier long field sweeps.

\begin{figure*}[t]
\includegraphics[width=\textwidth]{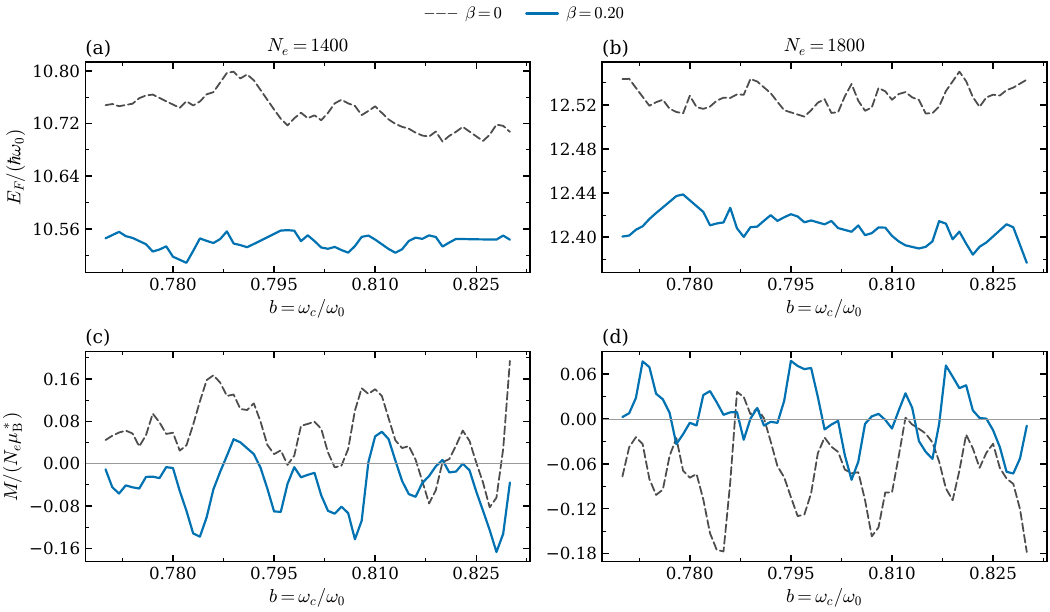}
\caption{Independently recomputed field window $0.77\le b\le0.83$
for $\kappa=50$, $\phi_{\rm AB}=0$, and $N_e=1400,1800$.
Panels (a,b) show the Fermi edge and (c,d) the orbital magnetization
per electron. Dashed curves are the exact $\beta=0$ reference;
solid curves are $\beta=0.20$, diagonalized at each field value
in a convergence-tested basis. Magnetization curves use identical
five-point local polynomial derivatives of the corresponding
many-body energies, with grid spacing $\Delta b=0.001$.
They are finite-window derivatives and are not assigned a unique
pointwise meaning at occupation cusps.}
\label{fig:EFM_highfill_AM}
\end{figure*}

Figure~\ref{fig:EFM_highfill_AM} shows the Fermi edge and the local
magnetic response alongside the exact $\beta=0$ Tan--Inkson
reference. The two many-body energy series ($\beta=0$ and $0.20$) are differentiated with the same five-point
local polynomial stencil on the $\Delta b=0.001$ grid, avoiding
artificial differences introduced by comparing a smoothed curve to an
unfiltered derivative. Since the spectrum exhibits rapid occupation
rearrangements, the plotted magnetization represents a \emph{finite-grid
response estimate}: at a cusp it averages nearby one-sided slopes
and should not be interpreted as the unique pointwise derivative.
Away from a cusp, the expectation-value form of the derivative is
provided by Hellmann--Feynman and agrees with sufficiently local
energy differences, as checked in Sec.~\ref{subsec:thermo_validation}.
The independently recomputed field window confirms that the
altermagnetic term shifts the occupied edge and changes the local
shell response, without relying on unverified full-range amplitudes.

\subsection{Current over a full Aharonov--Bohm flux period}
\label{subsec:highfill_flux}

At $b=0$, varying the independent flux tube does not change the
magnetic field in the occupied annulus. We independently computed the
$\beta=0.20$ many-body spectrum for
$0\le\phiAB\le1/2$ on 61 flux points, with $n_{\max}=25$,
$-125\le m\le125$, 360 Gauss--Legendre nodes, and $x_{\max}=30$.
The negative-flux half follows from complex conjugation of the fixed-spin orbital Hamiltonian at $b=0$: the quadratic kinetic terms are unchanged while $\phiAB$ changes sign. Consequently, the many-body ground-state energy is even in $\phiAB$, without requiring ordinary time reversal to be a symmetry of the fixed altermagnetic order.
Increasing $n_{\max}$ to 29 changes the $N_e=1800$
energy at $\phiAB=0$, $0.25$, and $0.5$ by at most
$1.1\times10^{-6}\hbar\omega_0$; extending the angular range to
$-155\le m\le155$ changes no displayed energies.

\begin{figure*}[t]
\includegraphics[width=\textwidth]{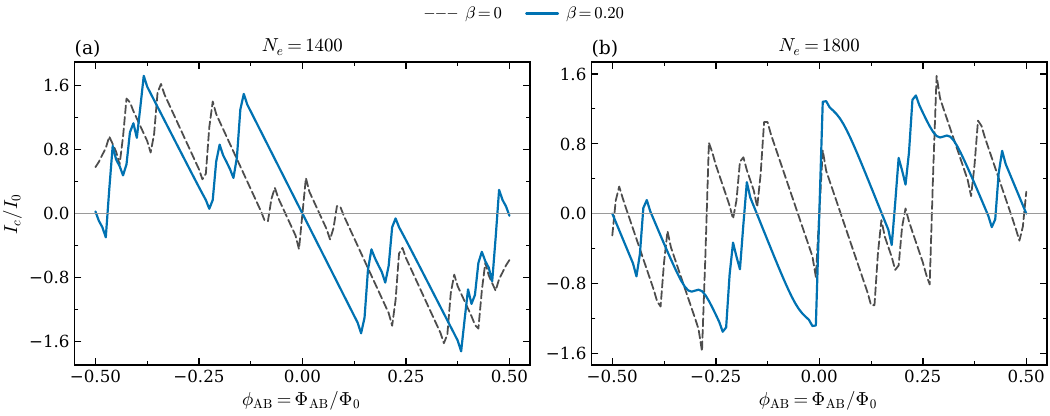}
\caption{Independently recomputed persistent charge current across
one AB period for $\kappa=50$, $b=0$, and $T=0$.
Panels (a,b) give $N_e=1400$ and $1800$. The dashed $\beta=0$
curves use the exact Tan--Inkson spectrum; the solid $\beta=0.20$
curves use independently diagonalized many-body energies.
The symmetric 121-point period is constructed from 61 positive-flux
calculations using the orbital complex-conjugation symmetry at $b=0$; both currents use the same
five-point derivative stencil. Values at ground-state cusps represent
symmetric grid derivatives, not unique local equilibrium slopes.}
\label{fig:Iflux_validated}
\end{figure*}

The persistent current is the flux derivative of this recomputed
fixed-$N_e$ energy. Figure~\ref{fig:Iflux_validated} compares its
finite-grid estimate with the exactly solvable Tan--Inkson reference
using the same five-point differentiation rule on
$\Delta\phiAB=1/120$. The altermagnetic term shifts the shell
rearrangements and redistributes the current over the period,
without changing the exact flux periodicity or lifting the twofold
spin degeneracy. At $\phiAB=0.25$, the finite-grid values are
approximately $-0.33126 I_0$ and $1.15657 I_0$ for $N_e=1400$
and $1800$, respectively. Direct local finite differences with
$\delta\phiAB=2\times10^{-4}$ give $-0.33127 I_0$ and
$1.15660 I_0$, confirming the values away from occupation cusps.
The current at a cusp must instead be specified by one-sided limits.

We emphasize the different status of two field scales. In the
conventional $\beta=0$ model, the slow dHvA-type cycles of
Fig.~\ref{fig:dhva_fillings} follow the successive loss of occupied
radial subbands, whereas the short field-scale oscillations reflect
discrete angular shell rearrangements and have an AB-type orbital
origin. The latter are not strictly topological AB oscillations when
the \emph{homogeneous} field itself is scanned through the conducting
annulus. Our converged $\beta=0.20$ calculations establish local
field-response changes and a full genuine AB-flux response, but they
do not independently resolve a global altermagnetic Fourier spectrum
or the amplitudes of dHvA-type envelopes over $0.8\le b\le6.5$.

\section{Conclusions and outlook}
\label{sec:conclusions}

For a finite-width Tan--Inkson ring, the quadratic $d$-wave term mixes angular momenta by $\Delta m=\pm2$ while leaving angular parity conserved. We have examined this coupling in a perpendicular field and under an independently varied AB flux. The circular geometry also makes the crystallographic angle spectrally irrelevant and maps the two spin sectors into one another by a quarter-turn rotation, producing exact spin isospectrality in the minimal model.

The spectral calculations were checked against exact Tan--Inkson
energies, flux periodicity, rotational and spin symmetries, basis
convergence, and an independent Cartesian finite-difference
calculation. In the high-occupation regime, a separate banded-basis
solver confirms representative energies and local Hellmann--Feynman
derivatives. At low filling, same-parity crossings turn into
anticrossings with gaps linear in $|\beta|$, while crossings between
opposite angular-parity sectors remain protected. Near the Fermi edge
of a highly occupied annulus, a projected pair is embedded in a
multichannel reconstructed spectrum.

The equilibrium responses reflect this structure. For four electrons,
the $m=-1\leftrightarrow+1$ anticrossing removes the zero-flux
open-shell cusp; the finite-temperature current and magnetization
have been recalculated using the \emph{exact canonical partition
function} at fixed integer particle number. This distinction matters
quantitatively at mesoscopic fillings, where fixing only the mean
number can give substantially different currents. For $N_e=1400$
and $1800$, an independently converged field window shows shifts of
the Fermi edge and orbital response due to altermagnetism, while a
separately recomputed full AB period confirms a reorganized persistent
charge current. The exact $\beta=0$ limit exhibits broad
dHvA-type subband-depopulation cycles as well as faster finite-ring
shell oscillations. Full-range harmonic weights of the $\beta=0.20$ response lie
beyond the convergence-tested intervals reported here. Throughout, circular symmetry preserves
spin isospectrality and rules out a persistent equilibrium spin
current in the minimal model.

The remaining question is how device anisotropy modifies the spin isospectrality protected by circular confinement. As discussed in Sec.~\ref{subsec:what_remains}, the vanishing spin response of the circular ring reflects a unitary equivalence between the two spin sectors rather than the absence of an altermagnetic effect: their spatial orbital textures are related by a quarter turn, although corresponding angular-channel weights are identical. Elliptic confinement or attached leads could introduce a fixed device axis and make the crystal orientation $\theta_A$ observable; whether this generates measurable spin asymmetry depends on the detailed perturbation and device coupling. The $\Delta m=\pm2$ matrix elements quantified here establish reference orbital-hybridization scales, not a predetermined spin response. To address conductance or magnetoresistance, the ring must instead be coupled to leads and treated as an open device, as in the geometry-controlled transport problem of Ref.~\cite{LiraSilvaSantangelo2026PhysE}. With such a framework, one could test whether device anisotropy converts the orbital texture found here into a spin-resolved interference signal.

\begin{acknowledgments}
The author acknowledges support from the Conselho Nacional de
Desenvolvimento Cient\'ifico e Tecnol\'ogico (CNPq), under Grant
No.~305427/2026-1; the Funda\c{c}\~ao de Amparo \`a Pesquisa e ao
Desenvolvimento Cient\'ifico e Tecnol\'ogico do Maranh\~ao (FAPEMA),
under Grant No.~UNIVERSAL-06395/22; and the Coordena\c{c}\~ao de
Aperfei\c{c}oamento de Pessoal de N\'ivel Superior (CAPES), Brazil,
under Finance Code 001.

\end{acknowledgments}

\section*{Declaration of AI-assisted tools}
AI-assisted tools were used as auxiliary resources during manuscript preparation and the analytical and numerical workflow. These included ChatGPT (OpenAI), with brief initial use of GPT Astra followed mainly by GPT Pro and GPT-5.6 Sol, and Claude Opus 5.5 (Anthropic). Their use encompassed language refinement, algebraic and numerical cross-checks, assistance with code development and debugging, and organization of the analysis. The author directed the workflow, reviewed the mathematical derivations and computational outputs, checked the physical interpretations and conclusions, and assumes full responsibility for the scientific content.


\begin{thebibliography}{129}%
	\makeatletter
	\providecommand \@ifxundefined [1]{%
		\@ifx{#1\undefined}
	}%
	\providecommand \@ifnum [1]{%
		\ifnum #1\expandafter \@firstoftwo
		\else \expandafter \@secondoftwo
		\fi
	}%
	\providecommand \@ifx [1]{%
		\ifx #1\expandafter \@firstoftwo
		\else \expandafter \@secondoftwo
		\fi
	}%
	\providecommand \natexlab [1]{#1}%
	\providecommand \enquote  [1]{``#1''}%
	\providecommand \bibnamefont  [1]{#1}%
	\providecommand \bibfnamefont [1]{#1}%
	\providecommand \citenamefont [1]{#1}%
	\providecommand \href@noop [0]{\@secondoftwo}%
	\providecommand \href [0]{\begingroup \@sanitize@url \@href}%
	\providecommand \@href[1]{\@@startlink{#1}\@@href}%
	\providecommand \@@href[1]{\endgroup#1\@@endlink}%
	\providecommand \@sanitize@url [0]{\catcode `\\12\catcode `\$12\catcode
		`\&12\catcode `\#12\catcode `\^12\catcode `\_12\catcode `\%12\relax}%
	\providecommand \@@startlink[1]{}%
	\providecommand \@@endlink[0]{}%
	\providecommand \url  [0]{\begingroup\@sanitize@url \@url }%
	\providecommand \@url [1]{\endgroup\@href {#1}{\urlprefix }}%
	\providecommand \urlprefix  [0]{URL }%
	\providecommand \Eprint [0]{\href }%
	\providecommand \doibase [0]{https://doi.org/}%
	\providecommand \selectlanguage [0]{\@gobble}%
	\providecommand \bibinfo  [0]{\@secondoftwo}%
	\providecommand \bibfield  [0]{\@secondoftwo}%
	\providecommand \translation [1]{[#1]}%
	\providecommand \BibitemOpen [0]{}%
	\providecommand \bibitemStop [0]{}%
	\providecommand \bibitemNoStop [0]{.\EOS\space}%
	\providecommand \EOS [0]{\spacefactor3000\relax}%
	\providecommand \BibitemShut  [1]{\csname bibitem#1\endcsname}%
	\let\auto@bib@innerbib\@empty
	\bibitem [{\citenamefont {Jungwirth}\ \emph {et~al.}(2016)\citenamefont
		{Jungwirth}, \citenamefont {Marti}, \citenamefont {Wadley},\ and\
		\citenamefont {Wunderlich}}]{Jungwirth2016NatNano}%
	\BibitemOpen
	\bibfield  {author} {\bibinfo {author} {\bibfnamefont {T.}~\bibnamefont
			{Jungwirth}}, \bibinfo {author} {\bibfnamefont {X.}~\bibnamefont {Marti}},
		\bibinfo {author} {\bibfnamefont {P.}~\bibnamefont {Wadley}},\ and\ \bibinfo
		{author} {\bibfnamefont {J.}~\bibnamefont {Wunderlich}},\ }\bibfield  {title}
	{\bibinfo {title} {{Antiferromagnetic spintronics}},\ }\href
	{https://doi.org/10.1038/nnano.2016.18} {\bibfield  {journal} {\bibinfo
			{journal} {Nat. Nanotechnol.}\ }\textbf {\bibinfo {volume} {11}},\ \bibinfo
		{pages} {231} (\bibinfo {year} {2016})}\BibitemShut {NoStop}%
	\bibitem [{\citenamefont {Baltz}\ \emph {et~al.}(2018)\citenamefont {Baltz},
		\citenamefont {Manchon}, \citenamefont {Tsoi}, \citenamefont {Moriyama},
		\citenamefont {Ono},\ and\ \citenamefont {Tserkovnyak}}]{Baltz2018RMP}%
	\BibitemOpen
	\bibfield  {author} {\bibinfo {author} {\bibfnamefont {V.}~\bibnamefont
			{Baltz}}, \bibinfo {author} {\bibfnamefont {A.}~\bibnamefont {Manchon}},
		\bibinfo {author} {\bibfnamefont {M.}~\bibnamefont {Tsoi}}, \bibinfo {author}
		{\bibfnamefont {T.}~\bibnamefont {Moriyama}}, \bibinfo {author}
		{\bibfnamefont {T.}~\bibnamefont {Ono}},\ and\ \bibinfo {author}
		{\bibfnamefont {Y.}~\bibnamefont {Tserkovnyak}},\ }\bibfield  {title}
	{\bibinfo {title} {{Antiferromagnetic spintronics}},\ }\href
	{https://doi.org/10.1103/RevModPhys.90.015005} {\bibfield  {journal}
		{\bibinfo  {journal} {Rev. Mod. Phys.}\ }\textbf {\bibinfo {volume} {90}},\
		\bibinfo {pages} {015005} (\bibinfo {year} {2018})}\BibitemShut {NoStop}%
	\bibitem [{\citenamefont {Marti}\ \emph {et~al.}(2014)\citenamefont {Marti},
		\citenamefont {Fina}, \citenamefont {Frontera}, \citenamefont {Liu},
		\citenamefont {Wadley}, \citenamefont {He}, \citenamefont {Paull},
		\citenamefont {Clarkson}, \citenamefont {Kudrnovsk\'{y}}, \citenamefont
		{Turek}, \citenamefont {Kune\v{s}}, \citenamefont {Yi}, \citenamefont {Chu},
		\citenamefont {Nelson}, \citenamefont {You}, \citenamefont {Arenholz},
		\citenamefont {Salahuddin}, \citenamefont {Fontcuberta}, \citenamefont
		{Jungwirth},\ and\ \citenamefont {Ramesh}}]{Marti2014NatMater}%
	\BibitemOpen
	\bibfield  {author} {\bibinfo {author} {\bibfnamefont {X.}~\bibnamefont
			{Marti}}, \bibinfo {author} {\bibfnamefont {I.}~\bibnamefont {Fina}},
		\bibinfo {author} {\bibfnamefont {C.}~\bibnamefont {Frontera}}, \bibinfo
		{author} {\bibfnamefont {J.}~\bibnamefont {Liu}}, \bibinfo {author}
		{\bibfnamefont {P.}~\bibnamefont {Wadley}}, \bibinfo {author} {\bibfnamefont
			{Q.}~\bibnamefont {He}}, \bibinfo {author} {\bibfnamefont {R.~J.}\
			\bibnamefont {Paull}}, \bibinfo {author} {\bibfnamefont {J.~D.}\ \bibnamefont
			{Clarkson}}, \bibinfo {author} {\bibfnamefont {J.}~\bibnamefont
			{Kudrnovsk\'{y}}}, \bibinfo {author} {\bibfnamefont {I.}~\bibnamefont
			{Turek}}, \bibinfo {author} {\bibfnamefont {J.}~\bibnamefont {Kune\v{s}}},
		\bibinfo {author} {\bibfnamefont {D.}~\bibnamefont {Yi}}, \bibinfo {author}
		{\bibfnamefont {J.-H.}\ \bibnamefont {Chu}}, \bibinfo {author} {\bibfnamefont
			{C.~T.}\ \bibnamefont {Nelson}}, \bibinfo {author} {\bibfnamefont
			{L.}~\bibnamefont {You}}, \bibinfo {author} {\bibfnamefont {E.}~\bibnamefont
			{Arenholz}}, \bibinfo {author} {\bibfnamefont {S.}~\bibnamefont
			{Salahuddin}}, \bibinfo {author} {\bibfnamefont {J.}~\bibnamefont
			{Fontcuberta}}, \bibinfo {author} {\bibfnamefont {T.}~\bibnamefont
			{Jungwirth}},\ and\ \bibinfo {author} {\bibfnamefont {R.}~\bibnamefont
			{Ramesh}},\ }\bibfield  {title} {\bibinfo {title} {{Room-temperature
				antiferromagnetic memory resistor}},\ }\href
	{https://doi.org/10.1038/nmat3861} {\bibfield  {journal} {\bibinfo  {journal}
			{Nat. Mater.}\ }\textbf {\bibinfo {volume} {13}},\ \bibinfo {pages} {367}
		(\bibinfo {year} {2014})}\BibitemShut {NoStop}%
	\bibitem [{\citenamefont {\v{Z}elezn\'{y}}\ \emph {et~al.}(2014)\citenamefont
		{\v{Z}elezn\'{y}}, \citenamefont {Gao}, \citenamefont {V\'{y}born\'{y}},
		\citenamefont {Zemen}, \citenamefont {Ma\v{s}ek}, \citenamefont {Manchon},
		\citenamefont {Wunderlich}, \citenamefont {Sinova},\ and\ \citenamefont
		{Jungwirth}}]{Zelezny2014PRL}%
	\BibitemOpen
	\bibfield  {author} {\bibinfo {author} {\bibfnamefont {J.}~\bibnamefont
			{\v{Z}elezn\'{y}}}, \bibinfo {author} {\bibfnamefont {H.}~\bibnamefont
			{Gao}}, \bibinfo {author} {\bibfnamefont {K.}~\bibnamefont
			{V\'{y}born\'{y}}}, \bibinfo {author} {\bibfnamefont {J.}~\bibnamefont
			{Zemen}}, \bibinfo {author} {\bibfnamefont {J.}~\bibnamefont {Ma\v{s}ek}},
		\bibinfo {author} {\bibfnamefont {A.}~\bibnamefont {Manchon}}, \bibinfo
		{author} {\bibfnamefont {J.}~\bibnamefont {Wunderlich}}, \bibinfo {author}
		{\bibfnamefont {J.}~\bibnamefont {Sinova}},\ and\ \bibinfo {author}
		{\bibfnamefont {T.}~\bibnamefont {Jungwirth}},\ }\bibfield  {title} {\bibinfo
		{title} {{Relativistic N\'{e}el-Order Fields Induced by Electrical Current in
				Antiferromagnets}},\ }\href {https://doi.org/10.1103/PhysRevLett.113.157201}
	{\bibfield  {journal} {\bibinfo  {journal} {Phys. Rev. Lett.}\ }\textbf
		{\bibinfo {volume} {113}},\ \bibinfo {pages} {157201} (\bibinfo {year}
		{2014})}\BibitemShut {NoStop}%
	\bibitem [{\citenamefont {Wadley}\ \emph {et~al.}(2016)\citenamefont {Wadley},
		\citenamefont {Howells}, \citenamefont {\v{Z}elezn\'{y}}, \citenamefont
		{Andrews}, \citenamefont {Hills}, \citenamefont {Campion}, \citenamefont
		{Nov\'{a}k}, \citenamefont {Olejn\'{i}k}, \citenamefont {Maccherozzi},
		\citenamefont {Dhesi}, \citenamefont {Martin}, \citenamefont {Wagner},
		\citenamefont {Wunderlich}, \citenamefont {Freimuth}, \citenamefont
		{Mokrousov}, \citenamefont {Kune\v{s}}, \citenamefont {Chauhan},
		\citenamefont {Grzybowski}, \citenamefont {Rushforth}, \citenamefont
		{Edmonds}, \citenamefont {Gallagher},\ and\ \citenamefont
		{Jungwirth}}]{Wadley2016Science}%
	\BibitemOpen
	\bibfield  {author} {\bibinfo {author} {\bibfnamefont {P.}~\bibnamefont
			{Wadley}}, \bibinfo {author} {\bibfnamefont {B.}~\bibnamefont {Howells}},
		\bibinfo {author} {\bibfnamefont {J.}~\bibnamefont {\v{Z}elezn\'{y}}},
		\bibinfo {author} {\bibfnamefont {C.}~\bibnamefont {Andrews}}, \bibinfo
		{author} {\bibfnamefont {V.}~\bibnamefont {Hills}}, \bibinfo {author}
		{\bibfnamefont {R.~P.}\ \bibnamefont {Campion}}, \bibinfo {author}
		{\bibfnamefont {V.}~\bibnamefont {Nov\'{a}k}}, \bibinfo {author}
		{\bibfnamefont {K.}~\bibnamefont {Olejn\'{i}k}}, \bibinfo {author}
		{\bibfnamefont {F.}~\bibnamefont {Maccherozzi}}, \bibinfo {author}
		{\bibfnamefont {S.~S.}\ \bibnamefont {Dhesi}}, \bibinfo {author}
		{\bibfnamefont {S.~Y.}\ \bibnamefont {Martin}}, \bibinfo {author}
		{\bibfnamefont {T.}~\bibnamefont {Wagner}}, \bibinfo {author} {\bibfnamefont
			{J.}~\bibnamefont {Wunderlich}}, \bibinfo {author} {\bibfnamefont
			{F.}~\bibnamefont {Freimuth}}, \bibinfo {author} {\bibfnamefont
			{Y.}~\bibnamefont {Mokrousov}}, \bibinfo {author} {\bibfnamefont
			{J.}~\bibnamefont {Kune\v{s}}}, \bibinfo {author} {\bibfnamefont {J.~S.}\
			\bibnamefont {Chauhan}}, \bibinfo {author} {\bibfnamefont {M.~J.}\
			\bibnamefont {Grzybowski}}, \bibinfo {author} {\bibfnamefont {A.~W.}\
			\bibnamefont {Rushforth}}, \bibinfo {author} {\bibfnamefont {K.~W.}\
			\bibnamefont {Edmonds}}, \bibinfo {author} {\bibfnamefont {B.~L.}\
			\bibnamefont {Gallagher}},\ and\ \bibinfo {author} {\bibfnamefont
			{T.}~\bibnamefont {Jungwirth}},\ }\bibfield  {title} {\bibinfo {title}
		{{Electrical switching of an antiferromagnet}},\ }\href
	{https://doi.org/10.1126/science.aab1031} {\bibfield  {journal} {\bibinfo
			{journal} {Science}\ }\textbf {\bibinfo {volume} {351}},\ \bibinfo {pages}
		{587} (\bibinfo {year} {2016})}\BibitemShut {NoStop}%
	\bibitem [{\citenamefont {Manchon}\ \emph {et~al.}(2019)\citenamefont
		{Manchon}, \citenamefont {\v{Z}elezn\'{y}}, \citenamefont {Miron},
		\citenamefont {Jungwirth}, \citenamefont {Sinova}, \citenamefont {Thiaville},
		\citenamefont {Garello},\ and\ \citenamefont {Gambardella}}]{Manchon2019RMP}%
	\BibitemOpen
	\bibfield  {author} {\bibinfo {author} {\bibfnamefont {A.}~\bibnamefont
			{Manchon}}, \bibinfo {author} {\bibfnamefont {J.}~\bibnamefont
			{\v{Z}elezn\'{y}}}, \bibinfo {author} {\bibfnamefont {I.~M.}\ \bibnamefont
			{Miron}}, \bibinfo {author} {\bibfnamefont {T.}~\bibnamefont {Jungwirth}},
		\bibinfo {author} {\bibfnamefont {J.}~\bibnamefont {Sinova}}, \bibinfo
		{author} {\bibfnamefont {A.}~\bibnamefont {Thiaville}}, \bibinfo {author}
		{\bibfnamefont {K.}~\bibnamefont {Garello}},\ and\ \bibinfo {author}
		{\bibfnamefont {P.}~\bibnamefont {Gambardella}},\ }\bibfield  {title}
	{\bibinfo {title} {{Current-induced spin-orbit torques in ferromagnetic and
				antiferromagnetic systems}},\ }\href
	{https://doi.org/10.1103/RevModPhys.91.035004} {\bibfield  {journal}
		{\bibinfo  {journal} {Rev. Mod. Phys.}\ }\textbf {\bibinfo {volume} {91}},\
		\bibinfo {pages} {035004} (\bibinfo {year} {2019})}\BibitemShut {NoStop}%
	\bibitem [{\citenamefont {Nakatsuji}\ \emph {et~al.}(2015)\citenamefont
		{Nakatsuji}, \citenamefont {Kiyohara},\ and\ \citenamefont
		{Higo}}]{Nakatsuji2015Nature}%
	\BibitemOpen
	\bibfield  {author} {\bibinfo {author} {\bibfnamefont {S.}~\bibnamefont
			{Nakatsuji}}, \bibinfo {author} {\bibfnamefont {N.}~\bibnamefont
			{Kiyohara}},\ and\ \bibinfo {author} {\bibfnamefont {T.}~\bibnamefont
			{Higo}},\ }\bibfield  {title} {\bibinfo {title} {{Large anomalous Hall effect
				in a non-collinear antiferromagnet at room temperature}},\ }\href
	{https://doi.org/10.1038/nature15723} {\bibfield  {journal} {\bibinfo
			{journal} {Nature}\ }\textbf {\bibinfo {volume} {527}},\ \bibinfo {pages}
		{212} (\bibinfo {year} {2015})}\BibitemShut {NoStop}%
	\bibitem [{\citenamefont {Nayak}\ \emph {et~al.}(2016)\citenamefont {Nayak},
		\citenamefont {Fischer}, \citenamefont {Sun}, \citenamefont {Yan},
		\citenamefont {Karel}, \citenamefont {Komarek}, \citenamefont {Shekhar},
		\citenamefont {Kumar}, \citenamefont {Schnelle}, \citenamefont {K\"{u}bler},
		\citenamefont {Felser},\ and\ \citenamefont {Parkin}}]{Nayak2016SciAdv}%
	\BibitemOpen
	\bibfield  {author} {\bibinfo {author} {\bibfnamefont {A.~K.}\ \bibnamefont
			{Nayak}}, \bibinfo {author} {\bibfnamefont {J.~E.}\ \bibnamefont {Fischer}},
		\bibinfo {author} {\bibfnamefont {Y.}~\bibnamefont {Sun}}, \bibinfo {author}
		{\bibfnamefont {B.}~\bibnamefont {Yan}}, \bibinfo {author} {\bibfnamefont
			{J.}~\bibnamefont {Karel}}, \bibinfo {author} {\bibfnamefont {A.~C.}\
			\bibnamefont {Komarek}}, \bibinfo {author} {\bibfnamefont {C.}~\bibnamefont
			{Shekhar}}, \bibinfo {author} {\bibfnamefont {N.}~\bibnamefont {Kumar}},
		\bibinfo {author} {\bibfnamefont {W.}~\bibnamefont {Schnelle}}, \bibinfo
		{author} {\bibfnamefont {J.}~\bibnamefont {K\"{u}bler}}, \bibinfo {author}
		{\bibfnamefont {C.}~\bibnamefont {Felser}},\ and\ \bibinfo {author}
		{\bibfnamefont {S.~S.~P.}\ \bibnamefont {Parkin}},\ }\bibfield  {title}
	{\bibinfo {title} {{Large anomalous Hall effect driven by a nonvanishing
				Berry curvature in the noncolinear antiferromagnet Mn3Ge}},\ }\href
	{https://doi.org/10.1126/sciadv.1501870} {\bibfield  {journal} {\bibinfo
			{journal} {Sci. Adv.}\ }\textbf {\bibinfo {volume} {2}},\ \bibinfo {pages}
		{e1501870} (\bibinfo {year} {2016})}\BibitemShut {NoStop}%
	\bibitem [{\citenamefont {Kiyohara}\ \emph {et~al.}(2016)\citenamefont
		{Kiyohara}, \citenamefont {Tomita},\ and\ \citenamefont
		{Nakatsuji}}]{Kiyohara2016PRA}%
	\BibitemOpen
	\bibfield  {author} {\bibinfo {author} {\bibfnamefont {N.}~\bibnamefont
			{Kiyohara}}, \bibinfo {author} {\bibfnamefont {T.}~\bibnamefont {Tomita}},\
		and\ \bibinfo {author} {\bibfnamefont {S.}~\bibnamefont {Nakatsuji}},\
	}\bibfield  {title} {\bibinfo {title} {{Giant Anomalous Hall Effect in the
				Chiral Antiferromagnet {Mn}$_3${Ge}}},\ }\href
	{https://doi.org/10.1103/PhysRevApplied.5.064009} {\bibfield  {journal}
		{\bibinfo  {journal} {Phys. Rev. Applied}\ }\textbf {\bibinfo {volume} {5}},\
		\bibinfo {pages} {064009} (\bibinfo {year} {2016})}\BibitemShut {NoStop}%
	\bibitem [{\citenamefont {Nagaosa}\ \emph {et~al.}(2010)\citenamefont
		{Nagaosa}, \citenamefont {Sinova}, \citenamefont {Onoda}, \citenamefont
		{MacDonald},\ and\ \citenamefont {Ong}}]{Nagaosa2010RMP}%
	\BibitemOpen
	\bibfield  {author} {\bibinfo {author} {\bibfnamefont {N.}~\bibnamefont
			{Nagaosa}}, \bibinfo {author} {\bibfnamefont {J.}~\bibnamefont {Sinova}},
		\bibinfo {author} {\bibfnamefont {S.}~\bibnamefont {Onoda}}, \bibinfo
		{author} {\bibfnamefont {A.~H.}\ \bibnamefont {MacDonald}},\ and\ \bibinfo
		{author} {\bibfnamefont {N.~P.}\ \bibnamefont {Ong}},\ }\bibfield  {title}
	{\bibinfo {title} {{Anomalous Hall effect}},\ }\href
	{https://doi.org/10.1103/RevModPhys.82.1539} {\bibfield  {journal} {\bibinfo
			{journal} {Rev. Mod. Phys.}\ }\textbf {\bibinfo {volume} {82}},\ \bibinfo
		{pages} {1539} (\bibinfo {year} {2010})}\BibitemShut {NoStop}%
	\bibitem [{\citenamefont {Sinova}\ \emph {et~al.}(2015)\citenamefont {Sinova},
		\citenamefont {Valenzuela}, \citenamefont {Wunderlich}, \citenamefont
		{Back},\ and\ \citenamefont {Jungwirth}}]{Sinova2015RMP}%
	\BibitemOpen
	\bibfield  {author} {\bibinfo {author} {\bibfnamefont {J.}~\bibnamefont
			{Sinova}}, \bibinfo {author} {\bibfnamefont {S.~O.}\ \bibnamefont
			{Valenzuela}}, \bibinfo {author} {\bibfnamefont {J.}~\bibnamefont
			{Wunderlich}}, \bibinfo {author} {\bibfnamefont {C.~H.}\ \bibnamefont
			{Back}},\ and\ \bibinfo {author} {\bibfnamefont {T.}~\bibnamefont
			{Jungwirth}},\ }\bibfield  {title} {\bibinfo {title} {{Spin Hall effects}},\
	}\href {https://doi.org/10.1103/RevModPhys.87.1213} {\bibfield  {journal}
		{\bibinfo  {journal} {Rev. Mod. Phys.}\ }\textbf {\bibinfo {volume} {87}},\
		\bibinfo {pages} {1213} (\bibinfo {year} {2015})}\BibitemShut {NoStop}%
	\bibitem [{\citenamefont {Chen}\ \emph {et~al.}(2014)\citenamefont {Chen},
		\citenamefont {Niu},\ and\ \citenamefont {MacDonald}}]{Chen2014PRL}%
	\BibitemOpen
	\bibfield  {author} {\bibinfo {author} {\bibfnamefont {H.}~\bibnamefont
			{Chen}}, \bibinfo {author} {\bibfnamefont {Q.}~\bibnamefont {Niu}},\ and\
		\bibinfo {author} {\bibfnamefont {A.~H.}\ \bibnamefont {MacDonald}},\
	}\bibfield  {title} {\bibinfo {title} {{Anomalous Hall Effect Arising from
				Noncollinear Antiferromagnetism}},\ }\href
	{https://doi.org/10.1103/PhysRevLett.112.017205} {\bibfield  {journal}
		{\bibinfo  {journal} {Phys. Rev. Lett.}\ }\textbf {\bibinfo {volume} {112}},\
		\bibinfo {pages} {017205} (\bibinfo {year} {2014})}\BibitemShut {NoStop}%
	\bibitem [{\citenamefont {Smejkal}\ \emph {et~al.}(2017)\citenamefont
		{Smejkal}, \citenamefont {Zelezny}, \citenamefont {Sinova},\ and\
		\citenamefont {Jungwirth}}]{Smejkal2017PRL}%
	\BibitemOpen
	\bibfield  {author} {\bibinfo {author} {\bibfnamefont {L.}~\bibnamefont
			{Smejkal}}, \bibinfo {author} {\bibfnamefont {J.}~\bibnamefont {Zelezny}},
		\bibinfo {author} {\bibfnamefont {J.}~\bibnamefont {Sinova}},\ and\ \bibinfo
		{author} {\bibfnamefont {T.}~\bibnamefont {Jungwirth}},\ }\bibfield  {title}
	{\bibinfo {title} {{Electric Control of Dirac Quasiparticles by Spin-Orbit
				Torque in an Antiferromagnet}},\ }\href
	{https://doi.org/10.1103/PhysRevLett.118.106402} {\bibfield  {journal}
		{\bibinfo  {journal} {Phys. Rev. Lett.}\ }\textbf {\bibinfo {volume} {118}},\
		\bibinfo {pages} {106402} (\bibinfo {year} {2017})}\BibitemShut {NoStop}%
	\bibitem [{\citenamefont {Smejkal}\ \emph {et~al.}(2022)\citenamefont
		{Smejkal}, \citenamefont {MacDonald}, \citenamefont {Sinova}, \citenamefont
		{Nakatsuji},\ and\ \citenamefont {Jungwirth}}]{Smejkal2022NRMHall}%
	\BibitemOpen
	\bibfield  {author} {\bibinfo {author} {\bibfnamefont {L.}~\bibnamefont
			{Smejkal}}, \bibinfo {author} {\bibfnamefont {A.~H.}\ \bibnamefont
			{MacDonald}}, \bibinfo {author} {\bibfnamefont {J.}~\bibnamefont {Sinova}},
		\bibinfo {author} {\bibfnamefont {S.}~\bibnamefont {Nakatsuji}},\ and\
		\bibinfo {author} {\bibfnamefont {T.}~\bibnamefont {Jungwirth}},\ }\bibfield
	{title} {\bibinfo {title} {{Anomalous Hall antiferromagnets}},\ }\href
	{https://doi.org/10.1038/s41578-022-00430-3} {\bibfield  {journal} {\bibinfo
			{journal} {Nat. Rev. Mater.}\ }\textbf {\bibinfo {volume} {7}},\ \bibinfo
		{pages} {482} (\bibinfo {year} {2022})}\BibitemShut {NoStop}%
	\bibitem [{\citenamefont {Naka}\ \emph {et~al.}(2019)\citenamefont {Naka},
		\citenamefont {Hayami}, \citenamefont {Kusunose}, \citenamefont {Yanagi},
		\citenamefont {Motome},\ and\ \citenamefont {Seo}}]{Naka2019NatCommun}%
	\BibitemOpen
	\bibfield  {author} {\bibinfo {author} {\bibfnamefont {M.}~\bibnamefont
			{Naka}}, \bibinfo {author} {\bibfnamefont {S.}~\bibnamefont {Hayami}},
		\bibinfo {author} {\bibfnamefont {H.}~\bibnamefont {Kusunose}}, \bibinfo
		{author} {\bibfnamefont {Y.}~\bibnamefont {Yanagi}}, \bibinfo {author}
		{\bibfnamefont {Y.}~\bibnamefont {Motome}},\ and\ \bibinfo {author}
		{\bibfnamefont {H.}~\bibnamefont {Seo}},\ }\bibfield  {title} {\bibinfo
		{title} {{Spin current generation in organic antiferromagnets}},\ }\href
	{https://doi.org/10.1038/s41467-019-12229-y} {\bibfield  {journal} {\bibinfo
			{journal} {Nat. Commun.}\ }\textbf {\bibinfo {volume} {10}},\ \bibinfo
		{pages} {4305} (\bibinfo {year} {2019})}\BibitemShut {NoStop}%
	\bibitem [{\citenamefont {Smejkal}\ \emph {et~al.}(2020)\citenamefont
		{Smejkal}, \citenamefont {Gonzalez-Hernandez}, \citenamefont {Jungwirth},\
		and\ \citenamefont {Sinova}}]{Smejkal2020SciAdv}%
	\BibitemOpen
	\bibfield  {author} {\bibinfo {author} {\bibfnamefont {L.}~\bibnamefont
			{Smejkal}}, \bibinfo {author} {\bibfnamefont {R.}~\bibnamefont
			{Gonzalez-Hernandez}}, \bibinfo {author} {\bibfnamefont {T.}~\bibnamefont
			{Jungwirth}},\ and\ \bibinfo {author} {\bibfnamefont {J.}~\bibnamefont
			{Sinova}},\ }\bibfield  {title} {\bibinfo {title} {{Crystal time-reversal
				symmetry breaking and spontaneous Hall effect in collinear
				antiferromagnets}},\ }\href {https://doi.org/10.1126/sciadv.aaz8809}
	{\bibfield  {journal} {\bibinfo  {journal} {Sci. Adv.}\ }\textbf {\bibinfo
			{volume} {6}},\ \bibinfo {pages} {eaaz8809} (\bibinfo {year}
		{2020})}\BibitemShut {NoStop}%
	\bibitem [{\citenamefont {Ahn}\ \emph {et~al.}(2019)\citenamefont {Ahn},
		\citenamefont {Hariki}, \citenamefont {Lee},\ and\ \citenamefont
		{Kunes}}]{Ahn2019PRB}%
	\BibitemOpen
	\bibfield  {author} {\bibinfo {author} {\bibfnamefont {K.~H.}\ \bibnamefont
			{Ahn}}, \bibinfo {author} {\bibfnamefont {A.}~\bibnamefont {Hariki}},
		\bibinfo {author} {\bibfnamefont {K.~W.}\ \bibnamefont {Lee}},\ and\ \bibinfo
		{author} {\bibfnamefont {J.}~\bibnamefont {Kunes}},\ }\bibfield  {title}
	{\bibinfo {title} {{Antiferromagnetism in RuO2 as d-wave Pomeranchuk
				instability}},\ }\href {https://doi.org/10.1103/PhysRevB.99.184432}
	{\bibfield  {journal} {\bibinfo  {journal} {Phys. Rev. B}\ }\textbf {\bibinfo
			{volume} {99}},\ \bibinfo {pages} {184432} (\bibinfo {year}
		{2019})}\BibitemShut {NoStop}%
	\bibitem [{\citenamefont {Yuan}\ \emph {et~al.}(2020)\citenamefont {Yuan},
		\citenamefont {Wang}, \citenamefont {Luo}, \citenamefont {Rashba},\ and\
		\citenamefont {Zunger}}]{Yuan2020PRB}%
	\BibitemOpen
	\bibfield  {author} {\bibinfo {author} {\bibfnamefont {L.~D.}\ \bibnamefont
			{Yuan}}, \bibinfo {author} {\bibfnamefont {Z.}~\bibnamefont {Wang}}, \bibinfo
		{author} {\bibfnamefont {J.~W.}\ \bibnamefont {Luo}}, \bibinfo {author}
		{\bibfnamefont {E.~I.}\ \bibnamefont {Rashba}},\ and\ \bibinfo {author}
		{\bibfnamefont {A.}~\bibnamefont {Zunger}},\ }\bibfield  {title} {\bibinfo
		{title} {{Giant momentum-dependent spin splitting in centrosymmetric low-Z
				antiferromagnets}},\ }\href {https://doi.org/10.1103/PhysRevB.102.014422}
	{\bibfield  {journal} {\bibinfo  {journal} {Phys. Rev. B}\ }\textbf {\bibinfo
			{volume} {102}},\ \bibinfo {pages} {014422} (\bibinfo {year}
		{2020})}\BibitemShut {NoStop}%
	\bibitem [{\citenamefont {Hayami}\ \emph {et~al.}(2020)\citenamefont {Hayami},
		\citenamefont {Yanagi},\ and\ \citenamefont {Kusunose}}]{Hayami2020PRB}%
	\BibitemOpen
	\bibfield  {author} {\bibinfo {author} {\bibfnamefont {S.}~\bibnamefont
			{Hayami}}, \bibinfo {author} {\bibfnamefont {Y.}~\bibnamefont {Yanagi}},\
		and\ \bibinfo {author} {\bibfnamefont {H.}~\bibnamefont {Kusunose}},\
	}\bibfield  {title} {\bibinfo {title} {{Bottom-up design of spin-split and
				reshaped electronic band structures in antiferromagnets without spin-orbit
				coupling: Procedure on the basis of augmented multipoles}},\ }\href
	{https://doi.org/10.1103/PhysRevB.102.144441} {\bibfield  {journal} {\bibinfo
			{journal} {Phys. Rev. B}\ }\textbf {\bibinfo {volume} {102}},\ \bibinfo
		{pages} {144441} (\bibinfo {year} {2020})}\BibitemShut {NoStop}%
	\bibitem [{\citenamefont {Ma}\ \emph {et~al.}(2021)\citenamefont {Ma},
		\citenamefont {Hu}, \citenamefont {Li}, \citenamefont {Liu}, \citenamefont
		{Yao}, \citenamefont {Jia},\ and\ \citenamefont {Liu}}]{Ma2021NatCommun}%
	\BibitemOpen
	\bibfield  {author} {\bibinfo {author} {\bibfnamefont {H.-Y.}\ \bibnamefont
			{Ma}}, \bibinfo {author} {\bibfnamefont {M.}~\bibnamefont {Hu}}, \bibinfo
		{author} {\bibfnamefont {N.}~\bibnamefont {Li}}, \bibinfo {author}
		{\bibfnamefont {J.}~\bibnamefont {Liu}}, \bibinfo {author} {\bibfnamefont
			{W.}~\bibnamefont {Yao}}, \bibinfo {author} {\bibfnamefont {J.-F.}\
			\bibnamefont {Jia}},\ and\ \bibinfo {author} {\bibfnamefont {J.}~\bibnamefont
			{Liu}},\ }\bibfield  {title} {\bibinfo {title} {{Multifunctional
				antiferromagnetic materials with giant piezomagnetism and noncollinear spin
				current}},\ }\href {https://doi.org/10.1038/s41467-021-23127-7} {\bibfield
		{journal} {\bibinfo  {journal} {Nat. Commun.}\ }\textbf {\bibinfo {volume}
			{12}},\ \bibinfo {pages} {2846} (\bibinfo {year} {2021})}\BibitemShut
	{NoStop}%
	\bibitem [{\citenamefont {Mazin}\ \emph {et~al.}(2021)\citenamefont {Mazin},
		\citenamefont {Koepernik}, \citenamefont {Johannes}, \citenamefont
		{Gonzalez-Hernandez},\ and\ \citenamefont {Smejkal}}]{Mazin2021PNAS}%
	\BibitemOpen
	\bibfield  {author} {\bibinfo {author} {\bibfnamefont {I.~I.}\ \bibnamefont
			{Mazin}}, \bibinfo {author} {\bibfnamefont {K.}~\bibnamefont {Koepernik}},
		\bibinfo {author} {\bibfnamefont {M.~D.}\ \bibnamefont {Johannes}}, \bibinfo
		{author} {\bibfnamefont {R.}~\bibnamefont {Gonzalez-Hernandez}},\ and\
		\bibinfo {author} {\bibfnamefont {L.}~\bibnamefont {Smejkal}},\ }\bibfield
	{title} {\bibinfo {title} {{Prediction of unconventional magnetism in doped
				{FeSb}$_2$}},\ }\href {https://doi.org/10.1073/pnas.2108924118} {\bibfield
		{journal} {\bibinfo  {journal} {Proc. Natl. Acad. Sci. USA}\ }\textbf
		{\bibinfo {volume} {118}},\ \bibinfo {pages} {e2108924118} (\bibinfo {year}
		{2021})}\BibitemShut {NoStop}%
	\bibitem [{\citenamefont {Liu}\ \emph {et~al.}(2022)\citenamefont {Liu},
		\citenamefont {Li}, \citenamefont {Han}, \citenamefont {Wan},\ and\
		\citenamefont {Liu}}]{Liu2022PRXSpinGroup}%
	\BibitemOpen
	\bibfield  {author} {\bibinfo {author} {\bibfnamefont {P.}~\bibnamefont
			{Liu}}, \bibinfo {author} {\bibfnamefont {J.}~\bibnamefont {Li}}, \bibinfo
		{author} {\bibfnamefont {J.}~\bibnamefont {Han}}, \bibinfo {author}
		{\bibfnamefont {X.}~\bibnamefont {Wan}},\ and\ \bibinfo {author}
		{\bibfnamefont {Q.}~\bibnamefont {Liu}},\ }\bibfield  {title} {\bibinfo
		{title} {{Spin-Group Symmetry in Magnetic Materials with Negligible
				Spin-Orbit Coupling}},\ }\href {https://doi.org/10.1103/PhysRevX.12.021016}
	{\bibfield  {journal} {\bibinfo  {journal} {Phys. Rev. X}\ }\textbf {\bibinfo
			{volume} {12}},\ \bibinfo {pages} {021016} (\bibinfo {year}
		{2022})}\BibitemShut {NoStop}%
	\bibitem [{\citenamefont {\v{S}mejkal}\ \emph
		{et~al.}(2022{\natexlab{a}})\citenamefont {\v{S}mejkal}, \citenamefont
		{Sinova},\ and\ \citenamefont {Jungwirth}}]{Smejkal2022PRXPhase}%
	\BibitemOpen
	\bibfield  {author} {\bibinfo {author} {\bibfnamefont {L.}~\bibnamefont
			{\v{S}mejkal}}, \bibinfo {author} {\bibfnamefont {J.}~\bibnamefont
			{Sinova}},\ and\ \bibinfo {author} {\bibfnamefont {T.}~\bibnamefont
			{Jungwirth}},\ }\bibfield  {title} {\bibinfo {title} {{Beyond Conventional
				Ferromagnetism and Antiferromagnetism: A Phase with Nonrelativistic Spin and
				Crystal Rotation Symmetry}},\ }\href
	{https://doi.org/10.1103/PhysRevX.12.031042} {\bibfield  {journal} {\bibinfo
			{journal} {Phys. Rev. X}\ }\textbf {\bibinfo {volume} {12}},\ \bibinfo
		{pages} {031042} (\bibinfo {year} {2022}{\natexlab{a}})}\BibitemShut
	{NoStop}%
	\bibitem [{\citenamefont {\v{S}mejkal}\ \emph
		{et~al.}(2022{\natexlab{b}})\citenamefont {\v{S}mejkal}, \citenamefont
		{Sinova},\ and\ \citenamefont {Jungwirth}}]{Smejkal2022PRXLandscape}%
	\BibitemOpen
	\bibfield  {author} {\bibinfo {author} {\bibfnamefont {L.}~\bibnamefont
			{\v{S}mejkal}}, \bibinfo {author} {\bibfnamefont {J.}~\bibnamefont
			{Sinova}},\ and\ \bibinfo {author} {\bibfnamefont {T.}~\bibnamefont
			{Jungwirth}},\ }\bibfield  {title} {\bibinfo {title} {{Emerging Research
				Landscape of Altermagnetism}},\ }\href
	{https://doi.org/10.1103/PhysRevX.12.040501} {\bibfield  {journal} {\bibinfo
			{journal} {Phys. Rev. X}\ }\textbf {\bibinfo {volume} {12}},\ \bibinfo
		{pages} {040501} (\bibinfo {year} {2022}{\natexlab{b}})}\BibitemShut
	{NoStop}%
	\bibitem [{\citenamefont {Mazin}(2022)}]{Mazin2022PRX}%
	\BibitemOpen
	\bibfield  {author} {\bibinfo {author} {\bibfnamefont {I.~I.}\ \bibnamefont
			{Mazin}},\ }\bibfield  {title} {\bibinfo {title} {{Editorial:
				Altermagnetism---A New Punch Line of Fundamental Magnetism}},\ }\href
	{https://doi.org/10.1103/PhysRevX.12.040002} {\bibfield  {journal} {\bibinfo
			{journal} {Phys. Rev. X}\ }\textbf {\bibinfo {volume} {12}},\ \bibinfo
		{pages} {040002} (\bibinfo {year} {2022})}\BibitemShut {NoStop}%
	\bibitem [{\citenamefont {Gonz\'{a}lez-Hern\'{a}ndez}\ \emph
		{et~al.}(2021)\citenamefont {Gonz\'{a}lez-Hern\'{a}ndez}, \citenamefont
		{\v{S}mejkal}, \citenamefont {V\'{y}born\'{y}}, \citenamefont {Yahagi},
		\citenamefont {Sinova}, \citenamefont {Jungwirth},\ and\ \citenamefont
		{\v{Z}elezn\'{y}}}]{GonzalezHernandez2021PRL}%
	\BibitemOpen
	\bibfield  {author} {\bibinfo {author} {\bibfnamefont {R.}~\bibnamefont
			{Gonz\'{a}lez-Hern\'{a}ndez}}, \bibinfo {author} {\bibfnamefont
			{L.}~\bibnamefont {\v{S}mejkal}}, \bibinfo {author} {\bibfnamefont
			{K.}~\bibnamefont {V\'{y}born\'{y}}}, \bibinfo {author} {\bibfnamefont
			{Y.}~\bibnamefont {Yahagi}}, \bibinfo {author} {\bibfnamefont
			{J.}~\bibnamefont {Sinova}}, \bibinfo {author} {\bibfnamefont
			{T.}~\bibnamefont {Jungwirth}},\ and\ \bibinfo {author} {\bibfnamefont
			{J.}~\bibnamefont {\v{Z}elezn\'{y}}},\ }\bibfield  {title} {\bibinfo {title}
		{{Efficient Electrical Spin Splitter Based on Nonrelativistic Collinear
				Antiferromagnetism}},\ }\href
	{https://doi.org/10.1103/PhysRevLett.126.127701} {\bibfield  {journal}
		{\bibinfo  {journal} {Phys. Rev. Lett.}\ }\textbf {\bibinfo {volume} {126}},\
		\bibinfo {pages} {127701} (\bibinfo {year} {2021})}\BibitemShut {NoStop}%
	\bibitem [{\citenamefont {Smejkal}\ \emph {et~al.}(2022)\citenamefont
		{Smejkal}, \citenamefont {Hellenes}, \citenamefont {Gonzalez-Hernandez},
		\citenamefont {Sinova},\ and\ \citenamefont {Jungwirth}}]{Smejkal2022PRXTMR}%
	\BibitemOpen
	\bibfield  {author} {\bibinfo {author} {\bibfnamefont {L.}~\bibnamefont
			{Smejkal}}, \bibinfo {author} {\bibfnamefont {A.~B.}\ \bibnamefont
			{Hellenes}}, \bibinfo {author} {\bibfnamefont {R.}~\bibnamefont
			{Gonzalez-Hernandez}}, \bibinfo {author} {\bibfnamefont {J.}~\bibnamefont
			{Sinova}},\ and\ \bibinfo {author} {\bibfnamefont {T.}~\bibnamefont
			{Jungwirth}},\ }\bibfield  {title} {\bibinfo {title} {{Giant and Tunneling
				Magnetoresistance in Unconventional Collinear Antiferromagnets with
				Nonrelativistic Spin-Momentum Coupling}},\ }\href
	{https://doi.org/10.1103/PhysRevX.12.011028} {\bibfield  {journal} {\bibinfo
			{journal} {Phys. Rev. X}\ }\textbf {\bibinfo {volume} {12}},\ \bibinfo
		{pages} {011028} (\bibinfo {year} {2022})}\BibitemShut {NoStop}%
	\bibitem [{\citenamefont {Shao}\ \emph {et~al.}(2021)\citenamefont {Shao},
		\citenamefont {Zhang}, \citenamefont {Li}, \citenamefont {Eom},\ and\
		\citenamefont {Tsymbal}}]{Shao2021NatCommun}%
	\BibitemOpen
	\bibfield  {author} {\bibinfo {author} {\bibfnamefont {D.-F.}\ \bibnamefont
			{Shao}}, \bibinfo {author} {\bibfnamefont {S.-H.}\ \bibnamefont {Zhang}},
		\bibinfo {author} {\bibfnamefont {M.}~\bibnamefont {Li}}, \bibinfo {author}
		{\bibfnamefont {C.-B.}\ \bibnamefont {Eom}},\ and\ \bibinfo {author}
		{\bibfnamefont {E.~Y.}\ \bibnamefont {Tsymbal}},\ }\bibfield  {title}
	{\bibinfo {title} {{Spin-neutral currents for spintronics}},\ }\href
	{https://doi.org/10.1038/s41467-021-26915-3} {\bibfield  {journal} {\bibinfo
			{journal} {Nat. Commun.}\ }\textbf {\bibinfo {volume} {12}},\ \bibinfo
		{pages} {7061} (\bibinfo {year} {2021})}\BibitemShut {NoStop}%
	\bibitem [{\citenamefont {Bhowal}\ and\ \citenamefont
		{Spaldin}(2024)}]{Bhowal2024PRX}%
	\BibitemOpen
	\bibfield  {author} {\bibinfo {author} {\bibfnamefont {S.}~\bibnamefont
			{Bhowal}}\ and\ \bibinfo {author} {\bibfnamefont {N.~A.}\ \bibnamefont
			{Spaldin}},\ }\bibfield  {title} {\bibinfo {title} {{Ferroically Ordered
				Magnetic Octupoles in d-Wave Altermagnets}},\ }\href
	{https://doi.org/10.1103/PhysRevX.14.011019} {\bibfield  {journal} {\bibinfo
			{journal} {Phys. Rev. X}\ }\textbf {\bibinfo {volume} {14}},\ \bibinfo
		{pages} {011019} (\bibinfo {year} {2024})}\BibitemShut {NoStop}%
	\bibitem [{\citenamefont {McClarty}\ and\ \citenamefont
		{Rau}(2024)}]{McClarty2024PRL}%
	\BibitemOpen
	\bibfield  {author} {\bibinfo {author} {\bibfnamefont {P.~A.}\ \bibnamefont
			{McClarty}}\ and\ \bibinfo {author} {\bibfnamefont {J.~G.}\ \bibnamefont
			{Rau}},\ }\bibfield  {title} {\bibinfo {title} {{Landau Theory of
				Altermagnetism}},\ }\href {https://doi.org/10.1103/PhysRevLett.132.176702}
	{\bibfield  {journal} {\bibinfo  {journal} {Phys. Rev. Lett.}\ }\textbf
		{\bibinfo {volume} {132}},\ \bibinfo {pages} {176702} (\bibinfo {year}
		{2024})}\BibitemShut {NoStop}%
	\bibitem [{\citenamefont {Jaeschke-Ubiergo}\ \emph {et~al.}(2024)\citenamefont
		{Jaeschke-Ubiergo}, \citenamefont {Bharadwaj}, \citenamefont {Jungwirth},
		\citenamefont {\v{S}mejkal},\ and\ \citenamefont {Sinova}}]{Jaeschke2024PRB}%
	\BibitemOpen
	\bibfield  {author} {\bibinfo {author} {\bibfnamefont {R.}~\bibnamefont
			{Jaeschke-Ubiergo}}, \bibinfo {author} {\bibfnamefont {V.~K.}\ \bibnamefont
			{Bharadwaj}}, \bibinfo {author} {\bibfnamefont {T.}~\bibnamefont
			{Jungwirth}}, \bibinfo {author} {\bibfnamefont {L.}~\bibnamefont
			{\v{S}mejkal}},\ and\ \bibinfo {author} {\bibfnamefont {J.}~\bibnamefont
			{Sinova}},\ }\bibfield  {title} {\bibinfo {title} {{Supercell
				altermagnets}},\ }\href {https://doi.org/10.1103/PhysRevB.109.094425}
	{\bibfield  {journal} {\bibinfo  {journal} {Phys. Rev. B}\ }\textbf {\bibinfo
			{volume} {109}},\ \bibinfo {pages} {094425} (\bibinfo {year}
		{2024})}\BibitemShut {NoStop}%
	\bibitem [{\citenamefont {Das}\ \emph {et~al.}(2024)\citenamefont {Das},
		\citenamefont {Leeb}, \citenamefont {Knolle},\ and\ \citenamefont
		{Knap}}]{Das2024PRL}%
	\BibitemOpen
	\bibfield  {author} {\bibinfo {author} {\bibfnamefont {P.}~\bibnamefont
			{Das}}, \bibinfo {author} {\bibfnamefont {V.}~\bibnamefont {Leeb}}, \bibinfo
		{author} {\bibfnamefont {J.}~\bibnamefont {Knolle}},\ and\ \bibinfo {author}
		{\bibfnamefont {M.}~\bibnamefont {Knap}},\ }\bibfield  {title} {\bibinfo
		{title} {{Realizing Altermagnetism in Fermi-Hubbard Models with Ultracold
				Atoms}},\ }\href {https://doi.org/10.1103/PhysRevLett.132.263402} {\bibfield
		{journal} {\bibinfo  {journal} {Phys. Rev. Lett.}\ }\textbf {\bibinfo
			{volume} {132}},\ \bibinfo {pages} {263402} (\bibinfo {year}
		{2024})}\BibitemShut {NoStop}%
	\bibitem [{\citenamefont {Liu}\ \emph {et~al.}(2024{\natexlab{a}})\citenamefont
		{Liu}, \citenamefont {Yu},\ and\ \citenamefont {Liu}}]{Liu2024PRLTwisted}%
	\BibitemOpen
	\bibfield  {author} {\bibinfo {author} {\bibfnamefont {D.}~\bibnamefont
			{Liu}}, \bibinfo {author} {\bibfnamefont {J.}~\bibnamefont {Yu}},\ and\
		\bibinfo {author} {\bibfnamefont {Q.}~\bibnamefont {Liu}},\ }\bibfield
	{title} {\bibinfo {title} {{Twisted Magnetic Van der Waals Bilayers: An Ideal
				Platform for Altermagnetism}},\ }\href
	{https://doi.org/10.1103/PhysRevLett.133.206702} {\bibfield  {journal}
		{\bibinfo  {journal} {Phys. Rev. Lett.}\ }\textbf {\bibinfo {volume} {133}},\
		\bibinfo {pages} {206702} (\bibinfo {year} {2024}{\natexlab{a}})}\BibitemShut
	{NoStop}%
	\bibitem [{\citenamefont {Bai}\ \emph {et~al.}(2024)\citenamefont {Bai},
		\citenamefont {Feng}, \citenamefont {Liu}, \citenamefont {Smejkal},
		\citenamefont {Mokrousov},\ and\ \citenamefont {Yao}}]{Bai2024AFM}%
	\BibitemOpen
	\bibfield  {author} {\bibinfo {author} {\bibfnamefont {L.}~\bibnamefont
			{Bai}}, \bibinfo {author} {\bibfnamefont {W.}~\bibnamefont {Feng}}, \bibinfo
		{author} {\bibfnamefont {S.}~\bibnamefont {Liu}}, \bibinfo {author}
		{\bibfnamefont {L.}~\bibnamefont {Smejkal}}, \bibinfo {author} {\bibfnamefont
			{Y.}~\bibnamefont {Mokrousov}},\ and\ \bibinfo {author} {\bibfnamefont
			{Y.}~\bibnamefont {Yao}},\ }\bibfield  {title} {\bibinfo {title}
		{{Altermagnetism: Exploring New Frontiers in Magnetism and Spintronics}},\
	}\href {https://doi.org/10.1002/adfm.202409327} {\bibfield  {journal}
		{\bibinfo  {journal} {Adv. Funct. Mater.}\ }\textbf {\bibinfo {volume}
			{34}},\ \bibinfo {pages} {2409327} (\bibinfo {year} {2024})}\BibitemShut
	{NoStop}%
	\bibitem [{\citenamefont {Berlijn}\ \emph {et~al.}(2017)\citenamefont
		{Berlijn}, \citenamefont {Snijders}, \citenamefont {Delaire}, \citenamefont
		{Zhou}, \citenamefont {Maier}, \citenamefont {Cao}, \citenamefont {Chi},
		\citenamefont {Matsuda}, \citenamefont {Wang}, \citenamefont {Koehler},
		\citenamefont {Kent},\ and\ \citenamefont {Weitering}}]{Berlijn2017PRL}%
	\BibitemOpen
	\bibfield  {author} {\bibinfo {author} {\bibfnamefont {T.}~\bibnamefont
			{Berlijn}}, \bibinfo {author} {\bibfnamefont {P.~C.}\ \bibnamefont
			{Snijders}}, \bibinfo {author} {\bibfnamefont {O.}~\bibnamefont {Delaire}},
		\bibinfo {author} {\bibfnamefont {H.-D.}\ \bibnamefont {Zhou}}, \bibinfo
		{author} {\bibfnamefont {T.~A.}\ \bibnamefont {Maier}}, \bibinfo {author}
		{\bibfnamefont {H.-B.}\ \bibnamefont {Cao}}, \bibinfo {author} {\bibfnamefont
			{S.-X.}\ \bibnamefont {Chi}}, \bibinfo {author} {\bibfnamefont
			{M.}~\bibnamefont {Matsuda}}, \bibinfo {author} {\bibfnamefont
			{Y.}~\bibnamefont {Wang}}, \bibinfo {author} {\bibfnamefont {M.~R.}\
			\bibnamefont {Koehler}}, \bibinfo {author} {\bibfnamefont {P.~R.~C.}\
			\bibnamefont {Kent}},\ and\ \bibinfo {author} {\bibfnamefont {H.~H.}\
			\bibnamefont {Weitering}},\ }\bibfield  {title} {\bibinfo {title} {{Itinerant
				Antiferromagnetism in RuO2}},\ }\href
	{https://doi.org/10.1103/PhysRevLett.118.077201} {\bibfield  {journal}
		{\bibinfo  {journal} {Phys. Rev. Lett.}\ }\textbf {\bibinfo {volume} {118}},\
		\bibinfo {pages} {077201} (\bibinfo {year} {2017})}\BibitemShut {NoStop}%
	\bibitem [{\citenamefont {Zhu}\ \emph {et~al.}(2019)\citenamefont {Zhu},
		\citenamefont {Strempfer}, \citenamefont {Rao}, \citenamefont {Occhialini},
		\citenamefont {Pelliciari}, \citenamefont {Choi}, \citenamefont {Kawaguchi},
		\citenamefont {You}, \citenamefont {Mitchell}, \citenamefont {Shao-Horn},\
		and\ \citenamefont {Comin}}]{Zhu2019PRL}%
	\BibitemOpen
	\bibfield  {author} {\bibinfo {author} {\bibfnamefont {Z.~H.}\ \bibnamefont
			{Zhu}}, \bibinfo {author} {\bibfnamefont {J.}~\bibnamefont {Strempfer}},
		\bibinfo {author} {\bibfnamefont {R.~R.}\ \bibnamefont {Rao}}, \bibinfo
		{author} {\bibfnamefont {C.~A.}\ \bibnamefont {Occhialini}}, \bibinfo
		{author} {\bibfnamefont {J.}~\bibnamefont {Pelliciari}}, \bibinfo {author}
		{\bibfnamefont {Y.}~\bibnamefont {Choi}}, \bibinfo {author} {\bibfnamefont
			{T.}~\bibnamefont {Kawaguchi}}, \bibinfo {author} {\bibfnamefont
			{H.}~\bibnamefont {You}}, \bibinfo {author} {\bibfnamefont {J.~F.}\
			\bibnamefont {Mitchell}}, \bibinfo {author} {\bibfnamefont {Y.}~\bibnamefont
			{Shao-Horn}},\ and\ \bibinfo {author} {\bibfnamefont {R.}~\bibnamefont
			{Comin}},\ }\bibfield  {title} {\bibinfo {title} {{Anomalous
				Antiferromagnetism in Metallic RuO2 Determined by Resonant X-ray
				Scattering}},\ }\href {https://doi.org/10.1103/PhysRevLett.122.017202}
	{\bibfield  {journal} {\bibinfo  {journal} {Phys. Rev. Lett.}\ }\textbf
		{\bibinfo {volume} {122}},\ \bibinfo {pages} {017202} (\bibinfo {year}
		{2019})}\BibitemShut {NoStop}%
	\bibitem [{\citenamefont {Feng}\ \emph {et~al.}(2022)\citenamefont {Feng},
		\citenamefont {Zhou}, \citenamefont {\v{S}mejkal}, \citenamefont {Wu},
		\citenamefont {Zhu}, \citenamefont {Guo}, \citenamefont
		{Gonz\'{a}lez-Hern\'{a}ndez}, \citenamefont {Wang}, \citenamefont {Yan},
		\citenamefont {Qin}, \citenamefont {Zhang}, \citenamefont {Wu}, \citenamefont
		{Chen}, \citenamefont {Meng}, \citenamefont {Liu}, \citenamefont {Xia},
		\citenamefont {Sinova}, \citenamefont {Jungwirth},\ and\ \citenamefont
		{Liu}}]{Feng2022NatElectron}%
	\BibitemOpen
	\bibfield  {author} {\bibinfo {author} {\bibfnamefont {Z.}~\bibnamefont
			{Feng}}, \bibinfo {author} {\bibfnamefont {X.}~\bibnamefont {Zhou}}, \bibinfo
		{author} {\bibfnamefont {L.}~\bibnamefont {\v{S}mejkal}}, \bibinfo {author}
		{\bibfnamefont {L.}~\bibnamefont {Wu}}, \bibinfo {author} {\bibfnamefont
			{Z.}~\bibnamefont {Zhu}}, \bibinfo {author} {\bibfnamefont {H.}~\bibnamefont
			{Guo}}, \bibinfo {author} {\bibfnamefont {R.}~\bibnamefont
			{Gonz\'{a}lez-Hern\'{a}ndez}}, \bibinfo {author} {\bibfnamefont
			{X.}~\bibnamefont {Wang}}, \bibinfo {author} {\bibfnamefont {H.}~\bibnamefont
			{Yan}}, \bibinfo {author} {\bibfnamefont {P.}~\bibnamefont {Qin}}, \bibinfo
		{author} {\bibfnamefont {X.}~\bibnamefont {Zhang}}, \bibinfo {author}
		{\bibfnamefont {H.}~\bibnamefont {Wu}}, \bibinfo {author} {\bibfnamefont
			{H.}~\bibnamefont {Chen}}, \bibinfo {author} {\bibfnamefont {Z.}~\bibnamefont
			{Meng}}, \bibinfo {author} {\bibfnamefont {L.}~\bibnamefont {Liu}}, \bibinfo
		{author} {\bibfnamefont {Z.}~\bibnamefont {Xia}}, \bibinfo {author}
		{\bibfnamefont {J.}~\bibnamefont {Sinova}}, \bibinfo {author} {\bibfnamefont
			{T.}~\bibnamefont {Jungwirth}},\ and\ \bibinfo {author} {\bibfnamefont
			{Z.}~\bibnamefont {Liu}},\ }\bibfield  {title} {\bibinfo {title} {{An
				anomalous Hall effect in altermagnetic ruthenium dioxide}},\ }\href
	{https://doi.org/10.1038/s41928-022-00866-z} {\bibfield  {journal} {\bibinfo
			{journal} {Nat. Electron.}\ }\textbf {\bibinfo {volume} {5}},\ \bibinfo
		{pages} {735} (\bibinfo {year} {2022})}\BibitemShut {NoStop}%
	\bibitem [{\citenamefont {Bai}\ \emph {et~al.}(2022)\citenamefont {Bai},
		\citenamefont {Han}, \citenamefont {Feng}, \citenamefont {Zhou},
		\citenamefont {Su}, \citenamefont {Wang}, \citenamefont {Liao}, \citenamefont
		{Zhu}, \citenamefont {Chen}, \citenamefont {Pan}, \citenamefont {Fan},\ and\
		\citenamefont {Song}}]{Bai2022PRL}%
	\BibitemOpen
	\bibfield  {author} {\bibinfo {author} {\bibfnamefont {H.}~\bibnamefont
			{Bai}}, \bibinfo {author} {\bibfnamefont {L.}~\bibnamefont {Han}}, \bibinfo
		{author} {\bibfnamefont {X.~Y.}\ \bibnamefont {Feng}}, \bibinfo {author}
		{\bibfnamefont {Y.~J.}\ \bibnamefont {Zhou}}, \bibinfo {author}
		{\bibfnamefont {R.~X.}\ \bibnamefont {Su}}, \bibinfo {author} {\bibfnamefont
			{Q.}~\bibnamefont {Wang}}, \bibinfo {author} {\bibfnamefont {L.~Y.}\
			\bibnamefont {Liao}}, \bibinfo {author} {\bibfnamefont {W.~X.}\ \bibnamefont
			{Zhu}}, \bibinfo {author} {\bibfnamefont {X.~Z.}\ \bibnamefont {Chen}},
		\bibinfo {author} {\bibfnamefont {F.}~\bibnamefont {Pan}}, \bibinfo {author}
		{\bibfnamefont {X.~L.}\ \bibnamefont {Fan}},\ and\ \bibinfo {author}
		{\bibfnamefont {C.}~\bibnamefont {Song}},\ }\bibfield  {title} {\bibinfo
		{title} {{Observation of Spin Splitting Torque in a Collinear Antiferromagnet
				RuO2}},\ }\href {https://doi.org/10.1103/PhysRevLett.128.197202} {\bibfield
		{journal} {\bibinfo  {journal} {Phys. Rev. Lett.}\ }\textbf {\bibinfo
			{volume} {128}},\ \bibinfo {pages} {197202} (\bibinfo {year}
		{2022})}\BibitemShut {NoStop}%
	\bibitem [{\citenamefont {Karube}\ \emph {et~al.}(2022)\citenamefont {Karube},
		\citenamefont {Tanaka}, \citenamefont {Sugawara}, \citenamefont {Kadoguchi},
		\citenamefont {Kohda},\ and\ \citenamefont {Nitta}}]{Karube2022PRL}%
	\BibitemOpen
	\bibfield  {author} {\bibinfo {author} {\bibfnamefont {S.}~\bibnamefont
			{Karube}}, \bibinfo {author} {\bibfnamefont {T.}~\bibnamefont {Tanaka}},
		\bibinfo {author} {\bibfnamefont {D.}~\bibnamefont {Sugawara}}, \bibinfo
		{author} {\bibfnamefont {N.}~\bibnamefont {Kadoguchi}}, \bibinfo {author}
		{\bibfnamefont {M.}~\bibnamefont {Kohda}},\ and\ \bibinfo {author}
		{\bibfnamefont {J.}~\bibnamefont {Nitta}},\ }\bibfield  {title} {\bibinfo
		{title} {{Observation of Spin-Splitter Torque in Collinear Antiferromagnetic
				RuO2}},\ }\href {https://doi.org/10.1103/PhysRevLett.129.137201} {\bibfield
		{journal} {\bibinfo  {journal} {Phys. Rev. Lett.}\ }\textbf {\bibinfo
			{volume} {129}},\ \bibinfo {pages} {137201} (\bibinfo {year}
		{2022})}\BibitemShut {NoStop}%
	\bibitem [{\citenamefont {Hiraishi}\ \emph {et~al.}(2024)\citenamefont
		{Hiraishi}, \citenamefont {Okabe}, \citenamefont {Koda}, \citenamefont
		{Kadono}, \citenamefont {Muroi}, \citenamefont {Hirai},\ and\ \citenamefont
		{Hiroi}}]{Hiraishi2024PRL}%
	\BibitemOpen
	\bibfield  {author} {\bibinfo {author} {\bibfnamefont {M.}~\bibnamefont
			{Hiraishi}}, \bibinfo {author} {\bibfnamefont {H.}~\bibnamefont {Okabe}},
		\bibinfo {author} {\bibfnamefont {A.}~\bibnamefont {Koda}}, \bibinfo {author}
		{\bibfnamefont {R.}~\bibnamefont {Kadono}}, \bibinfo {author} {\bibfnamefont
			{T.}~\bibnamefont {Muroi}}, \bibinfo {author} {\bibfnamefont
			{D.}~\bibnamefont {Hirai}},\ and\ \bibinfo {author} {\bibfnamefont
			{Z.}~\bibnamefont {Hiroi}},\ }\bibfield  {title} {\bibinfo {title}
		{{Nonmagnetic Ground State in RuO2 Revealed by Muon Spin Rotation}},\ }\href
	{https://doi.org/10.1103/PhysRevLett.132.166702} {\bibfield  {journal}
		{\bibinfo  {journal} {Phys. Rev. Lett.}\ }\textbf {\bibinfo {volume} {132}},\
		\bibinfo {pages} {166702} (\bibinfo {year} {2024})}\BibitemShut {NoStop}%
	\bibitem [{\citenamefont {Liu}\ \emph {et~al.}(2024{\natexlab{b}})\citenamefont
		{Liu}, \citenamefont {Zhan}, \citenamefont {Li}, \citenamefont {Liu},
		\citenamefont {Cheng}, \citenamefont {Shi}, \citenamefont {Deng},
		\citenamefont {Zhang}, \citenamefont {Li}, \citenamefont {Ding},
		\citenamefont {Jiang}, \citenamefont {Ye}, \citenamefont {Liu}, \citenamefont
		{Jiang}, \citenamefont {Wang}, \citenamefont {Li}, \citenamefont {Xie},
		\citenamefont {Wang}, \citenamefont {Qiao}, \citenamefont {Wen},
		\citenamefont {Sun},\ and\ \citenamefont {Shen}}]{Liu2024PRLRuO2}%
	\BibitemOpen
	\bibfield  {author} {\bibinfo {author} {\bibfnamefont {J.}~\bibnamefont
			{Liu}}, \bibinfo {author} {\bibfnamefont {J.}~\bibnamefont {Zhan}}, \bibinfo
		{author} {\bibfnamefont {T.}~\bibnamefont {Li}}, \bibinfo {author}
		{\bibfnamefont {J.}~\bibnamefont {Liu}}, \bibinfo {author} {\bibfnamefont
			{S.}~\bibnamefont {Cheng}}, \bibinfo {author} {\bibfnamefont
			{Y.}~\bibnamefont {Shi}}, \bibinfo {author} {\bibfnamefont {L.}~\bibnamefont
			{Deng}}, \bibinfo {author} {\bibfnamefont {M.}~\bibnamefont {Zhang}},
		\bibinfo {author} {\bibfnamefont {C.}~\bibnamefont {Li}}, \bibinfo {author}
		{\bibfnamefont {J.}~\bibnamefont {Ding}}, \bibinfo {author} {\bibfnamefont
			{Q.}~\bibnamefont {Jiang}}, \bibinfo {author} {\bibfnamefont
			{M.}~\bibnamefont {Ye}}, \bibinfo {author} {\bibfnamefont {Z.}~\bibnamefont
			{Liu}}, \bibinfo {author} {\bibfnamefont {Z.}~\bibnamefont {Jiang}}, \bibinfo
		{author} {\bibfnamefont {S.}~\bibnamefont {Wang}}, \bibinfo {author}
		{\bibfnamefont {Q.}~\bibnamefont {Li}}, \bibinfo {author} {\bibfnamefont
			{Y.}~\bibnamefont {Xie}}, \bibinfo {author} {\bibfnamefont {Y.}~\bibnamefont
			{Wang}}, \bibinfo {author} {\bibfnamefont {S.}~\bibnamefont {Qiao}}, \bibinfo
		{author} {\bibfnamefont {J.}~\bibnamefont {Wen}}, \bibinfo {author}
		{\bibfnamefont {Y.}~\bibnamefont {Sun}},\ and\ \bibinfo {author}
		{\bibfnamefont {D.}~\bibnamefont {Shen}},\ }\bibfield  {title} {\bibinfo
		{title} {{Absence of Altermagnetic Spin Splitting Character in Rutile Oxide
				RuO2}},\ }\href {https://doi.org/10.1103/PhysRevLett.133.176401} {\bibfield
		{journal} {\bibinfo  {journal} {Phys. Rev. Lett.}\ }\textbf {\bibinfo
			{volume} {133}},\ \bibinfo {pages} {176401} (\bibinfo {year}
		{2024}{\natexlab{b}})}\BibitemShut {NoStop}%
	\bibitem [{\citenamefont {Fedchenko}\ \emph {et~al.}(2024)\citenamefont
		{Fedchenko}, \citenamefont {Min\'{a}r}, \citenamefont {Akashdeep},
		\citenamefont {D'Souza}, \citenamefont {Vasilyev}, \citenamefont {Tkach},
		\citenamefont {Odenbreit}, \citenamefont {Nguyen}, \citenamefont
		{Kutnyakhov}, \citenamefont {Wind}, \citenamefont {Wenthaus}, \citenamefont
		{Scholz}, \citenamefont {Rossnagel}, \citenamefont {Hoesch}, \citenamefont
		{Aeschlimann}, \citenamefont {Stadtm\"{u}ller}, \citenamefont {Kl\"{a}ui},
		\citenamefont {Sch\"{o}nhense}, \citenamefont {Jungwirth}, \citenamefont
		{Hellenes}, \citenamefont {Jakob}, \citenamefont {\v{S}mejkal}, \citenamefont
		{Sinova},\ and\ \citenamefont {Elmers}}]{Fedchenko2024SciAdv}%
	\BibitemOpen
	\bibfield  {author} {\bibinfo {author} {\bibfnamefont {O.}~\bibnamefont
			{Fedchenko}}, \bibinfo {author} {\bibfnamefont {J.}~\bibnamefont
			{Min\'{a}r}}, \bibinfo {author} {\bibfnamefont {A.}~\bibnamefont
			{Akashdeep}}, \bibinfo {author} {\bibfnamefont {S.~W.}\ \bibnamefont
			{D'Souza}}, \bibinfo {author} {\bibfnamefont {D.}~\bibnamefont {Vasilyev}},
		\bibinfo {author} {\bibfnamefont {O.}~\bibnamefont {Tkach}}, \bibinfo
		{author} {\bibfnamefont {L.}~\bibnamefont {Odenbreit}}, \bibinfo {author}
		{\bibfnamefont {Q.}~\bibnamefont {Nguyen}}, \bibinfo {author} {\bibfnamefont
			{D.}~\bibnamefont {Kutnyakhov}}, \bibinfo {author} {\bibfnamefont
			{N.}~\bibnamefont {Wind}}, \bibinfo {author} {\bibfnamefont {L.}~\bibnamefont
			{Wenthaus}}, \bibinfo {author} {\bibfnamefont {M.}~\bibnamefont {Scholz}},
		\bibinfo {author} {\bibfnamefont {K.}~\bibnamefont {Rossnagel}}, \bibinfo
		{author} {\bibfnamefont {M.}~\bibnamefont {Hoesch}}, \bibinfo {author}
		{\bibfnamefont {M.}~\bibnamefont {Aeschlimann}}, \bibinfo {author}
		{\bibfnamefont {B.}~\bibnamefont {Stadtm\"{u}ller}}, \bibinfo {author}
		{\bibfnamefont {M.}~\bibnamefont {Kl\"{a}ui}}, \bibinfo {author}
		{\bibfnamefont {G.}~\bibnamefont {Sch\"{o}nhense}}, \bibinfo {author}
		{\bibfnamefont {T.}~\bibnamefont {Jungwirth}}, \bibinfo {author}
		{\bibfnamefont {A.~B.}\ \bibnamefont {Hellenes}}, \bibinfo {author}
		{\bibfnamefont {G.}~\bibnamefont {Jakob}}, \bibinfo {author} {\bibfnamefont
			{L.}~\bibnamefont {\v{S}mejkal}}, \bibinfo {author} {\bibfnamefont
			{J.}~\bibnamefont {Sinova}},\ and\ \bibinfo {author} {\bibfnamefont {H.-J.}\
			\bibnamefont {Elmers}},\ }\bibfield  {title} {\bibinfo {title} {{Observation
				of time-reversal symmetry breaking in the band structure of altermagnetic
				RuO2}},\ }\href {https://doi.org/10.1126/sciadv.adj4883} {\bibfield
		{journal} {\bibinfo  {journal} {Sci. Adv.}\ }\textbf {\bibinfo {volume}
			{10}},\ \bibinfo {pages} {eadj4883} (\bibinfo {year} {2024})}\BibitemShut
	{NoStop}%
	\bibitem [{\citenamefont {Smolyanyuk}\ \emph {et~al.}(2024)\citenamefont
		{Smolyanyuk}, \citenamefont {Mazin}, \citenamefont {Garcia-Gassull},\ and\
		\citenamefont {Valenti}}]{Smolyanyuk2024PRB}%
	\BibitemOpen
	\bibfield  {author} {\bibinfo {author} {\bibfnamefont {A.}~\bibnamefont
			{Smolyanyuk}}, \bibinfo {author} {\bibfnamefont {I.~I.}\ \bibnamefont
			{Mazin}}, \bibinfo {author} {\bibfnamefont {L.}~\bibnamefont
			{Garcia-Gassull}},\ and\ \bibinfo {author} {\bibfnamefont {R.}~\bibnamefont
			{Valenti}},\ }\bibfield  {title} {\bibinfo {title} {{Fragility of the
				magnetic order in the prototypical altermagnet RuO2}},\ }\href
	{https://doi.org/10.1103/PhysRevB.109.134424} {\bibfield  {journal} {\bibinfo
			{journal} {Phys. Rev. B}\ }\textbf {\bibinfo {volume} {109}},\ \bibinfo
		{pages} {134424} (\bibinfo {year} {2024})}\BibitemShut {NoStop}%
	\bibitem [{\citenamefont {Krempask\'{y}}\ \emph {et~al.}(2024)\citenamefont
		{Krempask\'{y}}, \citenamefont {\v{S}mejkal}, \citenamefont {D'Souza},
		\citenamefont {Hajlaoui}, \citenamefont {Springholz}, \citenamefont
		{Uhl\'{i}\v{r}ov\'{a}}, \citenamefont {Alarab}, \citenamefont {Constantinou},
		\citenamefont {Strocov}, \citenamefont {Usanov}, \citenamefont {Pudelko},
		\citenamefont {Gonz\'{a}lez-Hern\'{a}ndez}, \citenamefont {Birk~Hellenes},
		\citenamefont {Jansa}, \citenamefont {Reichlov\'{a}}, \citenamefont
		{\v{S}ob\'{a}\v{n}}, \citenamefont {Gonzalez~Betancourt}, \citenamefont
		{Wadley}, \citenamefont {Sinova}, \citenamefont {Kriegner}, \citenamefont
		{Min\'{a}r}, \citenamefont {Dil},\ and\ \citenamefont
		{Jungwirth}}]{Krempasky2024Nature}%
	\BibitemOpen
	\bibfield  {author} {\bibinfo {author} {\bibfnamefont {J.}~\bibnamefont
			{Krempask\'{y}}}, \bibinfo {author} {\bibfnamefont {L.}~\bibnamefont
			{\v{S}mejkal}}, \bibinfo {author} {\bibfnamefont {S.~W.}\ \bibnamefont
			{D'Souza}}, \bibinfo {author} {\bibfnamefont {M.}~\bibnamefont {Hajlaoui}},
		\bibinfo {author} {\bibfnamefont {G.}~\bibnamefont {Springholz}}, \bibinfo
		{author} {\bibfnamefont {K.}~\bibnamefont {Uhl\'{i}\v{r}ov\'{a}}}, \bibinfo
		{author} {\bibfnamefont {F.}~\bibnamefont {Alarab}}, \bibinfo {author}
		{\bibfnamefont {P.~C.}\ \bibnamefont {Constantinou}}, \bibinfo {author}
		{\bibfnamefont {V.}~\bibnamefont {Strocov}}, \bibinfo {author} {\bibfnamefont
			{D.}~\bibnamefont {Usanov}}, \bibinfo {author} {\bibfnamefont {W.~R.}\
			\bibnamefont {Pudelko}}, \bibinfo {author} {\bibfnamefont {R.}~\bibnamefont
			{Gonz\'{a}lez-Hern\'{a}ndez}}, \bibinfo {author} {\bibfnamefont
			{A.}~\bibnamefont {Birk~Hellenes}}, \bibinfo {author} {\bibfnamefont
			{Z.}~\bibnamefont {Jansa}}, \bibinfo {author} {\bibfnamefont
			{H.}~\bibnamefont {Reichlov\'{a}}}, \bibinfo {author} {\bibfnamefont
			{Z.}~\bibnamefont {\v{S}ob\'{a}\v{n}}}, \bibinfo {author} {\bibfnamefont
			{R.~D.}\ \bibnamefont {Gonzalez~Betancourt}}, \bibinfo {author}
		{\bibfnamefont {P.}~\bibnamefont {Wadley}}, \bibinfo {author} {\bibfnamefont
			{J.}~\bibnamefont {Sinova}}, \bibinfo {author} {\bibfnamefont
			{D.}~\bibnamefont {Kriegner}}, \bibinfo {author} {\bibfnamefont
			{J.}~\bibnamefont {Min\'{a}r}}, \bibinfo {author} {\bibfnamefont {J.~H.}\
			\bibnamefont {Dil}},\ and\ \bibinfo {author} {\bibfnamefont {T.}~\bibnamefont
			{Jungwirth}},\ }\bibfield  {title} {\bibinfo {title} {{Altermagnetic lifting
				of Kramers spin degeneracy}},\ }\href
	{https://doi.org/10.1038/s41586-023-06907-7} {\bibfield  {journal} {\bibinfo
			{journal} {Nature}\ }\textbf {\bibinfo {volume} {626}},\ \bibinfo {pages}
		{517} (\bibinfo {year} {2024})}\BibitemShut {NoStop}%
	\bibitem [{\citenamefont {Reimers}\ \emph {et~al.}(2024)\citenamefont
		{Reimers}, \citenamefont {Odenbreit}, \citenamefont {\v{S}mejkal},
		\citenamefont {Strocov}, \citenamefont {Constantinou}, \citenamefont
		{Hellenes}, \citenamefont {Jaeschke~Ubiergo}, \citenamefont {Campos},
		\citenamefont {Bharadwaj}, \citenamefont {Chakraborty}, \citenamefont
		{Denneulin}, \citenamefont {Shi}, \citenamefont {Dunin-Borkowski},
		\citenamefont {Das}, \citenamefont {Kl\"{a}ui}, \citenamefont {Sinova},\ and\
		\citenamefont {Jourdan}}]{Reimers2024NatCommun}%
	\BibitemOpen
	\bibfield  {author} {\bibinfo {author} {\bibfnamefont {S.}~\bibnamefont
			{Reimers}}, \bibinfo {author} {\bibfnamefont {L.}~\bibnamefont {Odenbreit}},
		\bibinfo {author} {\bibfnamefont {L.}~\bibnamefont {\v{S}mejkal}}, \bibinfo
		{author} {\bibfnamefont {V.~N.}\ \bibnamefont {Strocov}}, \bibinfo {author}
		{\bibfnamefont {P.}~\bibnamefont {Constantinou}}, \bibinfo {author}
		{\bibfnamefont {A.~B.}\ \bibnamefont {Hellenes}}, \bibinfo {author}
		{\bibfnamefont {R.}~\bibnamefont {Jaeschke~Ubiergo}}, \bibinfo {author}
		{\bibfnamefont {W.~H.}\ \bibnamefont {Campos}}, \bibinfo {author}
		{\bibfnamefont {V.~K.}\ \bibnamefont {Bharadwaj}}, \bibinfo {author}
		{\bibfnamefont {A.}~\bibnamefont {Chakraborty}}, \bibinfo {author}
		{\bibfnamefont {T.}~\bibnamefont {Denneulin}}, \bibinfo {author}
		{\bibfnamefont {W.}~\bibnamefont {Shi}}, \bibinfo {author} {\bibfnamefont
			{R.~E.}\ \bibnamefont {Dunin-Borkowski}}, \bibinfo {author} {\bibfnamefont
			{S.}~\bibnamefont {Das}}, \bibinfo {author} {\bibfnamefont {M.}~\bibnamefont
			{Kl\"{a}ui}}, \bibinfo {author} {\bibfnamefont {J.}~\bibnamefont {Sinova}},\
		and\ \bibinfo {author} {\bibfnamefont {M.}~\bibnamefont {Jourdan}},\
	}\bibfield  {title} {\bibinfo {title} {{Direct observation of altermagnetic
				band splitting in CrSb thin films}},\ }\href
	{https://doi.org/10.1038/s41467-024-46476-5} {\bibfield  {journal} {\bibinfo
			{journal} {Nat. Commun.}\ }\textbf {\bibinfo {volume} {15}},\ \bibinfo
		{pages} {2116} (\bibinfo {year} {2024})}\BibitemShut {NoStop}%
	\bibitem [{\citenamefont {Lee}\ \emph {et~al.}(2024)\citenamefont {Lee},
		\citenamefont {Lee}, \citenamefont {Jung}, \citenamefont {Jung},
		\citenamefont {Kim}, \citenamefont {Lee}, \citenamefont {Seok}, \citenamefont
		{Kim}, \citenamefont {Park}, \citenamefont {\v{S}mejkal}, \citenamefont
		{Kang},\ and\ \citenamefont {Kim}}]{Lee2024PRL}%
	\BibitemOpen
	\bibfield  {author} {\bibinfo {author} {\bibfnamefont {S.}~\bibnamefont
			{Lee}}, \bibinfo {author} {\bibfnamefont {S.}~\bibnamefont {Lee}}, \bibinfo
		{author} {\bibfnamefont {S.}~\bibnamefont {Jung}}, \bibinfo {author}
		{\bibfnamefont {J.}~\bibnamefont {Jung}}, \bibinfo {author} {\bibfnamefont
			{D.}~\bibnamefont {Kim}}, \bibinfo {author} {\bibfnamefont {Y.}~\bibnamefont
			{Lee}}, \bibinfo {author} {\bibfnamefont {B.}~\bibnamefont {Seok}}, \bibinfo
		{author} {\bibfnamefont {J.}~\bibnamefont {Kim}}, \bibinfo {author}
		{\bibfnamefont {B.~G.}\ \bibnamefont {Park}}, \bibinfo {author}
		{\bibfnamefont {L.}~\bibnamefont {\v{S}mejkal}}, \bibinfo {author}
		{\bibfnamefont {C.-J.}\ \bibnamefont {Kang}},\ and\ \bibinfo {author}
		{\bibfnamefont {C.}~\bibnamefont {Kim}},\ }\bibfield  {title} {\bibinfo
		{title} {{Broken Kramers Degeneracy in Altermagnetic MnTe}},\ }\href
	{https://doi.org/10.1103/PhysRevLett.132.036702} {\bibfield  {journal}
		{\bibinfo  {journal} {Phys. Rev. Lett.}\ }\textbf {\bibinfo {volume} {132}},\
		\bibinfo {pages} {036702} (\bibinfo {year} {2024})}\BibitemShut {NoStop}%
	\bibitem [{\citenamefont {Osumi}\ \emph {et~al.}(2024)\citenamefont {Osumi},
		\citenamefont {Souma}, \citenamefont {Aoyama}, \citenamefont {Yamauchi},
		\citenamefont {Honma}, \citenamefont {Nakayama}, \citenamefont {Takahashi},
		\citenamefont {Ohgushi},\ and\ \citenamefont {Sato}}]{Osumi2024PRB}%
	\BibitemOpen
	\bibfield  {author} {\bibinfo {author} {\bibfnamefont {T.}~\bibnamefont
			{Osumi}}, \bibinfo {author} {\bibfnamefont {S.}~\bibnamefont {Souma}},
		\bibinfo {author} {\bibfnamefont {T.}~\bibnamefont {Aoyama}}, \bibinfo
		{author} {\bibfnamefont {K.}~\bibnamefont {Yamauchi}}, \bibinfo {author}
		{\bibfnamefont {A.}~\bibnamefont {Honma}}, \bibinfo {author} {\bibfnamefont
			{K.}~\bibnamefont {Nakayama}}, \bibinfo {author} {\bibfnamefont
			{T.}~\bibnamefont {Takahashi}}, \bibinfo {author} {\bibfnamefont
			{K.}~\bibnamefont {Ohgushi}},\ and\ \bibinfo {author} {\bibfnamefont
			{T.}~\bibnamefont {Sato}},\ }\bibfield  {title} {\bibinfo {title}
		{{Observation of a giant band splitting in altermagnetic MnTe}},\ }\href
	{https://doi.org/10.1103/PhysRevB.109.115102} {\bibfield  {journal} {\bibinfo
			{journal} {Phys. Rev. B}\ }\textbf {\bibinfo {volume} {109}},\ \bibinfo
		{pages} {115102} (\bibinfo {year} {2024})}\BibitemShut {NoStop}%
	\bibitem [{\citenamefont {Zeng}\ \emph {et~al.}(2024)\citenamefont {Zeng},
		\citenamefont {Zhu}, \citenamefont {Zhu}, \citenamefont {Liu}, \citenamefont
		{Ma}, \citenamefont {Hao}, \citenamefont {Liu}, \citenamefont {Qu},
		\citenamefont {Yang}, \citenamefont {Jiang}, \citenamefont {Yamagami},
		\citenamefont {Arita}, \citenamefont {Zhang}, \citenamefont {Shao},
		\citenamefont {Dai}, \citenamefont {Shimada}, \citenamefont {Liu},
		\citenamefont {Ye}, \citenamefont {Huang}, \citenamefont {Liu},\ and\
		\citenamefont {Liu}}]{Zeng2024AdvSci}%
	\BibitemOpen
	\bibfield  {author} {\bibinfo {author} {\bibfnamefont {M.}~\bibnamefont
			{Zeng}}, \bibinfo {author} {\bibfnamefont {M.-Y.}\ \bibnamefont {Zhu}},
		\bibinfo {author} {\bibfnamefont {Y.-P.}\ \bibnamefont {Zhu}}, \bibinfo
		{author} {\bibfnamefont {X.-R.}\ \bibnamefont {Liu}}, \bibinfo {author}
		{\bibfnamefont {X.-M.}\ \bibnamefont {Ma}}, \bibinfo {author} {\bibfnamefont
			{Y.-J.}\ \bibnamefont {Hao}}, \bibinfo {author} {\bibfnamefont
			{P.}~\bibnamefont {Liu}}, \bibinfo {author} {\bibfnamefont {G.}~\bibnamefont
			{Qu}}, \bibinfo {author} {\bibfnamefont {Y.}~\bibnamefont {Yang}}, \bibinfo
		{author} {\bibfnamefont {Z.}~\bibnamefont {Jiang}}, \bibinfo {author}
		{\bibfnamefont {K.}~\bibnamefont {Yamagami}}, \bibinfo {author}
		{\bibfnamefont {M.}~\bibnamefont {Arita}}, \bibinfo {author} {\bibfnamefont
			{X.}~\bibnamefont {Zhang}}, \bibinfo {author} {\bibfnamefont {T.-H.}\
			\bibnamefont {Shao}}, \bibinfo {author} {\bibfnamefont {Y.}~\bibnamefont
			{Dai}}, \bibinfo {author} {\bibfnamefont {K.}~\bibnamefont {Shimada}},
		\bibinfo {author} {\bibfnamefont {Z.}~\bibnamefont {Liu}}, \bibinfo {author}
		{\bibfnamefont {M.}~\bibnamefont {Ye}}, \bibinfo {author} {\bibfnamefont
			{Y.}~\bibnamefont {Huang}}, \bibinfo {author} {\bibfnamefont
			{Q.}~\bibnamefont {Liu}},\ and\ \bibinfo {author} {\bibfnamefont
			{C.}~\bibnamefont {Liu}},\ }\bibfield  {title} {\bibinfo {title}
		{{Observation of Spin Splitting in Room-Temperature Metallic Antiferromagnet
				CrSb}},\ }\href {https://doi.org/10.1002/advs.202406529} {\bibfield
		{journal} {\bibinfo  {journal} {Adv. Sci.}\ }\textbf {\bibinfo {volume}
			{11}},\ \bibinfo {pages} {2406529} (\bibinfo {year} {2024})}\BibitemShut
	{NoStop}%
	\bibitem [{\citenamefont {Ding}\ \emph {et~al.}(2024)\citenamefont {Ding},
		\citenamefont {Jiang}, \citenamefont {Chen}, \citenamefont {Tao},
		\citenamefont {Liu}, \citenamefont {Li}, \citenamefont {Liu}, \citenamefont
		{Sun}, \citenamefont {Cheng}, \citenamefont {Liu}, \citenamefont {Yang},
		\citenamefont {Zhang}, \citenamefont {Deng}, \citenamefont {Jing},
		\citenamefont {Huang}, \citenamefont {Shi}, \citenamefont {Ye}, \citenamefont
		{Qiao}, \citenamefont {Wang}, \citenamefont {Guo}, \citenamefont {Feng},\
		and\ \citenamefont {Shen}}]{Ding2024PRL}%
	\BibitemOpen
	\bibfield  {author} {\bibinfo {author} {\bibfnamefont {J.}~\bibnamefont
			{Ding}}, \bibinfo {author} {\bibfnamefont {Z.}~\bibnamefont {Jiang}},
		\bibinfo {author} {\bibfnamefont {X.}~\bibnamefont {Chen}}, \bibinfo {author}
		{\bibfnamefont {Z.}~\bibnamefont {Tao}}, \bibinfo {author} {\bibfnamefont
			{Z.}~\bibnamefont {Liu}}, \bibinfo {author} {\bibfnamefont {T.}~\bibnamefont
			{Li}}, \bibinfo {author} {\bibfnamefont {J.}~\bibnamefont {Liu}}, \bibinfo
		{author} {\bibfnamefont {J.}~\bibnamefont {Sun}}, \bibinfo {author}
		{\bibfnamefont {J.}~\bibnamefont {Cheng}}, \bibinfo {author} {\bibfnamefont
			{J.}~\bibnamefont {Liu}}, \bibinfo {author} {\bibfnamefont {Y.}~\bibnamefont
			{Yang}}, \bibinfo {author} {\bibfnamefont {R.}~\bibnamefont {Zhang}},
		\bibinfo {author} {\bibfnamefont {L.}~\bibnamefont {Deng}}, \bibinfo {author}
		{\bibfnamefont {W.}~\bibnamefont {Jing}}, \bibinfo {author} {\bibfnamefont
			{Y.}~\bibnamefont {Huang}}, \bibinfo {author} {\bibfnamefont
			{Y.}~\bibnamefont {Shi}}, \bibinfo {author} {\bibfnamefont {M.}~\bibnamefont
			{Ye}}, \bibinfo {author} {\bibfnamefont {S.}~\bibnamefont {Qiao}}, \bibinfo
		{author} {\bibfnamefont {Y.}~\bibnamefont {Wang}}, \bibinfo {author}
		{\bibfnamefont {Y.}~\bibnamefont {Guo}}, \bibinfo {author} {\bibfnamefont
			{D.}~\bibnamefont {Feng}},\ and\ \bibinfo {author} {\bibfnamefont
			{D.}~\bibnamefont {Shen}},\ }\bibfield  {title} {\bibinfo {title} {{Large
				Band Splitting in g-Wave Altermagnet CrSb}},\ }\href
	{https://doi.org/10.1103/PhysRevLett.133.206401} {\bibfield  {journal}
		{\bibinfo  {journal} {Phys. Rev. Lett.}\ }\textbf {\bibinfo {volume} {133}},\
		\bibinfo {pages} {206401} (\bibinfo {year} {2024})}\BibitemShut {NoStop}%
	\bibitem [{\citenamefont {Jiang}\ \emph {et~al.}(2025)\citenamefont {Jiang},
		\citenamefont {Hu}, \citenamefont {Bai}, \citenamefont {Song}, \citenamefont
		{Mu}, \citenamefont {Qu}, \citenamefont {Li}, \citenamefont {Zhu},
		\citenamefont {Pi}, \citenamefont {Wei}, \citenamefont {Sun}, \citenamefont
		{Huang}, \citenamefont {Zheng}, \citenamefont {Peng}, \citenamefont {He},
		\citenamefont {Li}, \citenamefont {Luo}, \citenamefont {Li}, \citenamefont
		{Chen}, \citenamefont {Li}, \citenamefont {Weng},\ and\ \citenamefont
		{Qian}}]{Jiang2025NPhys}%
	\BibitemOpen
	\bibfield  {author} {\bibinfo {author} {\bibfnamefont {B.}~\bibnamefont
			{Jiang}}, \bibinfo {author} {\bibfnamefont {M.}~\bibnamefont {Hu}}, \bibinfo
		{author} {\bibfnamefont {J.}~\bibnamefont {Bai}}, \bibinfo {author}
		{\bibfnamefont {Z.}~\bibnamefont {Song}}, \bibinfo {author} {\bibfnamefont
			{C.}~\bibnamefont {Mu}}, \bibinfo {author} {\bibfnamefont {G.}~\bibnamefont
			{Qu}}, \bibinfo {author} {\bibfnamefont {W.}~\bibnamefont {Li}}, \bibinfo
		{author} {\bibfnamefont {W.}~\bibnamefont {Zhu}}, \bibinfo {author}
		{\bibfnamefont {H.}~\bibnamefont {Pi}}, \bibinfo {author} {\bibfnamefont
			{Z.}~\bibnamefont {Wei}}, \bibinfo {author} {\bibfnamefont {Y.-J.}\
			\bibnamefont {Sun}}, \bibinfo {author} {\bibfnamefont {Y.}~\bibnamefont
			{Huang}}, \bibinfo {author} {\bibfnamefont {X.}~\bibnamefont {Zheng}},
		\bibinfo {author} {\bibfnamefont {Y.}~\bibnamefont {Peng}}, \bibinfo {author}
		{\bibfnamefont {L.}~\bibnamefont {He}}, \bibinfo {author} {\bibfnamefont
			{S.}~\bibnamefont {Li}}, \bibinfo {author} {\bibfnamefont {J.}~\bibnamefont
			{Luo}}, \bibinfo {author} {\bibfnamefont {Z.}~\bibnamefont {Li}}, \bibinfo
		{author} {\bibfnamefont {G.}~\bibnamefont {Chen}}, \bibinfo {author}
		{\bibfnamefont {H.}~\bibnamefont {Li}}, \bibinfo {author} {\bibfnamefont
			{H.}~\bibnamefont {Weng}},\ and\ \bibinfo {author} {\bibfnamefont
			{T.}~\bibnamefont {Qian}},\ }\bibfield  {title} {\bibinfo {title} {{A
				metallic room-temperature d-wave altermagnet}},\ }\href
	{https://doi.org/10.1038/s41567-025-02822-y} {\bibfield  {journal} {\bibinfo
			{journal} {Nat. Phys.}\ }\textbf {\bibinfo {volume} {21}},\ \bibinfo {pages}
		{754} (\bibinfo {year} {2025})}\BibitemShut {NoStop}%
	\bibitem [{\citenamefont {Reichlov\'{a}}\ \emph {et~al.}(2024)\citenamefont
		{Reichlov\'{a}}, \citenamefont {Seeger}, \citenamefont
		{Gonz\'{a}lez-Hern\'{a}ndez}, \citenamefont {Kounta}, \citenamefont
		{Schlitz}, \citenamefont {Kriegner}, \citenamefont {Ritzinger}, \citenamefont
		{Lammel}, \citenamefont {Leivisk\"{a}}, \citenamefont {Hellenes},
		\citenamefont {Olejn\'{i}k}, \citenamefont {Pet\v{r}\'{i}\v{c}ek},
		\citenamefont {Dole\v{z}al}, \citenamefont {Hor\'{a}k}, \citenamefont
		{Schmoranzerova}, \citenamefont {Badura}, \citenamefont {Bertaina},
		\citenamefont {Thomas}, \citenamefont {Baltz}, \citenamefont {Michez},
		\citenamefont {Sinova}, \citenamefont {G\"{o}nnenwein}, \citenamefont
		{Jungwirth},\ and\ \citenamefont {\v{S}mejkal}}]{Reichlova2024NatCommun}%
	\BibitemOpen
	\bibfield  {author} {\bibinfo {author} {\bibfnamefont {H.}~\bibnamefont
			{Reichlov\'{a}}}, \bibinfo {author} {\bibfnamefont {R.~L.}\ \bibnamefont
			{Seeger}}, \bibinfo {author} {\bibfnamefont {R.}~\bibnamefont
			{Gonz\'{a}lez-Hern\'{a}ndez}}, \bibinfo {author} {\bibfnamefont
			{I.}~\bibnamefont {Kounta}}, \bibinfo {author} {\bibfnamefont
			{R.}~\bibnamefont {Schlitz}}, \bibinfo {author} {\bibfnamefont
			{D.}~\bibnamefont {Kriegner}}, \bibinfo {author} {\bibfnamefont
			{P.}~\bibnamefont {Ritzinger}}, \bibinfo {author} {\bibfnamefont
			{M.}~\bibnamefont {Lammel}}, \bibinfo {author} {\bibfnamefont
			{M.}~\bibnamefont {Leivisk\"{a}}}, \bibinfo {author} {\bibfnamefont {A.~B.}\
			\bibnamefont {Hellenes}}, \bibinfo {author} {\bibfnamefont {K.}~\bibnamefont
			{Olejn\'{i}k}}, \bibinfo {author} {\bibfnamefont {V.}~\bibnamefont
			{Pet\v{r}\'{i}\v{c}ek}}, \bibinfo {author} {\bibfnamefont {P.}~\bibnamefont
			{Dole\v{z}al}}, \bibinfo {author} {\bibfnamefont {L.}~\bibnamefont
			{Hor\'{a}k}}, \bibinfo {author} {\bibfnamefont {E.}~\bibnamefont
			{Schmoranzerova}}, \bibinfo {author} {\bibfnamefont {A.}~\bibnamefont
			{Badura}}, \bibinfo {author} {\bibfnamefont {S.}~\bibnamefont {Bertaina}},
		\bibinfo {author} {\bibfnamefont {A.}~\bibnamefont {Thomas}}, \bibinfo
		{author} {\bibfnamefont {V.}~\bibnamefont {Baltz}}, \bibinfo {author}
		{\bibfnamefont {L.}~\bibnamefont {Michez}}, \bibinfo {author} {\bibfnamefont
			{J.}~\bibnamefont {Sinova}}, \bibinfo {author} {\bibfnamefont {S.~T.~B.}\
			\bibnamefont {G\"{o}nnenwein}}, \bibinfo {author} {\bibfnamefont
			{T.}~\bibnamefont {Jungwirth}},\ and\ \bibinfo {author} {\bibfnamefont
			{L.}~\bibnamefont {\v{S}mejkal}},\ }\bibfield  {title} {\bibinfo {title}
		{{Observation of a spontaneous anomalous Hall response in the Mn5Si3 d-wave
				altermagnet candidate}},\ }\href {https://doi.org/10.1038/s41467-024-48493-w}
	{\bibfield  {journal} {\bibinfo  {journal} {Nat. Commun.}\ }\textbf {\bibinfo
			{volume} {15}},\ \bibinfo {pages} {4961} (\bibinfo {year}
		{2024})}\BibitemShut {NoStop}%
	\bibitem [{\citenamefont {Fender}\ \emph {et~al.}(2025)\citenamefont {Fender},
		\citenamefont {Gonzalez},\ and\ \citenamefont {Bediako}}]{Fender2025JACS}%
	\BibitemOpen
	\bibfield  {author} {\bibinfo {author} {\bibfnamefont {S.~S.}\ \bibnamefont
			{Fender}}, \bibinfo {author} {\bibfnamefont {O.}~\bibnamefont {Gonzalez}},\
		and\ \bibinfo {author} {\bibfnamefont {D.~K.}\ \bibnamefont {Bediako}},\
	}\bibfield  {title} {\bibinfo {title} {{Altermagnetism: A Chemical
				Perspective}},\ }\href {https://doi.org/10.1021/jacs.4c14503} {\bibfield
		{journal} {\bibinfo  {journal} {J. Am. Chem. Soc.}\ }\textbf {\bibinfo
			{volume} {147}},\ \bibinfo {pages} {2257} (\bibinfo {year}
		{2025})}\BibitemShut {NoStop}%
	\bibitem [{\citenamefont {Song}\ \emph {et~al.}(2025)\citenamefont {Song},
		\citenamefont {Bai}, \citenamefont {Zhou}, \citenamefont {Han}, \citenamefont
		{Reichlova}, \citenamefont {Dil}, \citenamefont {Liu}, \citenamefont {Chen},\
		and\ \citenamefont {Pan}}]{Song2025NRM}%
	\BibitemOpen
	\bibfield  {author} {\bibinfo {author} {\bibfnamefont {C.}~\bibnamefont
			{Song}}, \bibinfo {author} {\bibfnamefont {H.}~\bibnamefont {Bai}}, \bibinfo
		{author} {\bibfnamefont {Z.}~\bibnamefont {Zhou}}, \bibinfo {author}
		{\bibfnamefont {L.}~\bibnamefont {Han}}, \bibinfo {author} {\bibfnamefont
			{H.}~\bibnamefont {Reichlova}}, \bibinfo {author} {\bibfnamefont {J.~H.}\
			\bibnamefont {Dil}}, \bibinfo {author} {\bibfnamefont {J.}~\bibnamefont
			{Liu}}, \bibinfo {author} {\bibfnamefont {X.}~\bibnamefont {Chen}},\ and\
		\bibinfo {author} {\bibfnamefont {F.}~\bibnamefont {Pan}},\ }\bibfield
	{title} {\bibinfo {title} {{Altermagnets as a new class of functional
				materials}},\ }\href {https://doi.org/10.1038/s41578-025-00779-1} {\bibfield
		{journal} {\bibinfo  {journal} {Nat. Rev. Mater.}\ }\textbf {\bibinfo
			{volume} {10}},\ \bibinfo {pages} {473} (\bibinfo {year} {2025})}\BibitemShut
	{NoStop}%
	\bibitem [{\citenamefont {Jungwirth}\ \emph {et~al.}(2025)\citenamefont
		{Jungwirth}, \citenamefont {Fernandes}, \citenamefont {Fradkin},
		\citenamefont {MacDonald}, \citenamefont {Sinova},\ and\ \citenamefont
		{\v{S}mejkal}}]{Jungwirth2025Newton}%
	\BibitemOpen
	\bibfield  {author} {\bibinfo {author} {\bibfnamefont {T.}~\bibnamefont
			{Jungwirth}}, \bibinfo {author} {\bibfnamefont {R.~M.}\ \bibnamefont
			{Fernandes}}, \bibinfo {author} {\bibfnamefont {E.}~\bibnamefont {Fradkin}},
		\bibinfo {author} {\bibfnamefont {A.~H.}\ \bibnamefont {MacDonald}}, \bibinfo
		{author} {\bibfnamefont {J.}~\bibnamefont {Sinova}},\ and\ \bibinfo {author}
		{\bibfnamefont {L.}~\bibnamefont {\v{S}mejkal}},\ }\bibfield  {title}
	{\bibinfo {title} {{Altermagnetism: An unconventional spin-ordered phase of
				matter}},\ }\href {https://doi.org/10.1016/j.newton.2025.100162} {\bibfield
		{journal} {\bibinfo  {journal} {Newton}\ }\textbf {\bibinfo {volume} {1}},\
		\bibinfo {pages} {100162} (\bibinfo {year} {2025})}\BibitemShut {NoStop}%
	\bibitem [{\citenamefont {Jungwirth}\ \emph
		{et~al.}(2026{\natexlab{a}})\citenamefont {Jungwirth}, \citenamefont
		{Sinova}, \citenamefont {Fernandes}, \citenamefont {Liu}, \citenamefont
		{Watanabe}, \citenamefont {Murakami}, \citenamefont {Nakatsuji},\ and\
		\citenamefont {\v{S}mejkal}}]{Jungwirth2026Nature}%
	\BibitemOpen
	\bibfield  {author} {\bibinfo {author} {\bibfnamefont {T.}~\bibnamefont
			{Jungwirth}}, \bibinfo {author} {\bibfnamefont {J.}~\bibnamefont {Sinova}},
		\bibinfo {author} {\bibfnamefont {R.~M.}\ \bibnamefont {Fernandes}}, \bibinfo
		{author} {\bibfnamefont {Q.}~\bibnamefont {Liu}}, \bibinfo {author}
		{\bibfnamefont {H.}~\bibnamefont {Watanabe}}, \bibinfo {author}
		{\bibfnamefont {S.}~\bibnamefont {Murakami}}, \bibinfo {author}
		{\bibfnamefont {S.}~\bibnamefont {Nakatsuji}},\ and\ \bibinfo {author}
		{\bibfnamefont {L.}~\bibnamefont {\v{S}mejkal}},\ }\bibfield  {title}
	{\bibinfo {title} {{Symmetry, microscopy and spectroscopy signatures of
				altermagnetism}},\ }\href {https://doi.org/10.1038/s41586-025-09883-2}
	{\bibfield  {journal} {\bibinfo  {journal} {Nature}\ }\textbf {\bibinfo
			{volume} {649}},\ \bibinfo {pages} {837} (\bibinfo {year}
		{2026}{\natexlab{a}})}\BibitemShut {NoStop}%
	\bibitem [{\citenamefont {Jungwirth}\ \emph
		{et~al.}(2026{\natexlab{b}})\citenamefont {Jungwirth}, \citenamefont
		{Sinova}, \citenamefont {Wadley}, \citenamefont {Kriegner}, \citenamefont
		{Reichlov\'a}, \citenamefont {Krizek}, \citenamefont {Ohno},\ and\
		\citenamefont {\v{S}mejkal}}]{Jungwirth2026NPhys}%
	\BibitemOpen
	\bibfield  {author} {\bibinfo {author} {\bibfnamefont {T.}~\bibnamefont
			{Jungwirth}}, \bibinfo {author} {\bibfnamefont {J.}~\bibnamefont {Sinova}},
		\bibinfo {author} {\bibfnamefont {P.}~\bibnamefont {Wadley}}, \bibinfo
		{author} {\bibfnamefont {D.}~\bibnamefont {Kriegner}}, \bibinfo {author}
		{\bibfnamefont {H.}~\bibnamefont {Reichlov\'a}}, \bibinfo {author}
		{\bibfnamefont {F.}~\bibnamefont {Krizek}}, \bibinfo {author} {\bibfnamefont
			{H.}~\bibnamefont {Ohno}},\ and\ \bibinfo {author} {\bibfnamefont
			{L.}~\bibnamefont {\v{S}mejkal}},\ }\bibfield  {title} {\bibinfo {title}
		{{Altermagnetic spintronics}},\ }\href
	{https://doi.org/10.1038/s41567-026-03337-w} {\bibfield  {journal} {\bibinfo
			{journal} {Nat. Phys.}\ }\textbf {\bibinfo {volume} {22}},\ \bibinfo {pages}
		{1012} (\bibinfo {year} {2026}{\natexlab{b}})}\BibitemShut {NoStop}%
	\bibitem [{\citenamefont {He}\ \emph {et~al.}(2023)\citenamefont {He},
		\citenamefont {Wang}, \citenamefont {Luo}, \citenamefont {Zeng},
		\citenamefont {Chen},\ and\ \citenamefont {Tang}}]{He2023PRL}%
	\BibitemOpen
	\bibfield  {author} {\bibinfo {author} {\bibfnamefont {R.}~\bibnamefont
			{He}}, \bibinfo {author} {\bibfnamefont {D.}~\bibnamefont {Wang}}, \bibinfo
		{author} {\bibfnamefont {N.}~\bibnamefont {Luo}}, \bibinfo {author}
		{\bibfnamefont {J.}~\bibnamefont {Zeng}}, \bibinfo {author} {\bibfnamefont
			{K.-Q.}\ \bibnamefont {Chen}},\ and\ \bibinfo {author} {\bibfnamefont
			{L.-M.}\ \bibnamefont {Tang}},\ }\bibfield  {title} {\bibinfo {title}
		{{Nonrelativistic Spin-Momentum Coupling in Antiferromagnetic Twisted
				Bilayers}},\ }\href {https://doi.org/10.1103/PhysRevLett.130.046401}
	{\bibfield  {journal} {\bibinfo  {journal} {Phys. Rev. Lett.}\ }\textbf
		{\bibinfo {volume} {130}},\ \bibinfo {pages} {046401} (\bibinfo {year}
		{2023})}\BibitemShut {NoStop}%
	\bibitem [{\citenamefont {Sheoran}\ and\ \citenamefont
		{Bhattacharya}(2024)}]{Sheoran2024PRM}%
	\BibitemOpen
	\bibfield  {author} {\bibinfo {author} {\bibfnamefont {S.}~\bibnamefont
			{Sheoran}}\ and\ \bibinfo {author} {\bibfnamefont {S.}~\bibnamefont
			{Bhattacharya}},\ }\bibfield  {title} {\bibinfo {title} {{Nonrelativistic
				spin splittings and altermagnetism in twisted bilayers of centrosymmetric
				antiferromagnets}},\ }\href
	{https://doi.org/10.1103/PhysRevMaterials.8.L051401} {\bibfield  {journal}
		{\bibinfo  {journal} {Phys. Rev. Materials}\ }\textbf {\bibinfo {volume}
			{8}},\ \bibinfo {pages} {L051401} (\bibinfo {year} {2024})}\BibitemShut
	{NoStop}%
	\bibitem [{\citenamefont {Hodt}\ and\ \citenamefont
		{Linder}(2024)}]{Hodt2024PRB}%
	\BibitemOpen
	\bibfield  {author} {\bibinfo {author} {\bibfnamefont {E.~W.}\ \bibnamefont
			{Hodt}}\ and\ \bibinfo {author} {\bibfnamefont {J.}~\bibnamefont {Linder}},\
	}\bibfield  {title} {\bibinfo {title} {{Spin pumping in an
				altermagnet/normal-metal bilayer}},\ }\href
	{https://doi.org/10.1103/PhysRevB.109.174438} {\bibfield  {journal} {\bibinfo
			{journal} {Phys. Rev. B}\ }\textbf {\bibinfo {volume} {109}},\ \bibinfo
		{pages} {174438} (\bibinfo {year} {2024})}\BibitemShut {NoStop}%
	\bibitem [{\citenamefont {Ma}\ and\ \citenamefont {Jia}(2024)}]{Ma2024PRB}%
	\BibitemOpen
	\bibfield  {author} {\bibinfo {author} {\bibfnamefont {H.-Y.}\ \bibnamefont
			{Ma}}\ and\ \bibinfo {author} {\bibfnamefont {J.-F.}\ \bibnamefont {Jia}},\
	}\bibfield  {title} {\bibinfo {title} {{Altermagnetic topological insulator
				and the selection rules}},\ }\href
	{https://doi.org/10.1103/PhysRevB.110.064426} {\bibfield  {journal} {\bibinfo
			{journal} {Phys. Rev. B}\ }\textbf {\bibinfo {volume} {110}},\ \bibinfo
		{pages} {064426} (\bibinfo {year} {2024})}\BibitemShut {NoStop}%
	\bibitem [{\citenamefont {Lv}\ \emph {et~al.}(2025)\citenamefont {Lv},
		\citenamefont {Xu}, \citenamefont {Liu}, \citenamefont {Fu},\ and\
		\citenamefont {Yu}}]{Lv2025SciRep}%
	\BibitemOpen
	\bibfield  {author} {\bibinfo {author} {\bibfnamefont {Q.}~\bibnamefont
			{Lv}}, \bibinfo {author} {\bibfnamefont {Y.}~\bibnamefont {Xu}}, \bibinfo
		{author} {\bibfnamefont {J.-F.}\ \bibnamefont {Liu}}, \bibinfo {author}
		{\bibfnamefont {P.-H.}\ \bibnamefont {Fu}},\ and\ \bibinfo {author}
		{\bibfnamefont {X.-L.}\ \bibnamefont {Yu}},\ }\bibfield  {title} {\bibinfo
		{title} {{Gate engineering Fabry--P\'{e}rot resonance in altermagnetic
				junctions}},\ }\href {https://doi.org/10.1038/s41598-025-29607-w} {\bibfield
		{journal} {\bibinfo  {journal} {Sci. Rep.}\ }\textbf {\bibinfo {volume}
			{15}},\ \bibinfo {pages} {45508} (\bibinfo {year} {2025})}\BibitemShut
	{NoStop}%
	\bibitem [{\citenamefont {Soori}(2026)}]{Soori2026PRM}%
	\BibitemOpen
	\bibfield  {author} {\bibinfo {author} {\bibfnamefont {A.}~\bibnamefont
			{Soori}},\ }\bibfield  {title} {\bibinfo {title} {{Geometry induced net spin
				polarization of d-wave altermagnets}},\ }\href
	{https://doi.org/10.1103/8ntv-rwyn} {\bibfield  {journal} {\bibinfo
			{journal} {Phys. Rev. Materials}\ }\textbf {\bibinfo {volume} {10}},\
		\bibinfo {pages} {044406} (\bibinfo {year} {2026})}\BibitemShut {NoStop}%
	\bibitem [{\citenamefont {Abadillo-Uriel}\ \emph {et~al.}(2026)\citenamefont
		{Abadillo-Uriel}, \citenamefont {Maiani}, \citenamefont {Cortijo},
		\citenamefont {Aguado},\ and\ \citenamefont
		{Seoane~Souto}}]{AbadilloUriel2026arXiv}%
	\BibitemOpen
	\bibfield  {author} {\bibinfo {author} {\bibfnamefont {J.~C.}\ \bibnamefont
			{Abadillo-Uriel}}, \bibinfo {author} {\bibfnamefont {A.}~\bibnamefont
			{Maiani}}, \bibinfo {author} {\bibfnamefont {A.}~\bibnamefont {Cortijo}},
		\bibinfo {author} {\bibfnamefont {R.}~\bibnamefont {Aguado}},\ and\ \bibinfo
		{author} {\bibfnamefont {R.}~\bibnamefont {Seoane~Souto}},\ }\bibfield
	{title} {\bibinfo {title} {{All-electrical dephasing-protected spin qubits in
				altermagnets}},\ }\bibfield  {journal} {\bibinfo  {journal}
		{arXiv:2606.26066}\ }\href {https://doi.org/10.48550/arXiv.2606.26066}
	{10.48550/arXiv.2606.26066} (\bibinfo {year} {2026}),\ \Eprint
	{https://arxiv.org/abs/2606.26066} {arXiv:2606.26066} \BibitemShut {NoStop}%
	\bibitem [{\citenamefont {Zhu}\ \emph {et~al.}(2026)\citenamefont {Zhu},
		\citenamefont {Huang}, \citenamefont {Chen}, \citenamefont {Cui},
		\citenamefont {Duan}, \citenamefont {Zhang}, \citenamefont {\v{Z}uti\'c},\
		and\ \citenamefont {Zhou}}]{Zhu2026PRL}%
	\BibitemOpen
	\bibfield  {author} {\bibinfo {author} {\bibfnamefont {Z.}~\bibnamefont
			{Zhu}}, \bibinfo {author} {\bibfnamefont {R.}~\bibnamefont {Huang}}, \bibinfo
		{author} {\bibfnamefont {X.}~\bibnamefont {Chen}}, \bibinfo {author}
		{\bibfnamefont {Z.}~\bibnamefont {Cui}}, \bibinfo {author} {\bibfnamefont
			{X.}~\bibnamefont {Duan}}, \bibinfo {author} {\bibfnamefont {J.}~\bibnamefont
			{Zhang}}, \bibinfo {author} {\bibfnamefont {I.}~\bibnamefont {\v{Z}uti\'c}},\
		and\ \bibinfo {author} {\bibfnamefont {T.}~\bibnamefont {Zhou}},\ }\bibfield
	{title} {\bibinfo {title} {{Altermagnetic Proximity Effect}},\ }\href
	{https://doi.org/10.1103/kqy8-myz1} {\bibfield  {journal} {\bibinfo
			{journal} {Phys. Rev. Lett.}\ }\textbf {\bibinfo {volume} {136}},\ \bibinfo
		{pages} {186702} (\bibinfo {year} {2026})}\BibitemShut {NoStop}%
	\bibitem [{\citenamefont {Chen}\ \emph {et~al.}(2026)\citenamefont {Chen},
		\citenamefont {Xie}, \citenamefont {Li}, \citenamefont {Zhang}, \citenamefont
		{Hou}, \citenamefont {Zhang}, \citenamefont {He}, \citenamefont {Liu},
		\citenamefont {Wang},\ and\ \citenamefont {Liu}}]{Chen2026NPhysTwisted}%
	\BibitemOpen
	\bibfield  {author} {\bibinfo {author} {\bibfnamefont {J.}~\bibnamefont
			{Chen}}, \bibinfo {author} {\bibfnamefont {X.}~\bibnamefont {Xie}}, \bibinfo
		{author} {\bibfnamefont {S.}~\bibnamefont {Li}}, \bibinfo {author}
		{\bibfnamefont {S.}~\bibnamefont {Zhang}}, \bibinfo {author} {\bibfnamefont
			{S.}~\bibnamefont {Hou}}, \bibinfo {author} {\bibfnamefont {X.}~\bibnamefont
			{Zhang}}, \bibinfo {author} {\bibfnamefont {J.}~\bibnamefont {He}}, \bibinfo
		{author} {\bibfnamefont {Z.}~\bibnamefont {Liu}}, \bibinfo {author}
		{\bibfnamefont {J.-T.}\ \bibnamefont {Wang}},\ and\ \bibinfo {author}
		{\bibfnamefont {Y.}~\bibnamefont {Liu}},\ }\bibfield  {title} {\bibinfo
		{title} {{Altermagnetism in twisted van der Waals homostructures}},\
	}\bibfield  {journal} {\bibinfo  {journal} {Nat. Phys.}\ }\href
	{https://doi.org/10.1038/s41567-026-03440-y} {10.1038/s41567-026-03440-y}
	(\bibinfo {year} {2026}),\ \bibinfo {note} {published online 2 September
		2026}\BibitemShut {NoStop}%
	\bibitem [{\citenamefont {Aharonov}\ and\ \citenamefont
		{Bohm}(1959)}]{AharonovBohm1959}%
	\BibitemOpen
	\bibfield  {author} {\bibinfo {author} {\bibfnamefont {Y.}~\bibnamefont
			{Aharonov}}\ and\ \bibinfo {author} {\bibfnamefont {D.}~\bibnamefont
			{Bohm}},\ }\bibfield  {title} {\bibinfo {title} {{Significance of
				Electromagnetic Potentials in the Quantum Theory}},\ }\href
	{https://doi.org/10.1103/PhysRev.115.485} {\bibfield  {journal} {\bibinfo
			{journal} {Phys. Rev.}\ }\textbf {\bibinfo {volume} {115}},\ \bibinfo {pages}
		{485} (\bibinfo {year} {1959})}\BibitemShut {NoStop}%
	\bibitem [{\citenamefont {B\"uttiker}\ \emph {et~al.}(1983)\citenamefont
		{B\"uttiker}, \citenamefont {Imry},\ and\ \citenamefont
		{Landauer}}]{Buttiker1983}%
	\BibitemOpen
	\bibfield  {author} {\bibinfo {author} {\bibfnamefont {M.}~\bibnamefont
			{B\"uttiker}}, \bibinfo {author} {\bibfnamefont {Y.}~\bibnamefont {Imry}},\
		and\ \bibinfo {author} {\bibfnamefont {R.}~\bibnamefont {Landauer}},\
	}\bibfield  {title} {\bibinfo {title} {{Josephson behavior in small normal
				one-dimensional rings}},\ }\href
	{https://doi.org/10.1016/0375-9601(83)90011-7} {\bibfield  {journal}
		{\bibinfo  {journal} {Phys. Lett. A}\ }\textbf {\bibinfo {volume} {96}},\
		\bibinfo {pages} {365} (\bibinfo {year} {1983})}\BibitemShut {NoStop}%
	\bibitem [{\citenamefont {Webb}\ \emph {et~al.}(1985)\citenamefont {Webb},
		\citenamefont {Washburn}, \citenamefont {Umbach},\ and\ \citenamefont
		{Laibowitz}}]{Webb1985PRL}%
	\BibitemOpen
	\bibfield  {author} {\bibinfo {author} {\bibfnamefont {R.~A.}\ \bibnamefont
			{Webb}}, \bibinfo {author} {\bibfnamefont {S.}~\bibnamefont {Washburn}},
		\bibinfo {author} {\bibfnamefont {C.~P.}\ \bibnamefont {Umbach}},\ and\
		\bibinfo {author} {\bibfnamefont {R.~B.}\ \bibnamefont {Laibowitz}},\
	}\bibfield  {title} {\bibinfo {title} {{Observation of h/e Aharonov-Bohm
				Oscillations in Normal-Metal Rings}},\ }\href
	{https://doi.org/10.1103/PhysRevLett.54.2696} {\bibfield  {journal} {\bibinfo
			{journal} {Phys. Rev. Lett.}\ }\textbf {\bibinfo {volume} {54}},\ \bibinfo
		{pages} {2696} (\bibinfo {year} {1985})}\BibitemShut {NoStop}%
	\bibitem [{\citenamefont {Levy}\ \emph {et~al.}(1990)\citenamefont {Levy},
		\citenamefont {Dolan}, \citenamefont {Dunsmuir},\ and\ \citenamefont
		{Bouchiat}}]{Levy1990PRL}%
	\BibitemOpen
	\bibfield  {author} {\bibinfo {author} {\bibfnamefont {L.~P.}\ \bibnamefont
			{Levy}}, \bibinfo {author} {\bibfnamefont {G.}~\bibnamefont {Dolan}},
		\bibinfo {author} {\bibfnamefont {J.}~\bibnamefont {Dunsmuir}},\ and\
		\bibinfo {author} {\bibfnamefont {H.}~\bibnamefont {Bouchiat}},\ }\bibfield
	{title} {\bibinfo {title} {{Magnetization of mesoscopic copper rings:
				Evidence for persistent currents}},\ }\href
	{https://doi.org/10.1103/PhysRevLett.64.2074} {\bibfield  {journal} {\bibinfo
			{journal} {Phys. Rev. Lett.}\ }\textbf {\bibinfo {volume} {64}},\ \bibinfo
		{pages} {2074} (\bibinfo {year} {1990})}\BibitemShut {NoStop}%
	\bibitem [{\citenamefont {Chandrasekhar}\ \emph {et~al.}(1991)\citenamefont
		{Chandrasekhar}, \citenamefont {Webb}, \citenamefont {Brady}, \citenamefont
		{Ketchen}, \citenamefont {Gallagher},\ and\ \citenamefont
		{Kleinsasser}}]{Chandrasekhar1991PRL}%
	\BibitemOpen
	\bibfield  {author} {\bibinfo {author} {\bibfnamefont {V.}~\bibnamefont
			{Chandrasekhar}}, \bibinfo {author} {\bibfnamefont {R.~A.}\ \bibnamefont
			{Webb}}, \bibinfo {author} {\bibfnamefont {M.~J.}\ \bibnamefont {Brady}},
		\bibinfo {author} {\bibfnamefont {M.~B.}\ \bibnamefont {Ketchen}}, \bibinfo
		{author} {\bibfnamefont {W.~J.}\ \bibnamefont {Gallagher}},\ and\ \bibinfo
		{author} {\bibfnamefont {A.}~\bibnamefont {Kleinsasser}},\ }\bibfield
	{title} {\bibinfo {title} {{Magnetic response of a single, isolated gold
				loop}},\ }\href {https://doi.org/10.1103/PhysRevLett.67.3578} {\bibfield
		{journal} {\bibinfo  {journal} {Phys. Rev. Lett.}\ }\textbf {\bibinfo
			{volume} {67}},\ \bibinfo {pages} {3578} (\bibinfo {year}
		{1991})}\BibitemShut {NoStop}%
	\bibitem [{\citenamefont {Timp}\ \emph {et~al.}(1987)\citenamefont {Timp},
		\citenamefont {Chang}, \citenamefont {Cunningham}, \citenamefont {Chang},
		\citenamefont {Mankiewich}, \citenamefont {Behringer},\ and\ \citenamefont
		{Howard}}]{Timp1987PRL}%
	\BibitemOpen
	\bibfield  {author} {\bibinfo {author} {\bibfnamefont {G.}~\bibnamefont
			{Timp}}, \bibinfo {author} {\bibfnamefont {A.~M.}\ \bibnamefont {Chang}},
		\bibinfo {author} {\bibfnamefont {J.~E.}\ \bibnamefont {Cunningham}},
		\bibinfo {author} {\bibfnamefont {T.~Y.}\ \bibnamefont {Chang}}, \bibinfo
		{author} {\bibfnamefont {P.}~\bibnamefont {Mankiewich}}, \bibinfo {author}
		{\bibfnamefont {R.}~\bibnamefont {Behringer}},\ and\ \bibinfo {author}
		{\bibfnamefont {R.~E.}\ \bibnamefont {Howard}},\ }\bibfield  {title}
	{\bibinfo {title} {{Observation of the Aharonov--Bohm effect for
				$\omega_c\tau > 1$}},\ }\href {https://doi.org/10.1103/PhysRevLett.58.2814}
	{\bibfield  {journal} {\bibinfo  {journal} {Phys. Rev. Lett.}\ }\textbf
		{\bibinfo {volume} {58}},\ \bibinfo {pages} {2814} (\bibinfo {year}
		{1987})}\BibitemShut {NoStop}%
	\bibitem [{\citenamefont {Timp}\ \emph {et~al.}(1989)\citenamefont {Timp},
		\citenamefont {Mankiewich}, \citenamefont {deVegvar}, \citenamefont
		{Behringer}, \citenamefont {Cunningham}, \citenamefont {Howard},
		\citenamefont {Baranger},\ and\ \citenamefont {Jain}}]{Timp1989PRB}%
	\BibitemOpen
	\bibfield  {author} {\bibinfo {author} {\bibfnamefont {G.}~\bibnamefont
			{Timp}}, \bibinfo {author} {\bibfnamefont {P.~M.}\ \bibnamefont
			{Mankiewich}}, \bibinfo {author} {\bibfnamefont {P.}~\bibnamefont
			{deVegvar}}, \bibinfo {author} {\bibfnamefont {R.}~\bibnamefont {Behringer}},
		\bibinfo {author} {\bibfnamefont {J.~E.}\ \bibnamefont {Cunningham}},
		\bibinfo {author} {\bibfnamefont {R.~E.}\ \bibnamefont {Howard}}, \bibinfo
		{author} {\bibfnamefont {H.~U.}\ \bibnamefont {Baranger}},\ and\ \bibinfo
		{author} {\bibfnamefont {J.~K.}\ \bibnamefont {Jain}},\ }\bibfield  {title}
	{\bibinfo {title} {{Suppression of the Aharonov-Bohm effect in the quantized
				Hall regime}},\ }\href {https://doi.org/10.1103/PhysRevB.39.6227} {\bibfield
		{journal} {\bibinfo  {journal} {Phys. Rev. B}\ }\textbf {\bibinfo {volume}
			{39}},\ \bibinfo {pages} {6227} (\bibinfo {year} {1989})}\BibitemShut
	{NoStop}%
	\bibitem [{\citenamefont {Mailly}\ \emph {et~al.}(1993)\citenamefont {Mailly},
		\citenamefont {Chapelier},\ and\ \citenamefont {Benoit}}]{Mailly1993PRL}%
	\BibitemOpen
	\bibfield  {author} {\bibinfo {author} {\bibfnamefont {D.}~\bibnamefont
			{Mailly}}, \bibinfo {author} {\bibfnamefont {C.}~\bibnamefont {Chapelier}},\
		and\ \bibinfo {author} {\bibfnamefont {A.}~\bibnamefont {Benoit}},\
	}\bibfield  {title} {\bibinfo {title} {{Experimental observation of
				persistent currents in GaAs-AlGaAs single loop}},\ }\href
	{https://doi.org/10.1103/PhysRevLett.70.2020} {\bibfield  {journal} {\bibinfo
			{journal} {Phys. Rev. Lett.}\ }\textbf {\bibinfo {volume} {70}},\ \bibinfo
		{pages} {2020} (\bibinfo {year} {1993})}\BibitemShut {NoStop}%
	\bibitem [{\citenamefont {Liu}\ \emph {et~al.}(1993)\citenamefont {Liu},
		\citenamefont {Gao}, \citenamefont {Ismail}, \citenamefont {Lee},
		\citenamefont {Hong},\ and\ \citenamefont {Washburn}}]{Liu1993PRB}%
	\BibitemOpen
	\bibfield  {author} {\bibinfo {author} {\bibfnamefont {J.}~\bibnamefont
			{Liu}}, \bibinfo {author} {\bibfnamefont {W.~X.}\ \bibnamefont {Gao}},
		\bibinfo {author} {\bibfnamefont {K.}~\bibnamefont {Ismail}}, \bibinfo
		{author} {\bibfnamefont {K.~Y.}\ \bibnamefont {Lee}}, \bibinfo {author}
		{\bibfnamefont {J.~M.}\ \bibnamefont {Hong}},\ and\ \bibinfo {author}
		{\bibfnamefont {S.}~\bibnamefont {Washburn}},\ }\bibfield  {title} {\bibinfo
		{title} {{Correlations between Aharonov--Bohm effects and one-dimensional
				subband populations in {GaAs}/{Al}$_x${Ga}$_{1-x}${As} rings}},\ }\href
	{https://doi.org/10.1103/PhysRevB.48.15148} {\bibfield  {journal} {\bibinfo
			{journal} {Phys. Rev. B}\ }\textbf {\bibinfo {volume} {48}},\ \bibinfo
		{pages} {15148} (\bibinfo {year} {1993})}\BibitemShut {NoStop}%
	\bibitem [{\citenamefont {Liu}\ \emph {et~al.}(1994)\citenamefont {Liu},
		\citenamefont {Gao}, \citenamefont {Ismail}, \citenamefont {Lee},
		\citenamefont {Hong},\ and\ \citenamefont {Washburn}}]{Liu1994PRB}%
	\BibitemOpen
	\bibfield  {author} {\bibinfo {author} {\bibfnamefont {J.}~\bibnamefont
			{Liu}}, \bibinfo {author} {\bibfnamefont {W.~X.}\ \bibnamefont {Gao}},
		\bibinfo {author} {\bibfnamefont {K.}~\bibnamefont {Ismail}}, \bibinfo
		{author} {\bibfnamefont {K.~Y.}\ \bibnamefont {Lee}}, \bibinfo {author}
		{\bibfnamefont {J.~M.}\ \bibnamefont {Hong}},\ and\ \bibinfo {author}
		{\bibfnamefont {S.}~\bibnamefont {Washburn}},\ }\bibfield  {title} {\bibinfo
		{title} {{Tunneling between edge states in clean multiple
				{GaAs}/{Al}$_x${Ga}$_{1-x}${As} rings and increase of the phase coherence
				length with magnetic field}},\ }\href
	{https://doi.org/10.1103/PhysRevB.50.17383} {\bibfield  {journal} {\bibinfo
			{journal} {Phys. Rev. B}\ }\textbf {\bibinfo {volume} {50}},\ \bibinfo
		{pages} {17383} (\bibinfo {year} {1994})}\BibitemShut {NoStop}%
	\bibitem [{\citenamefont {Loss}\ \emph {et~al.}(1990)\citenamefont {Loss},
		\citenamefont {Goldbart},\ and\ \citenamefont {Balatsky}}]{Loss1990PRL}%
	\BibitemOpen
	\bibfield  {author} {\bibinfo {author} {\bibfnamefont {D.}~\bibnamefont
			{Loss}}, \bibinfo {author} {\bibfnamefont {P.}~\bibnamefont {Goldbart}},\
		and\ \bibinfo {author} {\bibfnamefont {A.~V.}\ \bibnamefont {Balatsky}},\
	}\bibfield  {title} {\bibinfo {title} {{Berry's phase and persistent charge
				and spin currents in textured mesoscopic rings}},\ }\href
	{https://doi.org/10.1103/PhysRevLett.65.1655} {\bibfield  {journal} {\bibinfo
			{journal} {Phys. Rev. Lett.}\ }\textbf {\bibinfo {volume} {65}},\ \bibinfo
		{pages} {1655} (\bibinfo {year} {1990})}\BibitemShut {NoStop}%
	\bibitem [{\citenamefont {Stern}(1992)}]{Stern1992PRL}%
	\BibitemOpen
	\bibfield  {author} {\bibinfo {author} {\bibfnamefont {A.}~\bibnamefont
			{Stern}},\ }\bibfield  {title} {\bibinfo {title} {{Berry's phase, motive
				forces, and mesoscopic conductivity}},\ }\href
	{https://doi.org/10.1103/PhysRevLett.68.1022} {\bibfield  {journal} {\bibinfo
			{journal} {Phys. Rev. Lett.}\ }\textbf {\bibinfo {volume} {68}},\ \bibinfo
		{pages} {1022} (\bibinfo {year} {1992})}\BibitemShut {NoStop}%
	\bibitem [{\citenamefont {Aronov}\ and\ \citenamefont
		{Lyanda-Geller}(1993)}]{Aronov1993PRL}%
	\BibitemOpen
	\bibfield  {author} {\bibinfo {author} {\bibfnamefont {A.~G.}\ \bibnamefont
			{Aronov}}\ and\ \bibinfo {author} {\bibfnamefont {Y.~B.}\ \bibnamefont
			{Lyanda-Geller}},\ }\bibfield  {title} {\bibinfo {title} {{Spin-orbit Berry
				phase in conducting rings}},\ }\href
	{https://doi.org/10.1103/PhysRevLett.70.343} {\bibfield  {journal} {\bibinfo
			{journal} {Phys. Rev. Lett.}\ }\textbf {\bibinfo {volume} {70}},\ \bibinfo
		{pages} {343} (\bibinfo {year} {1993})}\BibitemShut {NoStop}%
	\bibitem [{\citenamefont {Qian}\ and\ \citenamefont {Su}(1994)}]{Qian1994PRL}%
	\BibitemOpen
	\bibfield  {author} {\bibinfo {author} {\bibfnamefont {T.-Z.}\ \bibnamefont
			{Qian}}\ and\ \bibinfo {author} {\bibfnamefont {Z.-B.}\ \bibnamefont {Su}},\
	}\bibfield  {title} {\bibinfo {title} {{Spin-orbit interaction and
				Aharonov-Anandan phase in mesoscopic rings}},\ }\href
	{https://doi.org/10.1103/PhysRevLett.72.2311} {\bibfield  {journal} {\bibinfo
			{journal} {Phys. Rev. Lett.}\ }\textbf {\bibinfo {volume} {72}},\ \bibinfo
		{pages} {2311} (\bibinfo {year} {1994})}\BibitemShut {NoStop}%
	\bibitem [{\citenamefont {Yi}\ \emph {et~al.}(1997)\citenamefont {Yi},
		\citenamefont {Qian},\ and\ \citenamefont {Su}}]{Yi1997PRB}%
	\BibitemOpen
	\bibfield  {author} {\bibinfo {author} {\bibfnamefont {Y.-S.}\ \bibnamefont
			{Yi}}, \bibinfo {author} {\bibfnamefont {T.-Z.}\ \bibnamefont {Qian}},\ and\
		\bibinfo {author} {\bibfnamefont {Z.-B.}\ \bibnamefont {Su}},\ }\bibfield
	{title} {\bibinfo {title} {{Spin precession and time-reversal symmetry
				breaking in quantum transport of electrons through mesoscopic rings}},\
	}\href {https://doi.org/10.1103/PhysRevB.55.10631} {\bibfield  {journal}
		{\bibinfo  {journal} {Phys. Rev. B}\ }\textbf {\bibinfo {volume} {55}},\
		\bibinfo {pages} {10631} (\bibinfo {year} {1997})}\BibitemShut {NoStop}%
	\bibitem [{\citenamefont {Morpurgo}\ \emph {et~al.}(1998)\citenamefont
		{Morpurgo}, \citenamefont {Heida}, \citenamefont {Klapwijk},\ and\
		\citenamefont {van Wees}}]{Morpurgo1998PRL}%
	\BibitemOpen
	\bibfield  {author} {\bibinfo {author} {\bibfnamefont {A.~F.}\ \bibnamefont
			{Morpurgo}}, \bibinfo {author} {\bibfnamefont {J.~P.}\ \bibnamefont {Heida}},
		\bibinfo {author} {\bibfnamefont {T.~M.}\ \bibnamefont {Klapwijk}},\ and\
		\bibinfo {author} {\bibfnamefont {B.~J.}\ \bibnamefont {van Wees}},\
	}\bibfield  {title} {\bibinfo {title} {{Ensemble-Average Spectrum of
				Aharonov-Bohm Conductance Oscillations: Evidence for Spin-Orbit-Induced
				Berry's Phase}},\ }\href {https://doi.org/10.1103/PhysRevLett.80.1050}
	{\bibfield  {journal} {\bibinfo  {journal} {Phys. Rev. Lett.}\ }\textbf
		{\bibinfo {volume} {80}},\ \bibinfo {pages} {1050} (\bibinfo {year}
		{1998})}\BibitemShut {NoStop}%
	\bibitem [{\citenamefont {Yau}\ \emph {et~al.}(2002)\citenamefont {Yau},
		\citenamefont {de~Poortere},\ and\ \citenamefont {Shayegan}}]{Yau2002PRL}%
	\BibitemOpen
	\bibfield  {author} {\bibinfo {author} {\bibfnamefont {J.-B.}\ \bibnamefont
			{Yau}}, \bibinfo {author} {\bibfnamefont {E.~P.}\ \bibnamefont
			{de~Poortere}},\ and\ \bibinfo {author} {\bibfnamefont {M.}~\bibnamefont
			{Shayegan}},\ }\bibfield  {title} {\bibinfo {title} {{Aharonov-Bohm
				Oscillations with Spin: Evidence for Berry's Phase}},\ }\href
	{https://doi.org/10.1103/PhysRevLett.88.146801} {\bibfield  {journal}
		{\bibinfo  {journal} {Phys. Rev. Lett.}\ }\textbf {\bibinfo {volume} {88}},\
		\bibinfo {pages} {146801} (\bibinfo {year} {2002})}\BibitemShut {NoStop}%
	\bibitem [{\citenamefont {Rabaud}\ \emph {et~al.}(2001)\citenamefont {Rabaud},
		\citenamefont {Saminadayar}, \citenamefont {Mailly}, \citenamefont
		{Hasselbach}, \citenamefont {Beno\^{i}t},\ and\ \citenamefont
		{Etienne}}]{Rabaud2001PRL}%
	\BibitemOpen
	\bibfield  {author} {\bibinfo {author} {\bibfnamefont {W.}~\bibnamefont
			{Rabaud}}, \bibinfo {author} {\bibfnamefont {L.}~\bibnamefont {Saminadayar}},
		\bibinfo {author} {\bibfnamefont {D.}~\bibnamefont {Mailly}}, \bibinfo
		{author} {\bibfnamefont {K.}~\bibnamefont {Hasselbach}}, \bibinfo {author}
		{\bibfnamefont {A.}~\bibnamefont {Beno\^{i}t}},\ and\ \bibinfo {author}
		{\bibfnamefont {B.}~\bibnamefont {Etienne}},\ }\bibfield  {title} {\bibinfo
		{title} {{Persistent Currents in Mesoscopic Connected Rings}},\ }\href
	{https://doi.org/10.1103/PhysRevLett.86.3124} {\bibfield  {journal} {\bibinfo
			{journal} {Phys. Rev. Lett.}\ }\textbf {\bibinfo {volume} {86}},\ \bibinfo
		{pages} {3124} (\bibinfo {year} {2001})}\BibitemShut {NoStop}%
	\bibitem [{\citenamefont {Bluhm}\ \emph {et~al.}(2009)\citenamefont {Bluhm},
		\citenamefont {Koshnick}, \citenamefont {Bert}, \citenamefont {Huber},\ and\
		\citenamefont {Moler}}]{Bluhm2009PRL}%
	\BibitemOpen
	\bibfield  {author} {\bibinfo {author} {\bibfnamefont {H.}~\bibnamefont
			{Bluhm}}, \bibinfo {author} {\bibfnamefont {N.~C.}\ \bibnamefont {Koshnick}},
		\bibinfo {author} {\bibfnamefont {J.~A.}\ \bibnamefont {Bert}}, \bibinfo
		{author} {\bibfnamefont {M.~E.}\ \bibnamefont {Huber}},\ and\ \bibinfo
		{author} {\bibfnamefont {K.~A.}\ \bibnamefont {Moler}},\ }\bibfield  {title}
	{\bibinfo {title} {{Persistent Currents in Normal Metal Rings}},\ }\href
	{https://doi.org/10.1103/PhysRevLett.102.136802} {\bibfield  {journal}
		{\bibinfo  {journal} {Phys. Rev. Lett.}\ }\textbf {\bibinfo {volume} {102}},\
		\bibinfo {pages} {136802} (\bibinfo {year} {2009})}\BibitemShut {NoStop}%
	\bibitem [{\citenamefont {Bleszynski-Jayich}\ \emph {et~al.}(2009)\citenamefont
		{Bleszynski-Jayich}, \citenamefont {Shanks}, \citenamefont {Peaudecerf},
		\citenamefont {Ginossar}, \citenamefont {von Oppen}, \citenamefont
		{Glazman},\ and\ \citenamefont {Harris}}]{Bleszynski2009Science}%
	\BibitemOpen
	\bibfield  {author} {\bibinfo {author} {\bibfnamefont {A.~C.}\ \bibnamefont
			{Bleszynski-Jayich}}, \bibinfo {author} {\bibfnamefont {W.~E.}\ \bibnamefont
			{Shanks}}, \bibinfo {author} {\bibfnamefont {B.}~\bibnamefont {Peaudecerf}},
		\bibinfo {author} {\bibfnamefont {E.}~\bibnamefont {Ginossar}}, \bibinfo
		{author} {\bibfnamefont {F.}~\bibnamefont {von Oppen}}, \bibinfo {author}
		{\bibfnamefont {L.}~\bibnamefont {Glazman}},\ and\ \bibinfo {author}
		{\bibfnamefont {J.~G.~E.}\ \bibnamefont {Harris}},\ }\bibfield  {title}
	{\bibinfo {title} {{Persistent currents in normal metal rings}},\ }\href
	{https://doi.org/10.1126/science.1178139} {\bibfield  {journal} {\bibinfo
			{journal} {Science}\ }\textbf {\bibinfo {volume} {326}},\ \bibinfo {pages}
		{272} (\bibinfo {year} {2009})}\BibitemShut {NoStop}%
	\bibitem [{\citenamefont {Lorke}\ \emph {et~al.}(2000)\citenamefont {Lorke},
		\citenamefont {Luyken}, \citenamefont {Govorov}, \citenamefont {Kotthaus},
		\citenamefont {Garcia},\ and\ \citenamefont {Petroff}}]{Lorke2000}%
	\BibitemOpen
	\bibfield  {author} {\bibinfo {author} {\bibfnamefont {A.}~\bibnamefont
			{Lorke}}, \bibinfo {author} {\bibfnamefont {R.~J.}\ \bibnamefont {Luyken}},
		\bibinfo {author} {\bibfnamefont {A.~O.}\ \bibnamefont {Govorov}}, \bibinfo
		{author} {\bibfnamefont {J.~P.}\ \bibnamefont {Kotthaus}}, \bibinfo {author}
		{\bibfnamefont {J.~M.}\ \bibnamefont {Garcia}},\ and\ \bibinfo {author}
		{\bibfnamefont {P.~M.}\ \bibnamefont {Petroff}},\ }\bibfield  {title}
	{\bibinfo {title} {{Spectroscopy of Nanoscopic Semiconductor Rings}},\ }\href
	{https://doi.org/10.1103/PhysRevLett.84.2223} {\bibfield  {journal} {\bibinfo
			{journal} {Phys. Rev. Lett.}\ }\textbf {\bibinfo {volume} {84}},\ \bibinfo
		{pages} {2223} (\bibinfo {year} {2000})}\BibitemShut {NoStop}%
	\bibitem [{\citenamefont {F\"{u}hrer}\ \emph {et~al.}(2001)\citenamefont
		{F\"{u}hrer}, \citenamefont {L\"{u}scher}, \citenamefont {Ihn}, \citenamefont
		{Heinzel}, \citenamefont {Ensslin}, \citenamefont {Wegscheider},\ and\
		\citenamefont {Bichler}}]{Fuhrer2001Nature}%
	\BibitemOpen
	\bibfield  {author} {\bibinfo {author} {\bibfnamefont {A.}~\bibnamefont
			{F\"{u}hrer}}, \bibinfo {author} {\bibfnamefont {S.}~\bibnamefont
			{L\"{u}scher}}, \bibinfo {author} {\bibfnamefont {T.}~\bibnamefont {Ihn}},
		\bibinfo {author} {\bibfnamefont {T.}~\bibnamefont {Heinzel}}, \bibinfo
		{author} {\bibfnamefont {K.}~\bibnamefont {Ensslin}}, \bibinfo {author}
		{\bibfnamefont {W.}~\bibnamefont {Wegscheider}},\ and\ \bibinfo {author}
		{\bibfnamefont {M.}~\bibnamefont {Bichler}},\ }\bibfield  {title} {\bibinfo
		{title} {{Energy spectra of quantum rings}},\ }\href
	{https://doi.org/10.1038/35101552} {\bibfield  {journal} {\bibinfo  {journal}
			{Nature}\ }\textbf {\bibinfo {volume} {413}},\ \bibinfo {pages} {822}
		(\bibinfo {year} {2001})}\BibitemShut {NoStop}%
	\bibitem [{\citenamefont {Warburton}\ \emph {et~al.}(2000)\citenamefont
		{Warburton}, \citenamefont {Sch\"{a}flein}, \citenamefont {Haft},
		\citenamefont {Bickel}, \citenamefont {Lorke}, \citenamefont {Karrai},
		\citenamefont {Garc\'{i}a}, \citenamefont {Schoenfeld},\ and\ \citenamefont
		{Petroff}}]{Warburton2000Nature}%
	\BibitemOpen
	\bibfield  {author} {\bibinfo {author} {\bibfnamefont {R.~J.}\ \bibnamefont
			{Warburton}}, \bibinfo {author} {\bibfnamefont {C.}~\bibnamefont
			{Sch\"{a}flein}}, \bibinfo {author} {\bibfnamefont {D.}~\bibnamefont {Haft}},
		\bibinfo {author} {\bibfnamefont {F.}~\bibnamefont {Bickel}}, \bibinfo
		{author} {\bibfnamefont {A.}~\bibnamefont {Lorke}}, \bibinfo {author}
		{\bibfnamefont {K.}~\bibnamefont {Karrai}}, \bibinfo {author} {\bibfnamefont
			{J.~M.}\ \bibnamefont {Garc\'{i}a}}, \bibinfo {author} {\bibfnamefont
			{W.}~\bibnamefont {Schoenfeld}},\ and\ \bibinfo {author} {\bibfnamefont
			{P.~M.}\ \bibnamefont {Petroff}},\ }\bibfield  {title} {\bibinfo {title}
		{{Optical emission from a charge-tunable quantum ring}},\ }\href
	{https://doi.org/10.1038/35016030} {\bibfield  {journal} {\bibinfo  {journal}
			{Nature}\ }\textbf {\bibinfo {volume} {405}},\ \bibinfo {pages} {926}
		(\bibinfo {year} {2000})}\BibitemShut {NoStop}%
	\bibitem [{\citenamefont {Kleemans}\ \emph {et~al.}(2007)\citenamefont
		{Kleemans}, \citenamefont {Bominaar-Silkens}, \citenamefont {Fomin},
		\citenamefont {Gladilin}, \citenamefont {Granados}, \citenamefont {Taboada},
		\citenamefont {Garc\'{i}a}, \citenamefont {Offermans}, \citenamefont
		{Zeitler}, \citenamefont {Christianen}, \citenamefont {Maan}, \citenamefont
		{Devreese},\ and\ \citenamefont {Koenraad}}]{Kleemans2007PRL}%
	\BibitemOpen
	\bibfield  {author} {\bibinfo {author} {\bibfnamefont {N.~A. J.~M.}\
			\bibnamefont {Kleemans}}, \bibinfo {author} {\bibfnamefont {I.~M.~A.}\
			\bibnamefont {Bominaar-Silkens}}, \bibinfo {author} {\bibfnamefont {V.~M.}\
			\bibnamefont {Fomin}}, \bibinfo {author} {\bibfnamefont {V.~N.}\ \bibnamefont
			{Gladilin}}, \bibinfo {author} {\bibfnamefont {D.}~\bibnamefont {Granados}},
		\bibinfo {author} {\bibfnamefont {A.~G.}\ \bibnamefont {Taboada}}, \bibinfo
		{author} {\bibfnamefont {J.~M.}\ \bibnamefont {Garc\'{i}a}}, \bibinfo
		{author} {\bibfnamefont {P.}~\bibnamefont {Offermans}}, \bibinfo {author}
		{\bibfnamefont {U.}~\bibnamefont {Zeitler}}, \bibinfo {author} {\bibfnamefont
			{P.~C.~M.}\ \bibnamefont {Christianen}}, \bibinfo {author} {\bibfnamefont
			{J.~C.}\ \bibnamefont {Maan}}, \bibinfo {author} {\bibfnamefont {J.~T.}\
			\bibnamefont {Devreese}},\ and\ \bibinfo {author} {\bibfnamefont {P.~M.}\
			\bibnamefont {Koenraad}},\ }\bibfield  {title} {\bibinfo {title}
		{{Oscillatory Persistent Currents in Self-Assembled Quantum Rings}},\ }\href
	{https://doi.org/10.1103/PhysRevLett.99.146808} {\bibfield  {journal}
		{\bibinfo  {journal} {Phys. Rev. Lett.}\ }\textbf {\bibinfo {volume} {99}},\
		\bibinfo {pages} {146808} (\bibinfo {year} {2007})}\BibitemShut {NoStop}%
	\bibitem [{\citenamefont {Tarucha}\ \emph {et~al.}(1996)\citenamefont
		{Tarucha}, \citenamefont {Austing}, \citenamefont {Honda}, \citenamefont
		{van~der Hage},\ and\ \citenamefont {Kouwenhoven}}]{Tarucha1996PRL}%
	\BibitemOpen
	\bibfield  {author} {\bibinfo {author} {\bibfnamefont {S.}~\bibnamefont
			{Tarucha}}, \bibinfo {author} {\bibfnamefont {D.~G.}\ \bibnamefont
			{Austing}}, \bibinfo {author} {\bibfnamefont {T.}~\bibnamefont {Honda}},
		\bibinfo {author} {\bibfnamefont {R.~J.}\ \bibnamefont {van~der Hage}},\ and\
		\bibinfo {author} {\bibfnamefont {L.~P.}\ \bibnamefont {Kouwenhoven}},\
	}\bibfield  {title} {\bibinfo {title} {{Shell Filling and Spin Effects in a
				Few Electron Quantum Dot}},\ }\href
	{https://doi.org/10.1103/PhysRevLett.77.3613} {\bibfield  {journal} {\bibinfo
			{journal} {Phys. Rev. Lett.}\ }\textbf {\bibinfo {volume} {77}},\ \bibinfo
		{pages} {3613} (\bibinfo {year} {1996})}\BibitemShut {NoStop}%
	\bibitem [{\citenamefont {Keyser}\ \emph {et~al.}(2003)\citenamefont {Keyser},
		\citenamefont {F\"{u}hner}, \citenamefont {Borck}, \citenamefont {Haug},
		\citenamefont {Bichler}, \citenamefont {Abstreiter},\ and\ \citenamefont
		{Wegscheider}}]{Keyser2003PRL}%
	\BibitemOpen
	\bibfield  {author} {\bibinfo {author} {\bibfnamefont {U.~F.}\ \bibnamefont
			{Keyser}}, \bibinfo {author} {\bibfnamefont {C.}~\bibnamefont {F\"{u}hner}},
		\bibinfo {author} {\bibfnamefont {S.}~\bibnamefont {Borck}}, \bibinfo
		{author} {\bibfnamefont {R.~J.}\ \bibnamefont {Haug}}, \bibinfo {author}
		{\bibfnamefont {M.}~\bibnamefont {Bichler}}, \bibinfo {author} {\bibfnamefont
			{G.}~\bibnamefont {Abstreiter}},\ and\ \bibinfo {author} {\bibfnamefont
			{W.}~\bibnamefont {Wegscheider}},\ }\bibfield  {title} {\bibinfo {title}
		{{Kondo Effect in a Few-Electron Quantum Ring}},\ }\href
	{https://doi.org/10.1103/PhysRevLett.90.196601} {\bibfield  {journal}
		{\bibinfo  {journal} {Phys. Rev. Lett.}\ }\textbf {\bibinfo {volume} {90}},\
		\bibinfo {pages} {196601} (\bibinfo {year} {2003})}\BibitemShut {NoStop}%
	\bibitem [{\citenamefont {Pedersen}\ \emph {et~al.}(2000)\citenamefont
		{Pedersen}, \citenamefont {Hansen}, \citenamefont {Kristensen}, \citenamefont
		{Sorensen},\ and\ \citenamefont {Lindelof}}]{Pedersen2000PRB}%
	\BibitemOpen
	\bibfield  {author} {\bibinfo {author} {\bibfnamefont {S.}~\bibnamefont
			{Pedersen}}, \bibinfo {author} {\bibfnamefont {A.~E.}\ \bibnamefont
			{Hansen}}, \bibinfo {author} {\bibfnamefont {A.}~\bibnamefont {Kristensen}},
		\bibinfo {author} {\bibfnamefont {C.~B.}\ \bibnamefont {Sorensen}},\ and\
		\bibinfo {author} {\bibfnamefont {P.~E.}\ \bibnamefont {Lindelof}},\
	}\bibfield  {title} {\bibinfo {title} {{Observation of quantum asymmetry in
				an Aharonov-Bohm ring}},\ }\href {https://doi.org/10.1103/PhysRevB.61.5457}
	{\bibfield  {journal} {\bibinfo  {journal} {Phys. Rev. B}\ }\textbf {\bibinfo
			{volume} {61}},\ \bibinfo {pages} {5457} (\bibinfo {year}
		{2000})}\BibitemShut {NoStop}%
	\bibitem [{\citenamefont {Viefers}\ \emph {et~al.}(2004)\citenamefont
		{Viefers}, \citenamefont {Koskinen}, \citenamefont {Singha~Deo},\ and\
		\citenamefont {Manninen}}]{Viefers2004}%
	\BibitemOpen
	\bibfield  {author} {\bibinfo {author} {\bibfnamefont {S.}~\bibnamefont
			{Viefers}}, \bibinfo {author} {\bibfnamefont {P.}~\bibnamefont {Koskinen}},
		\bibinfo {author} {\bibfnamefont {P.}~\bibnamefont {Singha~Deo}},\ and\
		\bibinfo {author} {\bibfnamefont {M.}~\bibnamefont {Manninen}},\ }\bibfield
	{title} {\bibinfo {title} {{Quantum rings for beginners: Energy spectra and
				persistent currents}},\ }\href {https://doi.org/10.1016/j.physe.2003.08.076}
	{\bibfield  {journal} {\bibinfo  {journal} {Physica E}\ }\textbf {\bibinfo
			{volume} {21}},\ \bibinfo {pages} {1} (\bibinfo {year} {2004})}\BibitemShut
	{NoStop}%
	\bibitem [{\citenamefont {Reimann}\ and\ \citenamefont
		{Manninen}(2002)}]{Reimann2002RMP}%
	\BibitemOpen
	\bibfield  {author} {\bibinfo {author} {\bibfnamefont {S.~M.}\ \bibnamefont
			{Reimann}}\ and\ \bibinfo {author} {\bibfnamefont {M.}~\bibnamefont
			{Manninen}},\ }\bibfield  {title} {\bibinfo {title} {{Electronic structure of
				quantum dots}},\ }\href {https://doi.org/10.1103/RevModPhys.74.1283}
	{\bibfield  {journal} {\bibinfo  {journal} {Rev. Mod. Phys.}\ }\textbf
		{\bibinfo {volume} {74}},\ \bibinfo {pages} {1283} (\bibinfo {year}
		{2002})}\BibitemShut {NoStop}%
	\bibitem [{\citenamefont {Maksym}(1996)}]{Maksym1996PRB}%
	\BibitemOpen
	\bibfield  {author} {\bibinfo {author} {\bibfnamefont {P.~A.}\ \bibnamefont
			{Maksym}},\ }\bibfield  {title} {\bibinfo {title} {{Eckardt frame theory of
				interacting electrons in quantum dots}},\ }\href
	{https://doi.org/10.1103/PhysRevB.53.10871} {\bibfield  {journal} {\bibinfo
			{journal} {Phys. Rev. B}\ }\textbf {\bibinfo {volume} {53}},\ \bibinfo
		{pages} {10871} (\bibinfo {year} {1996})}\BibitemShut {NoStop}%
	\bibitem [{\citenamefont {Koskinen}\ \emph {et~al.}(2001)\citenamefont
		{Koskinen}, \citenamefont {Manninen}, \citenamefont {Mottelson},\ and\
		\citenamefont {Reimann}}]{Koskinen2001PRB}%
	\BibitemOpen
	\bibfield  {author} {\bibinfo {author} {\bibfnamefont {M.}~\bibnamefont
			{Koskinen}}, \bibinfo {author} {\bibfnamefont {M.}~\bibnamefont {Manninen}},
		\bibinfo {author} {\bibfnamefont {B.}~\bibnamefont {Mottelson}},\ and\
		\bibinfo {author} {\bibfnamefont {S.~M.}\ \bibnamefont {Reimann}},\
	}\bibfield  {title} {\bibinfo {title} {{Rotational and vibrational spectra of
				quantum rings}},\ }\href {https://doi.org/10.1103/PhysRevB.63.205323}
	{\bibfield  {journal} {\bibinfo  {journal} {Phys. Rev. B}\ }\textbf {\bibinfo
			{volume} {63}},\ \bibinfo {pages} {205323} (\bibinfo {year}
		{2001})}\BibitemShut {NoStop}%
	\bibitem [{\citenamefont {Loss}\ and\ \citenamefont
		{Goldbart}(1991)}]{Loss1991PRB}%
	\BibitemOpen
	\bibfield  {author} {\bibinfo {author} {\bibfnamefont {D.}~\bibnamefont
			{Loss}}\ and\ \bibinfo {author} {\bibfnamefont {P.}~\bibnamefont
			{Goldbart}},\ }\bibfield  {title} {\bibinfo {title} {{Period and amplitude
				halving in mesoscopic rings with spin}},\ }\href
	{https://doi.org/10.1103/PhysRevB.43.13762} {\bibfield  {journal} {\bibinfo
			{journal} {Phys. Rev. B}\ }\textbf {\bibinfo {volume} {43}},\ \bibinfo
		{pages} {13762} (\bibinfo {year} {1991})}\BibitemShut {NoStop}%
	\bibitem [{\citenamefont {Reimann}\ \emph {et~al.}(1999)\citenamefont
		{Reimann}, \citenamefont {Koskinen},\ and\ \citenamefont
		{Manninen}}]{Reimann1999PRB}%
	\BibitemOpen
	\bibfield  {author} {\bibinfo {author} {\bibfnamefont {S.~M.}\ \bibnamefont
			{Reimann}}, \bibinfo {author} {\bibfnamefont {M.}~\bibnamefont {Koskinen}},\
		and\ \bibinfo {author} {\bibfnamefont {M.}~\bibnamefont {Manninen}},\
	}\bibfield  {title} {\bibinfo {title} {{End states due to a spin-Peierls
				transition in quantum wires}},\ }\href
	{https://doi.org/10.1103/PhysRevB.59.1613} {\bibfield  {journal} {\bibinfo
			{journal} {Phys. Rev. B}\ }\textbf {\bibinfo {volume} {59}},\ \bibinfo
		{pages} {1613} (\bibinfo {year} {1999})}\BibitemShut {NoStop}%
	\bibitem [{\citenamefont {Reimann}\ \emph {et~al.}(2000)\citenamefont
		{Reimann}, \citenamefont {Koskinen},\ and\ \citenamefont
		{Manninen}}]{Reimann2000PRB}%
	\BibitemOpen
	\bibfield  {author} {\bibinfo {author} {\bibfnamefont {S.~M.}\ \bibnamefont
			{Reimann}}, \bibinfo {author} {\bibfnamefont {M.}~\bibnamefont {Koskinen}},\
		and\ \bibinfo {author} {\bibfnamefont {M.}~\bibnamefont {Manninen}},\
	}\bibfield  {title} {\bibinfo {title} {{Formation of Wigner molecules in
				small quantum dots}},\ }\href {https://doi.org/10.1103/PhysRevB.62.8108}
	{\bibfield  {journal} {\bibinfo  {journal} {Phys. Rev. B}\ }\textbf {\bibinfo
			{volume} {62}},\ \bibinfo {pages} {8108} (\bibinfo {year}
		{2000})}\BibitemShut {NoStop}%
	\bibitem [{\citenamefont {Yannouleas}\ and\ \citenamefont
		{Landman}(2000)}]{Yannouleas2000PRL}%
	\BibitemOpen
	\bibfield  {author} {\bibinfo {author} {\bibfnamefont {C.}~\bibnamefont
			{Yannouleas}}\ and\ \bibinfo {author} {\bibfnamefont {U.}~\bibnamefont
			{Landman}},\ }\bibfield  {title} {\bibinfo {title} {{Collective and
				Independent-Particle Motion in Two-Electron Artificial Atoms}},\ }\href
	{https://doi.org/10.1103/PhysRevLett.85.1726} {\bibfield  {journal} {\bibinfo
			{journal} {Phys. Rev. Lett.}\ }\textbf {\bibinfo {volume} {85}},\ \bibinfo
		{pages} {1726} (\bibinfo {year} {2000})}\BibitemShut {NoStop}%
	\bibitem [{\citenamefont {Dineykhan}\ and\ \citenamefont
		{Nazmitdinov}(1997)}]{Dineykhan1997PRB}%
	\BibitemOpen
	\bibfield  {author} {\bibinfo {author} {\bibfnamefont {M.}~\bibnamefont
			{Dineykhan}}\ and\ \bibinfo {author} {\bibfnamefont {R.~G.}\ \bibnamefont
			{Nazmitdinov}},\ }\bibfield  {title} {\bibinfo {title} {{Two-electron quantum
				dot in a magnetic field: Analytical results}},\ }\href
	{https://doi.org/10.1103/PhysRevB.55.13707} {\bibfield  {journal} {\bibinfo
			{journal} {Phys. Rev. B}\ }\textbf {\bibinfo {volume} {55}},\ \bibinfo
		{pages} {13707} (\bibinfo {year} {1997})}\BibitemShut {NoStop}%
	\bibitem [{\citenamefont {Manninen}\ \emph {et~al.}(2001)\citenamefont
		{Manninen}, \citenamefont {Viefers}, \citenamefont {Koskinen},\ and\
		\citenamefont {Reimann}}]{Manninen2001PRB}%
	\BibitemOpen
	\bibfield  {author} {\bibinfo {author} {\bibfnamefont {M.}~\bibnamefont
			{Manninen}}, \bibinfo {author} {\bibfnamefont {S.}~\bibnamefont {Viefers}},
		\bibinfo {author} {\bibfnamefont {M.}~\bibnamefont {Koskinen}},\ and\
		\bibinfo {author} {\bibfnamefont {S.~M.}\ \bibnamefont {Reimann}},\
	}\bibfield  {title} {\bibinfo {title} {{Many-body spectrum and particle
				localization in quantum dots and finite rotating Bose condensates}},\ }\href
	{https://doi.org/10.1103/PhysRevB.64.245322} {\bibfield  {journal} {\bibinfo
			{journal} {Phys. Rev. B}\ }\textbf {\bibinfo {volume} {64}},\ \bibinfo
		{pages} {245322} (\bibinfo {year} {2001})}\BibitemShut {NoStop}%
	\bibitem [{\citenamefont {Chakraborty}\ and\ \citenamefont
		{Pietilainen}(1994)}]{Chakraborty1994PRB}%
	\BibitemOpen
	\bibfield  {author} {\bibinfo {author} {\bibfnamefont {T.}~\bibnamefont
			{Chakraborty}}\ and\ \bibinfo {author} {\bibfnamefont {P.}~\bibnamefont
			{Pietilainen}},\ }\bibfield  {title} {\bibinfo {title} {{Electron-electron
				interaction and the persistent current in a quantum ring}},\ }\href
	{https://doi.org/10.1103/PhysRevB.50.8460} {\bibfield  {journal} {\bibinfo
			{journal} {Phys. Rev. B}\ }\textbf {\bibinfo {volume} {50}},\ \bibinfo
		{pages} {8460} (\bibinfo {year} {1994})}\BibitemShut {NoStop}%
	\bibitem [{\citenamefont {Kusmartsev}\ \emph {et~al.}(1994)\citenamefont
		{Kusmartsev}, \citenamefont {Weisz}, \citenamefont {Kishore},\ and\
		\citenamefont {Takahashi}}]{Kusmartsev1994PRB}%
	\BibitemOpen
	\bibfield  {author} {\bibinfo {author} {\bibfnamefont {F.~V.}\ \bibnamefont
			{Kusmartsev}}, \bibinfo {author} {\bibfnamefont {J.~F.}\ \bibnamefont
			{Weisz}}, \bibinfo {author} {\bibfnamefont {R.}~\bibnamefont {Kishore}},\
		and\ \bibinfo {author} {\bibfnamefont {M.}~\bibnamefont {Takahashi}},\
	}\bibfield  {title} {\bibinfo {title} {{Strong correlations versus U-center
				pairing and fractional Aharonov-Bohm effect}},\ }\href
	{https://doi.org/10.1103/PhysRevB.49.16234} {\bibfield  {journal} {\bibinfo
			{journal} {Phys. Rev. B}\ }\textbf {\bibinfo {volume} {49}},\ \bibinfo
		{pages} {16234} (\bibinfo {year} {1994})}\BibitemShut {NoStop}%
	\bibitem [{\citenamefont {Viefers}\ \emph {et~al.}(2000)\citenamefont
		{Viefers}, \citenamefont {Singha~Deo}, \citenamefont {Reimann}, \citenamefont
		{Manninen},\ and\ \citenamefont {Koskinen}}]{Viefers2000PRB}%
	\BibitemOpen
	\bibfield  {author} {\bibinfo {author} {\bibfnamefont {S.}~\bibnamefont
			{Viefers}}, \bibinfo {author} {\bibfnamefont {P.}~\bibnamefont {Singha~Deo}},
		\bibinfo {author} {\bibfnamefont {S.~M.}\ \bibnamefont {Reimann}}, \bibinfo
		{author} {\bibfnamefont {M.}~\bibnamefont {Manninen}},\ and\ \bibinfo
		{author} {\bibfnamefont {M.}~\bibnamefont {Koskinen}},\ }\bibfield  {title}
	{\bibinfo {title} {{Current-spin-density-functional study of persistent
				currents in quantum rings}},\ }\href
	{https://doi.org/10.1103/PhysRevB.62.10668} {\bibfield  {journal} {\bibinfo
			{journal} {Phys. Rev. B}\ }\textbf {\bibinfo {volume} {62}},\ \bibinfo
		{pages} {10668} (\bibinfo {year} {2000})}\BibitemShut {NoStop}%
	\bibitem [{\citenamefont {Yerin}\ \emph {et~al.}(2021)\citenamefont {Yerin},
		\citenamefont {Gusynin}, \citenamefont {Sharapov},\ and\ \citenamefont
		{Varlamov}}]{YerinGusyninSharapovVarlamov2021PRB}%
	\BibitemOpen
	\bibfield  {author} {\bibinfo {author} {\bibfnamefont {Y.}~\bibnamefont
			{Yerin}}, \bibinfo {author} {\bibfnamefont {V.~P.}\ \bibnamefont {Gusynin}},
		\bibinfo {author} {\bibfnamefont {S.~G.}\ \bibnamefont {Sharapov}},\ and\
		\bibinfo {author} {\bibfnamefont {A.~A.}\ \bibnamefont {Varlamov}},\
	}\bibfield  {title} {\bibinfo {title} {{Genesis and fading away of persistent
				currents in a Corbino disk geometry}},\ }\href
	{https://doi.org/10.1103/PhysRevB.104.075415} {\bibfield  {journal} {\bibinfo
			{journal} {Phys. Rev. B}\ }\textbf {\bibinfo {volume} {104}},\ \bibinfo
		{pages} {075415} (\bibinfo {year} {2021})}\BibitemShut {NoStop}%
	\bibitem [{\citenamefont {Pereira}\ \emph {et~al.}(2019)\citenamefont
		{Pereira}, \citenamefont {Andrade}, \citenamefont {Filgueiras},\ and\
		\citenamefont {Silva}}]{PereiraAndradeSilva2019AnnPhys}%
	\BibitemOpen
	\bibfield  {author} {\bibinfo {author} {\bibfnamefont {L.~F.~C.}\
			\bibnamefont {Pereira}}, \bibinfo {author} {\bibfnamefont {F.~M.}\
			\bibnamefont {Andrade}}, \bibinfo {author} {\bibfnamefont {C.}~\bibnamefont
			{Filgueiras}},\ and\ \bibinfo {author} {\bibfnamefont {E.~O.}\ \bibnamefont
			{Silva}},\ }\bibfield  {title} {\bibinfo {title} {{Modifications of Electron
				States, Magnetization, and Persistent Current in a Quantum Dot by Controlled
				Curvature}},\ }\href {https://doi.org/10.1002/andp.201900254} {\bibfield
		{journal} {\bibinfo  {journal} {Ann. Phys. (Berlin)}\ }\textbf {\bibinfo
			{volume} {531}},\ \bibinfo {pages} {1900254} (\bibinfo {year}
		{2019})}\BibitemShut {NoStop}%
	\bibitem [{\citenamefont {Pereira}\ \emph {et~al.}(2021)\citenamefont
		{Pereira}, \citenamefont {Andrade}, \citenamefont {Filgueiras},\ and\
		\citenamefont {Silva}}]{PereiraAndradeSilva2021PhysE}%
	\BibitemOpen
	\bibfield  {author} {\bibinfo {author} {\bibfnamefont {L.~F.~C.}\
			\bibnamefont {Pereira}}, \bibinfo {author} {\bibfnamefont {F.~M.}\
			\bibnamefont {Andrade}}, \bibinfo {author} {\bibfnamefont {C.}~\bibnamefont
			{Filgueiras}},\ and\ \bibinfo {author} {\bibfnamefont {E.~O.}\ \bibnamefont
			{Silva}},\ }\bibfield  {title} {\bibinfo {title} {{Study of electronic
				properties, magnetization and persistent currents in a mesoscopic ring by
				controlled curvature}},\ }\href {https://doi.org/10.1016/j.physe.2021.114760}
	{\bibfield  {journal} {\bibinfo  {journal} {Physica E}\ }\textbf {\bibinfo
			{volume} {132}},\ \bibinfo {pages} {114760} (\bibinfo {year}
		{2021})}\BibitemShut {NoStop}%
	\bibitem [{\citenamefont {Pereira}\ \emph
		{et~al.}(2022{\natexlab{a}})\citenamefont {Pereira}, \citenamefont {Andrade},
		\citenamefont {Filgueiras},\ and\ \citenamefont
		{Silva}}]{PereiraAndradeSilva2022FewBody}%
	\BibitemOpen
	\bibfield  {author} {\bibinfo {author} {\bibfnamefont {L.~F.~C.}\
			\bibnamefont {Pereira}}, \bibinfo {author} {\bibfnamefont {F.~M.}\
			\bibnamefont {Andrade}}, \bibinfo {author} {\bibfnamefont {C.}~\bibnamefont
			{Filgueiras}},\ and\ \bibinfo {author} {\bibfnamefont {E.~O.}\ \bibnamefont
			{Silva}},\ }\bibfield  {title} {\bibinfo {title} {{Effects of Curvature on
				the Electronic States of a Two-Dimensional Mesoscopic Ring}},\ }\href
	{https://doi.org/10.1007/s00601-022-01767-9} {\bibfield  {journal} {\bibinfo
			{journal} {Few-Body Syst.}\ }\textbf {\bibinfo {volume} {63}},\ \bibinfo
		{pages} {64} (\bibinfo {year} {2022}{\natexlab{a}})}\BibitemShut {NoStop}%
	\bibitem [{\citenamefont {Pereira}\ \emph
		{et~al.}(2022{\natexlab{b}})\citenamefont {Pereira}, \citenamefont {Cunha},\
		and\ \citenamefont {Silva}}]{PereiraCunhaSilva2022FewBody}%
	\BibitemOpen
	\bibfield  {author} {\bibinfo {author} {\bibfnamefont {L.~F.~C.}\
			\bibnamefont {Pereira}}, \bibinfo {author} {\bibfnamefont {M.~M.}\
			\bibnamefont {Cunha}},\ and\ \bibinfo {author} {\bibfnamefont {E.~O.}\
			\bibnamefont {Silva}},\ }\bibfield  {title} {\bibinfo {title} {{1D Quantum
				Ring: A Toy Model Describing Noninertial Effects on Electronic States,
				Persistent Current and Magnetization}},\ }\href
	{https://doi.org/10.1007/s00601-022-01761-1} {\bibfield  {journal} {\bibinfo
			{journal} {Few-Body Syst.}\ }\textbf {\bibinfo {volume} {63}},\ \bibinfo
		{pages} {58} (\bibinfo {year} {2022}{\natexlab{b}})}\BibitemShut {NoStop}%
	\bibitem [{\citenamefont {Pereira}\ and\ \citenamefont
		{Silva}(2023)}]{PereiraSilva2023AnnPhys}%
	\BibitemOpen
	\bibfield  {author} {\bibinfo {author} {\bibfnamefont {L.~F.~C.}\
			\bibnamefont {Pereira}}\ and\ \bibinfo {author} {\bibfnamefont {E.~O.}\
			\bibnamefont {Silva}},\ }\bibfield  {title} {\bibinfo {title} {{Modification
				of Landau Levels in a 2D Ring Due to Rotation Effects and Edge States}},\
	}\href {https://doi.org/10.1002/andp.202200371} {\bibfield  {journal}
		{\bibinfo  {journal} {Ann. Phys. (Berlin)}\ }\textbf {\bibinfo {volume}
			{535}},\ \bibinfo {pages} {2200371} (\bibinfo {year} {2023})}\BibitemShut
	{NoStop}%
	\bibitem [{\citenamefont {Pereira}\ \emph {et~al.}(2026)\citenamefont
		{Pereira}, \citenamefont {Azevedo},\ and\ \citenamefont
		{Silva}}]{PereiraAzevedoSilva2026PhysE}%
	\BibitemOpen
	\bibfield  {author} {\bibinfo {author} {\bibfnamefont {C.~M.~O.}\
			\bibnamefont {Pereira}}, \bibinfo {author} {\bibfnamefont {F.~d.~S.}\
			\bibnamefont {Azevedo}},\ and\ \bibinfo {author} {\bibfnamefont {E.~O.}\
			\bibnamefont {Silva}},\ }\bibfield  {title} {\bibinfo {title} {{Exact
				treatment of rotation-induced modifications in two-dimensional quantum
				rings}},\ }\href {https://doi.org/10.1016/j.physe.2025.116416} {\bibfield
		{journal} {\bibinfo  {journal} {Physica E}\ }\textbf {\bibinfo {volume}
			{176}},\ \bibinfo {pages} {116416} (\bibinfo {year} {2026})}\BibitemShut
	{NoStop}%
	\bibitem [{\citenamefont {Pereira}\ \emph {et~al.}(2024)\citenamefont
		{Pereira}, \citenamefont {de~Lira},\ and\ \citenamefont
		{Silva}}]{PereiraLiraSilva2024Encyclopedia}%
	\BibitemOpen
	\bibfield  {author} {\bibinfo {author} {\bibfnamefont {L.~F.~C.}\
			\bibnamefont {Pereira}}, \bibinfo {author} {\bibfnamefont {F.~A.~G.}\
			\bibnamefont {de~Lira}},\ and\ \bibinfo {author} {\bibfnamefont {E.~O.}\
			\bibnamefont {Silva}},\ }\bibfield  {title} {\bibinfo {title} {Quantum rings:
			Electronic properties},\ }in\ \href
	{https://doi.org/10.1016/B978-0-323-90800-9.00199-2} {\emph {\bibinfo
			{booktitle} {Encyclopedia of Condensed Matter Physics}}},\ Vol.~\bibinfo
	{volume} {3},\ \bibinfo {editor} {edited by\ \bibinfo {editor} {\bibfnamefont
			{T.}~\bibnamefont {Chakraborty}}}\ (\bibinfo  {publisher} {Academic Press},\
	\bibinfo {year} {2024})\ \bibinfo {edition} {2nd}\ ed.,\ pp.\ \bibinfo
	{pages} {415--425}\BibitemShut {NoStop}%
	\bibitem [{\citenamefont {Meijer}\ \emph {et~al.}(2002)\citenamefont {Meijer},
		\citenamefont {Morpurgo},\ and\ \citenamefont {Klapwijk}}]{Meijer2002PRB}%
	\BibitemOpen
	\bibfield  {author} {\bibinfo {author} {\bibfnamefont {F.~E.}\ \bibnamefont
			{Meijer}}, \bibinfo {author} {\bibfnamefont {A.~F.}\ \bibnamefont
			{Morpurgo}},\ and\ \bibinfo {author} {\bibfnamefont {T.~M.}\ \bibnamefont
			{Klapwijk}},\ }\bibfield  {title} {\bibinfo {title} {{One-dimensional ring in
				the presence of Rashba spin-orbit interaction: Derivation of the correct
				Hamiltonian}},\ }\href {https://doi.org/10.1103/PhysRevB.66.033107}
	{\bibfield  {journal} {\bibinfo  {journal} {Phys. Rev. B}\ }\textbf {\bibinfo
			{volume} {66}},\ \bibinfo {pages} {033107} (\bibinfo {year}
		{2002})}\BibitemShut {NoStop}%
	\bibitem [{\citenamefont {Frustaglia}\ and\ \citenamefont
		{Richter}(2004)}]{Frustaglia2004PRB}%
	\BibitemOpen
	\bibfield  {author} {\bibinfo {author} {\bibfnamefont {D.}~\bibnamefont
			{Frustaglia}}\ and\ \bibinfo {author} {\bibfnamefont {K.}~\bibnamefont
			{Richter}},\ }\bibfield  {title} {\bibinfo {title} {{Spin interference
				effects in ring conductors subject to Rashba coupling}},\ }\href
	{https://doi.org/10.1103/PhysRevB.69.235310} {\bibfield  {journal} {\bibinfo
			{journal} {Phys. Rev. B}\ }\textbf {\bibinfo {volume} {69}},\ \bibinfo
		{pages} {235310} (\bibinfo {year} {2004})}\BibitemShut {NoStop}%
	\bibitem [{\citenamefont {Molnar}\ \emph {et~al.}(2004)\citenamefont {Molnar},
		\citenamefont {Peeters},\ and\ \citenamefont {Vasilopoulos}}]{Molnar2004PRB}%
	\BibitemOpen
	\bibfield  {author} {\bibinfo {author} {\bibfnamefont {B.}~\bibnamefont
			{Molnar}}, \bibinfo {author} {\bibfnamefont {F.~M.}\ \bibnamefont
			{Peeters}},\ and\ \bibinfo {author} {\bibfnamefont {P.}~\bibnamefont
			{Vasilopoulos}},\ }\bibfield  {title} {\bibinfo {title} {{Spin-dependent
				magnetotransport through a ring due to spin-orbit interaction}},\ }\href
	{https://doi.org/10.1103/PhysRevB.69.155335} {\bibfield  {journal} {\bibinfo
			{journal} {Phys. Rev. B}\ }\textbf {\bibinfo {volume} {69}},\ \bibinfo
		{pages} {155335} (\bibinfo {year} {2004})}\BibitemShut {NoStop}%
	\bibitem [{\citenamefont {Splettstoesser}\ \emph {et~al.}(2003)\citenamefont
		{Splettstoesser}, \citenamefont {Governale},\ and\ \citenamefont
		{Zulicke}}]{Splettstoesser2003PRB}%
	\BibitemOpen
	\bibfield  {author} {\bibinfo {author} {\bibfnamefont {J.}~\bibnamefont
			{Splettstoesser}}, \bibinfo {author} {\bibfnamefont {M.}~\bibnamefont
			{Governale}},\ and\ \bibinfo {author} {\bibfnamefont {U.}~\bibnamefont
			{Zulicke}},\ }\bibfield  {title} {\bibinfo {title} {{Persistent current in
				ballistic mesoscopic rings with Rashba spin-orbit coupling}},\ }\href
	{https://doi.org/10.1103/PhysRevB.68.165341} {\bibfield  {journal} {\bibinfo
			{journal} {Phys. Rev. B}\ }\textbf {\bibinfo {volume} {68}},\ \bibinfo
		{pages} {165341} (\bibinfo {year} {2003})}\BibitemShut {NoStop}%
	\bibitem [{\citenamefont {Souma}\ and\ \citenamefont
		{Nikolic}(2005)}]{SoumaNikolic2005PRL}%
	\BibitemOpen
	\bibfield  {author} {\bibinfo {author} {\bibfnamefont {S.}~\bibnamefont
			{Souma}}\ and\ \bibinfo {author} {\bibfnamefont {B.~K.}\ \bibnamefont
			{Nikolic}},\ }\bibfield  {title} {\bibinfo {title} {{Spin Hall Current Driven
				by Quantum Interferences in Mesoscopic Rashba Rings}},\ }\href
	{https://doi.org/10.1103/PhysRevLett.94.106602} {\bibfield  {journal}
		{\bibinfo  {journal} {Phys. Rev. Lett.}\ }\textbf {\bibinfo {volume} {94}},\
		\bibinfo {pages} {106602} (\bibinfo {year} {2005})}\BibitemShut {NoStop}%
	\bibitem [{\citenamefont {K\"{o}nig}\ \emph {et~al.}(2006)\citenamefont
		{K\"{o}nig}, \citenamefont {Tschetschetkin}, \citenamefont {Hankiewicz},
		\citenamefont {Sinova}, \citenamefont {Hock}, \citenamefont {Daumer},
		\citenamefont {Sch\"{a}fer}, \citenamefont {Becker}, \citenamefont
		{Buhmann},\ and\ \citenamefont {Molenkamp}}]{Konig2006PRL}%
	\BibitemOpen
	\bibfield  {author} {\bibinfo {author} {\bibfnamefont {M.}~\bibnamefont
			{K\"{o}nig}}, \bibinfo {author} {\bibfnamefont {A.}~\bibnamefont
			{Tschetschetkin}}, \bibinfo {author} {\bibfnamefont {E.~M.}\ \bibnamefont
			{Hankiewicz}}, \bibinfo {author} {\bibfnamefont {J.}~\bibnamefont {Sinova}},
		\bibinfo {author} {\bibfnamefont {V.}~\bibnamefont {Hock}}, \bibinfo {author}
		{\bibfnamefont {V.}~\bibnamefont {Daumer}}, \bibinfo {author} {\bibfnamefont
			{M.}~\bibnamefont {Sch\"{a}fer}}, \bibinfo {author} {\bibfnamefont {C.~R.}\
			\bibnamefont {Becker}}, \bibinfo {author} {\bibfnamefont {H.}~\bibnamefont
			{Buhmann}},\ and\ \bibinfo {author} {\bibfnamefont {L.~W.}\ \bibnamefont
			{Molenkamp}},\ }\bibfield  {title} {\bibinfo {title} {{Direct Observation of
				the Aharonov-Casher Phase}},\ }\href
	{https://doi.org/10.1103/PhysRevLett.96.076804} {\bibfield  {journal}
		{\bibinfo  {journal} {Phys. Rev. Lett.}\ }\textbf {\bibinfo {volume} {96}},\
		\bibinfo {pages} {076804} (\bibinfo {year} {2006})}\BibitemShut {NoStop}%
	\bibitem [{\citenamefont {Grbi\'{c}}\ \emph {et~al.}(2007)\citenamefont
		{Grbi\'{c}}, \citenamefont {Leturcq}, \citenamefont {Ihn}, \citenamefont
		{Ensslin}, \citenamefont {Reuter},\ and\ \citenamefont
		{Wieck}}]{Grbic2007PRL}%
	\BibitemOpen
	\bibfield  {author} {\bibinfo {author} {\bibfnamefont {B.}~\bibnamefont
			{Grbi\'{c}}}, \bibinfo {author} {\bibfnamefont {R.}~\bibnamefont {Leturcq}},
		\bibinfo {author} {\bibfnamefont {T.}~\bibnamefont {Ihn}}, \bibinfo {author}
		{\bibfnamefont {K.}~\bibnamefont {Ensslin}}, \bibinfo {author} {\bibfnamefont
			{D.}~\bibnamefont {Reuter}},\ and\ \bibinfo {author} {\bibfnamefont {A.~D.}\
			\bibnamefont {Wieck}},\ }\bibfield  {title} {\bibinfo {title} {{Aharonov-Bohm
				Oscillations in the Presence of Strong Spin-Orbit Interactions}},\ }\href
	{https://doi.org/10.1103/PhysRevLett.99.176803} {\bibfield  {journal}
		{\bibinfo  {journal} {Phys. Rev. Lett.}\ }\textbf {\bibinfo {volume} {99}},\
		\bibinfo {pages} {176803} (\bibinfo {year} {2007})}\BibitemShut {NoStop}%
	\bibitem [{\citenamefont {Tan}\ and\ \citenamefont
		{Inkson}(1996)}]{TanInkson1996}%
	\BibitemOpen
	\bibfield  {author} {\bibinfo {author} {\bibfnamefont {W.-C.}\ \bibnamefont
			{Tan}}\ and\ \bibinfo {author} {\bibfnamefont {J.~C.}\ \bibnamefont
			{Inkson}},\ }\bibfield  {title} {\bibinfo {title} {{Landau quantization and
				the Aharonov--Bohm effect in a two-dimensional ring}},\ }\href
	{https://doi.org/10.1103/PhysRevB.53.6947} {\bibfield  {journal} {\bibinfo
			{journal} {Phys. Rev. B}\ }\textbf {\bibinfo {volume} {53}},\ \bibinfo
		{pages} {6947} (\bibinfo {year} {1996})}\BibitemShut {NoStop}%
	\bibitem [{\citenamefont {Tan}\ and\ \citenamefont
		{Inkson}(1999)}]{TanInkson1999}%
	\BibitemOpen
	\bibfield  {author} {\bibinfo {author} {\bibfnamefont {W.-C.}\ \bibnamefont
			{Tan}}\ and\ \bibinfo {author} {\bibfnamefont {J.~C.}\ \bibnamefont
			{Inkson}},\ }\bibfield  {title} {\bibinfo {title} {{Magnetization, persistent
				currents, and their relation in quantum rings and dots}},\ }\href
	{https://doi.org/10.1103/PhysRevB.60.5626} {\bibfield  {journal} {\bibinfo
			{journal} {Phys. Rev. B}\ }\textbf {\bibinfo {volume} {60}},\ \bibinfo
		{pages} {5626} (\bibinfo {year} {1999})}\BibitemShut {NoStop}%
	\bibitem [{\citenamefont {de~Lira}\ \emph {et~al.}(2024)\citenamefont
		{de~Lira}, \citenamefont {Pereira},\ and\ \citenamefont
		{Silva}}]{LiraPereiraSilva2024PhysE}%
	\BibitemOpen
	\bibfield  {author} {\bibinfo {author} {\bibfnamefont {F.~A.~G.}\
			\bibnamefont {de~Lira}}, \bibinfo {author} {\bibfnamefont {L.~F.~C.}\
			\bibnamefont {Pereira}},\ and\ \bibinfo {author} {\bibfnamefont {E.~O.}\
			\bibnamefont {Silva}},\ }\bibfield  {title} {\bibinfo {title} {{Study on the
				effects of anisotropic effective mass on electronic properties, magnetization
				and persistent current in semiconductor quantum ring with conical
				geometry}},\ }\href {https://doi.org/10.1016/j.physe.2024.115898} {\bibfield
		{journal} {\bibinfo  {journal} {Physica E}\ }\textbf {\bibinfo {volume}
			{158}},\ \bibinfo {pages} {115898} (\bibinfo {year} {2024})}\BibitemShut
	{NoStop}%
	\bibitem [{\citenamefont {Lima}\ \emph {et~al.}(2023)\citenamefont {Lima},
		\citenamefont {Azevedo}, \citenamefont {Pereira}, \citenamefont
		{Filgueiras},\ and\ \citenamefont {Silva}}]{LimaAzevedoPereira2023AnnPhys}%
	\BibitemOpen
	\bibfield  {author} {\bibinfo {author} {\bibfnamefont {D.~F.}\ \bibnamefont
			{Lima}}, \bibinfo {author} {\bibfnamefont {F.~d.~S.}\ \bibnamefont
			{Azevedo}}, \bibinfo {author} {\bibfnamefont {L.~F.~C.}\ \bibnamefont
			{Pereira}}, \bibinfo {author} {\bibfnamefont {C.}~\bibnamefont
			{Filgueiras}},\ and\ \bibinfo {author} {\bibfnamefont {E.~O.}\ \bibnamefont
			{Silva}},\ }\bibfield  {title} {\bibinfo {title} {{Optical and electronic
				properties of a two-dimensional quantum ring under rotating effects}},\
	}\href {https://doi.org/10.1016/j.aop.2023.169547} {\bibfield  {journal}
		{\bibinfo  {journal} {Ann. Phys. (N.Y.)}\ }\textbf {\bibinfo {volume}
			{459}},\ \bibinfo {pages} {169547} (\bibinfo {year} {2023})}\BibitemShut
	{NoStop}%
	\bibitem [{\citenamefont {de~Lira}\ \emph {et~al.}(2026)\citenamefont
		{de~Lira}, \citenamefont {Silva},\ and\ \citenamefont
		{Santangelo}}]{LiraSilvaSantangelo2026PhysE}%
	\BibitemOpen
	\bibfield  {author} {\bibinfo {author} {\bibfnamefont {F.~A.~G.}\
			\bibnamefont {de~Lira}}, \bibinfo {author} {\bibfnamefont {E.~O.}\
			\bibnamefont {Silva}},\ and\ \bibinfo {author} {\bibfnamefont {C.~D.}\
			\bibnamefont {Santangelo}},\ }\bibfield  {title} {\bibinfo {title}
		{{Magnetoresistance and electric current oscillations induced by geometry in
				a two-dimensional quantum ring}},\ }\href
	{https://doi.org/10.1016/j.physe.2026.116530} {\bibfield  {journal} {\bibinfo
			{journal} {Physica E}\ }\textbf {\bibinfo {volume} {181}},\ \bibinfo {pages}
		{116530} (\bibinfo {year} {2026})}\BibitemShut {NoStop}%
	\bibitem [{\citenamefont {de~Lira}\ and\ \citenamefont
		{Silva}(2023)}]{LiraSilva2023PhysE}%
	\BibitemOpen
	\bibfield  {author} {\bibinfo {author} {\bibfnamefont {F.~A.~G.}\
			\bibnamefont {de~Lira}}\ and\ \bibinfo {author} {\bibfnamefont {E.~O.}\
			\bibnamefont {Silva}},\ }\bibfield  {title} {\bibinfo {title} {{Theoretical
				approach for the description of a single quantum dot using geometry}},\
	}\href {https://doi.org/10.1016/j.physe.2022.115617} {\bibfield  {journal}
		{\bibinfo  {journal} {Physica E}\ }\textbf {\bibinfo {volume} {147}},\
		\bibinfo {pages} {115617} (\bibinfo {year} {2023})}\BibitemShut {NoStop}%
	\bibitem [{\citenamefont {Pereira}\ and\ \citenamefont
		{Silva}(2024)}]{PereiraSilva2024QuantumRep}%
	\BibitemOpen
	\bibfield  {author} {\bibinfo {author} {\bibfnamefont {L.~F.~C.}\
			\bibnamefont {Pereira}}\ and\ \bibinfo {author} {\bibfnamefont {E.~O.}\
			\bibnamefont {Silva}},\ }\bibfield  {title} {\bibinfo {title} {{Thermodynamic
				Properties of an Electron Gas in a Two-Dimensional Quantum Dot: An Approach
				Using Density of States}},\ }\href {https://doi.org/10.3390/quantum6040040}
	{\bibfield  {journal} {\bibinfo  {journal} {Quantum Rep.}\ }\textbf {\bibinfo
			{volume} {6}},\ \bibinfo {pages} {664} (\bibinfo {year} {2024})}\BibitemShut
	{NoStop}%
	\bibitem [{\citenamefont {Pereira}\ and\ \citenamefont
		{Silva}(2025)}]{PereiraSilva2025AnnPhysThermo}%
	\BibitemOpen
	\bibfield  {author} {\bibinfo {author} {\bibfnamefont {L.~F.~C.}\
			\bibnamefont {Pereira}}\ and\ \bibinfo {author} {\bibfnamefont {E.~O.}\
			\bibnamefont {Silva}},\ }\bibfield  {title} {\bibinfo {title} {{Effects of
				rotation on the thermo-magnetic properties of a quantum dot}},\ }\href
	{https://doi.org/10.1016/j.aop.2025.170098} {\bibfield  {journal} {\bibinfo
			{journal} {Ann. Phys. (N.Y.)}\ }\textbf {\bibinfo {volume} {480}},\ \bibinfo
		{pages} {170098} (\bibinfo {year} {2025})}\BibitemShut {NoStop}%
	\bibitem [{\citenamefont {Sivan}\ and\ \citenamefont
		{Imry}(1988)}]{SivanImry1988PRL}%
	\BibitemOpen
	\bibfield  {author} {\bibinfo {author} {\bibfnamefont {U.}~\bibnamefont
			{Sivan}}\ and\ \bibinfo {author} {\bibfnamefont {Y.}~\bibnamefont {Imry}},\
	}\bibfield  {title} {\bibinfo {title} {{de Haas--van Alphen and
				Aharonov--Bohm-type Persistent Current Oscillations in Singly Connected
				Quantum Dots}},\ }\href {https://doi.org/10.1103/PhysRevLett.61.1001}
	{\bibfield  {journal} {\bibinfo  {journal} {Phys. Rev. Lett.}\ }\textbf
		{\bibinfo {volume} {61}},\ \bibinfo {pages} {1001} (\bibinfo {year}
		{1988})}\BibitemShut {NoStop}%
\end{thebibliography}
\end{document}